\documentstyle{mn}
 \def\LOADED{\relax}

 \ifx\BoxedEPSFLoaded\LOADED
  \immediate\write16{}
  \immediate\write16{  BoxedEPSF macros already defined!}
  \immediate\write16{}
   
 \fi
 
 \let\BoxedEPSFLoaded\LOADED
 
 \chardef\CatAt\the\catcode`\@
 \chardef\CatColon\the\catcode`\:
 \catcode`\@=11
 \catcode`\:=12

 \let\wlog@ld\wlog 
 \def\wlog#1{\relax} 

 \newif\ifIN@
 \newdimen\XShift@ \newdimen\YShift@ 
 \newtoks\Realtoks
 
 \newdimen\Wd@ \newdimen\Ht@
 \newdimen\Wd@@ \newdimen\Ht@@
 \newdimen\TT@
 \newdimen\LT@
 \newdimen\BT@
 \newdimen\RT@
 \newdimen\XSlide@ \newdimen\YSlide@ 
 \newdimen\TheScale  
 \newdimen\FigScale  
 \newdimen\ForcedDim@@

 \newtoks\EPSFDirectorytoks@
 \newtoks\EPSFNametoks@
 \newtoks\BdBoxtoks@
  
 \newif\ifNotIn@
 \newif\ifForcedDim@
 \newif\ifForcedHeight@
 \newif\ifPSOrigin

 \newread\EPSFile@

 \newif\ifIN@\def\IN@{\expandafter\INN@\expandafter}
  \long\def\INN@0#1@#2@{\long\def\NI@##1#1##2##3\ENDNI@
    {\ifx\m@rker##2\IN@false\else\IN@true\fi}%
     \expandafter\NI@#2@@#1\m@rker\ENDNI@}
  \def\m@rker{\m@@rker}

  \newtoks\Initialtoks@  \newtoks\Terminaltoks@
  \def\SPLIT@{\expandafter\SPLITT@\expandafter}
  \def\SPLITT@0#1@#2@{\def\TTILPS@##1#1##2@{%
     \Initialtoks@{##1}\Terminaltoks@{##2}}\expandafter\TTILPS@#2@}

  \newtoks\Trimtoks@

 \def\ForeTrim@{\expandafter\ForeTrim@@\expandafter}
 \def\ForePrim@0 #1@{\Trimtoks@{#1}}
 \def\ForeTrim@@0#1@{\IN@0\m@rker. @\m@rker.#1@%
     \ifIN@\ForePrim@0#1@%
     \else\Trimtoks@\expandafter{#1}\fi}

  \def\Trim@0#1@{%
      \ForeTrim@0#1@%
      \IN@0 @\the\Trimtoks@ @%
        \ifIN@ 
             \SPLIT@0 @\the\Trimtoks@ @\Trimtoks@\Initialtoks@
             \IN@0\the\Terminaltoks@ @ @%
                 \ifIN@
                 \else \Trimtoks@ {FigNameWithSpace}%
                 \fi
        \fi
      }

   \newtoks\pt@ks
   \def \getpt@ks 0.0#1@{\pt@ks{#1}}
  \newtoks\Realtoks
  \def\Real#1{%
    \dimen2=#1%
      \SPLIT@0\the\pt@ks @\the\dimen2@
       \Realtoks=\Initialtoks@
            }

   \newdimen\Product
   \def\Mult#1#2{%
     \dimen4=#1\relax
     \dimen6=#2%
     \Real{\dimen4}%
     \Product=\the\Realtoks\dimen6%
        }

 \newdimen\Inverse
 \newdimen\hmxdim@ \hmxdim@=8192pt
 \def\Invert#1{%
  \Inverse=\hmxdim@
  \dimen0=#1%
  \divide\Inverse \dimen0%
  \multiply\Inverse 8}

   \def\Rescale#1#2#3{
              \divide #1 by 100\relax
              \dimen2=#3\divide\dimen2 by 100 \Invert{\dimen2}%
              \Mult{#1}{#2}%
              \Mult\Product\Inverse 
              #1=\Product}

  \def\Scale#1{\dimen0=\TheScale %
      \divide #1 by  1280 
      \divide \dimen0 by 5120 %
      \multiply#1 by \dimen0 
      \divide#1 by 10   
     }
 
 \newbox\scrunchbox

 \def\Scrunched#1{{\setbox\scrunchbox\hbox{#1}%
   \wd\scrunchbox=0pt
   \ht\scrunchbox=0pt
   \dp\scrunchbox=0pt
   \box\scrunchbox}}

 \def\Shifted@#1{%
   \vbox {\kern-\YShift@
       \hbox {\kern\XShift@\hbox{#1}\kern-\XShift@}%
           \kern\YShift@}}

 \def\cBoxedEPSF#1{{}\leavevmode 
   \ReadNameAndScale@{#1}%
   \SetEPSFSpec@
   \ReadEPSFile@ \ReadBdB@x  
     \TrimFigDims@ 
     \CalculateFigScale@  
     \ScaleFigDims@
     \SetInkShift@
   \hbox{$\mathsurround=0pt\relax
         \vcenter{\hbox{%
             \FrameSpider{\hskip-.4pt\vrule}%
             \vbox to \Ht@{\offinterlineskip\parindent=\z@%
                \FrameSpider{\vskip-.4pt\hrule}\vfil 
                \hbox to \Wd@{\hfil}%
                \vfil
                \InkShift@{\EPSFSpecial{\EPSFSpec@}{\FigSc@leReal}}%
             \FrameSpider{\hrule\vskip-.4pt}}%
         \FrameSpider{\vrule\hskip-.4pt}}}%
     $}%
    \CleanRegisters@ 
    }
 
 \def\tBoxedEPSF#1{\setbox4\hbox{\cBoxedEPSF{#1}}%
     \setbox4\hbox{\raise -\ht4 \hbox{\box4}}%
     \box4
      }

 \def\bBoxedEPSF#1{\setbox4\hbox{\cBoxedEPSF{#1}}%
     \setbox4\hbox{\raise \dp4 \hbox{\box4}}%
     \box4
      }

  \let\BoxedEPSF\cBoxedEPSF

  \def\gLinefigure[#1scaled#2]_#3{%
        \BoxedEPSF{#3 scaled #2}}
    
  \def\EPSFxsize{\afterassignment\ForceW@\ForcedDim@@}
      \def\ForceW@{\ForcedDim@true\ForcedHeight@false}
  
  \def\EPSFysize{\afterassignment\ForceH@\ForcedDim@@}
      \def\ForceH@{\ForcedDim@true\ForcedHeight@true}

 \def\ReadNameAndScale@#1{\IN@0 scaled@#1@
   \ifIN@\ReadNameAndScale@@0#1@%
   \else \ReadNameAndScale@@0#1 scaled\DefaultMilScale @
   \fi}
  
 \def\ReadNameAndScale@@0#1scaled#2@{
    \Trim@0#1@%
    \EPSFNametoks@\expandafter{\the\Trimtoks@}%
    \FigScale=#2 pt%
     }
 
 \def\SetDefaultEPSFScale#1{%
      \global\def\DefaultMilScale{#1}}

 \SetDefaultEPSFScale{1000}

 \def \SetBogusBbox@{%
     \global\BdBoxtoks@{ BoundingBox:0 0 100 100 }%
     \global\def\BdBoxLine@{ BoundingBox:0 0 100 100 }%
     }

 \def\ReadEPSFile@{
     \openin\EPSFile@\EPSFSpec@
     \relax  
  \ifeof\EPSFile@
     \SetBogusBbox@
     \immediate\write16{}%
     \message{ *** EPS FILE  }%
     \message\expandafter{\the\EPSFNametoks@}%
     \message{ NOT FOUND!  }%
     \immediate\write16{}\relax%
  \else
   \begingroup
   \catcode`\%=12\catcode`\:=12\catcode`\\=12 
   \NotIn@true 
    \loop   
      \ifeof\EPSFile@\NotIn@false 
        \SetBogusBbox@
        \immediate\write16{}%
        \message{ *** BoundingBox not found in }%
        \message\expandafter{\the\EPSFNametoks@\space *** }%
        \immediate\write16{}%
      \else\global\read\EPSFile@ to \BdBoxLine@
      \fi
      \global\BdBoxtoks@\expandafter{\BdBoxLine@}%
      \IN@0BoundingBox:@\the\BdBoxtoks@ @%
      \ifIN@\NotIn@false\fi%
    \ifNotIn@\repeat
   \endgroup\relax
  \fi
  \closein\EPSFile@ 
   }

  \def\ReadBdB@x{
   \expandafter\ReadBdB@x@\the\BdBoxtoks@ @}
  
  \def\ReadBdB@x@#1BoundingBox:#2@{
    \ForeTrim@0#2@%
    \expandafter\ReadBdB@x@@\the\Trimtoks@ @%
   }
    
  \newtoks\LLXtoks@  
  \newtoks\LLYtoks@  

  \def\ReadBdB@x@@#1 #2 #3 #4@{
      \Wd@=#3bp\advance\Wd@ by -#1bp%
      \Ht@=#4bp\advance\Ht@ by-#2bp%
       \Wd@@=\Wd@ \Ht@@=\Ht@ 
       \LLXtoks@={#1}\LLYtoks@={#2}
      \ifPSOrigin\XShift@=-#1bp\YShift@=-#2bp\fi 
     }

   \def\G@bbl@#1{}
   \bgroup
     \global\edef\OtherB@ckslash{\expandafter\G@bbl@\string\\}
   \egroup

  \def\SetEPSFDirectory{
           \bgroup\catcode`\:=12\relax
           \SetEPSFDirectory@}

 \def\SetEPSFDirectory@#1{
    \Trim@0#1@
    \global\EPSFDirectorytoks@\expandafter{\the\Trimtoks@ }\relax
    \egroup}

 \def\SetEPSFSpec@{%
     \bgroup
     \let\\=\OtherB@ckslash
     \global\edef\EPSFSpec@{\the\EPSFDirectorytoks@\the\EPSFNametoks@}%
     \global\edef\EPSFSpec@{\EPSFSpec@}%
     \egroup}

 \def\TrimTop#1{\advance\TT@ by #1}
 \def\TrimLeft#1{\advance\LT@ by #1}
 \def\TrimBottom#1{\advance\BT@ by #1}
 \def\TrimRight#1{\advance\RT@ by #1}

 \def\TrimFigDims@{%
    \advance\Wd@ by -\LT@ 
    \advance\Wd@ by -\RT@ \RT@=\z@
    \advance\Ht@ by -\TT@ \TT@=\z@
    \advance\Ht@ by -\BT@ 
    }

  \def\ForceWidth#1{\ForcedDim@true
       \ForcedDim@@#1\ForcedHeight@false}
  
  \def\ForceHeight#1{\ForcedDim@true
       \ForcedDim@@=#1\ForcedHeight@true}
  
  \def\epsfxsize{\afterassignment\ForceW@\ForcedDim@@}
      \def\ForceW@{\ForcedDim@true\ForcedHeight@false}
  
  \def\epsfysize{\afterassignment\ForceH@\ForcedDim@@}
      \def\ForceH@{\ForcedDim@true\ForcedHeight@true}
  
  \def\CalculateFigScale@{%
     \ifForcedDim@\FigScale=1000pt
           \ifForcedHeight@
                \Rescale\FigScale\ForcedDim@@\Ht@
           \else
                \Rescale\FigScale\ForcedDim@@\Wd@
           \fi
     \fi
     \Real{\FigScale}%
     \edef\FigSc@leReal{\the\Realtoks}%
     }
   
  \def\ScaleFigDims@{\TheScale=\FigScale
      \ifForcedDim@
           \ifForcedHeight@ \Ht@=\ForcedDim@@  \Scale\Wd@
           \else \Wd@=\ForcedDim@@ \Scale\Ht@
           \fi
      \else \Scale\Wd@\Scale\Ht@        
      \fi
      \ForcedDim@false
      \Scale\LT@\Scale\BT@  
      \Scale\XShift@\Scale\YShift@
      }
      
 \def\ShowReservedBoxes{\gdef\FrameSpider##1{##1}}
 \let\HideDisplacementBoxes\HideReservedBoxes  

  \ShowReservedBoxes
 
 \def\hSlide#1{\advance\XSlide@ by #1}
 \def\vSlide#1{\advance\YSlide@ by #1}
 
  \def\SetInkShift@{%
            \advance\XShift@ by -\LT@
            \advance\XShift@ by \XSlide@
            \advance\YShift@ by -\BT@
            \advance\YShift@ by -\YSlide@
             }
  \def\InkShift@#1{\Shifted@{\Scrunched{#1}}}
 
  \def\CleanRegisters@{%
      \globaldefs=1\relax
        \XShift@=\z@\YShift@=\z@\XSlide@=\z@\YSlide@=\z@
        \TT@=\z@\LT@=\z@\BT@=\z@\RT@=\z@
      \globaldefs=0\relax}

 \def\SetTexturesEPSFSpecial{\PSOriginfalse
  \gdef\EPSFSpecial##1##2{\relax
    \edef\specialthis{##2}%
    \SPLIT@0.@\specialthis.@\relax
    \special{illustration ##1 scaled
                        \the\Initialtoks@}}}
 
  \def\SetUnixCoopEPSFSpecial{\PSOrigintrue 
   \gdef\EPSFSpecial##1##2{%
      \dimen4=##2pt
      \divide\dimen4 by 1000\relax
      \Real{\dimen4}
      \edef\Aux@{\the\Realtoks}%
      \includegraphics{##1\space}}}

  \def\SetBechtolsheimRokickiEPSFSpecial{\PSOrigintrue 
   \gdef\EPSFSpecial##1##2{%
      \dimen4=##2pt
      \divide\dimen4 by 1000\relax
      \Real{\dimen4}
      \edef\Aux@{\the\Realtoks}%
      \special{ps: psfiginit}%
      \special{ps: literal 1 1 0 0 1 1 startTexFig
           \the\mag\space 1000 div \Aux@\space mul 
           \the\mag\space 1000 div \Aux@\space mul scale}%
      \special{ps: include  ##1}%
      \special{ps: literal endTexFig}%
        }}

  \def\SetLisEPSFSpecial{\PSOrigintrue 
   \gdef\EPSFSpecial##1##2{%
      \dimen4=##2pt
      \divide\dimen4 by 1000\relax
      \Real{\dimen4}
      \edef\Aux@{\the\Realtoks}%
      \special{pstext="1 1 0 0 1 1 startTexFig\space
           \the\mag\space 1000 div \Aux@\space mul 
           \the\mag\space 1000 div \Aux@\space mul scale}%
      \includegraphics{##1}%
      \special{pstext=endTexFig}%
        }}

  \def\SetRokickiEPSFSpecial{\PSOrigintrue 
   \gdef\EPSFSpecial##1##2{%
      \dimen4=##2pt
      \divide\dimen4 by 10\relax
      \Real{\dimen4}
      \edef\Aux@{\the\Realtoks}%
      \includegraphics{##1}}}

  \def\SetInlineRokickiEPSFSpecial{\PSOrigintrue 
   \gdef\EPSFSpecial##1##2{%
      \dimen4=##2pt
      \divide\dimen4 by 1000\relax
      \Real{\dimen4}
      \edef\Aux@{\the\Realtoks}%
      \special{ps::[begin] 1 1 0 0 1 1 startTexFig\space
           \the\mag\space 1000 div \Aux@\space mul 
           \the\mag\space 1000 div \Aux@\space mul scale}%
      \special{ps: plotfile ##1}%
      \special{ps::[end] endTexFig}%
        }}

  \def\SetOzTeXEPSFSpecial{\PSOriginfalse 
  \gdef\EPSFSpecial##1##2{
     \special{##1\space 
       ##2 1000 div \the\mag\space 1000 div mul
       ##2 1000 div \the\mag\space 1000 div mul scale
       \the\LLXtoks@\space neg \the\LLYtoks@\space neg translate
             }}} 
 
 \def\SetArborEPSFSpecial{\PSOriginfalse 
   \gdef\EPSFSpecial##1##2{%
     \edef\specialthis{##2}%
     \SPLIT@0.@\specialthis.@\relax 
     \special{ps: epsfile ##1\space \the\Initialtoks@}}}

 \def\SetClarkEPSFSpecial{\PSOriginfalse 
   \gdef\EPSFSpecial##1##2{%
     \Rescale {\Wd@@}{##2pt}{1000pt}%
     \Rescale {\Ht@@}{##2pt}{1000pt}%
     \special{dvitops: import 
           ##1\space\the\Wd@@\space\the\Ht@@}}}

 \def\SetStandardEPSFSpecial{%
   \gdef\EPSFSpecial##1##2{%
     \immediate\write16{}
     \immediate\write16{%
       **** Sorry! There is still no standard for \string%
       \special \space EPSF integration *****}%
     \immediate\write16{%
      --- So you will have to identify your driver using a command}%
     \immediate\write16{%
      --- of the form \string\Set...EPSFSpecial, in order to get}%
     \immediate\write16{%
      --- your graphics to print.  See BoxedEPSF.doc.}%
     \immediate\write16{}
     \KillEPSFSpecial
     }}

  \def\KillEPSFSpecial{\gdef\EPSFSpecial##1##2{}}

  \SetStandardEPSFSpecial 
 
 \let\wlog\wlog@ld 

 \catcode`\@=\CatAt
 \catcode`\:=\CatColon

\newcommand{\mc}{\multicolumn}
\newcommand{\arc}{$^{\prime\prime}$}
\newcommand{\aip}{{\small ${\cal AIPS}$}}
\newcommand{\gtsim}{\mbox{{\raisebox{-0.4ex}{$\stackrel{>}{{\scriptstyle\sim}}
$}}}}
\newcommand{\ltsim}{\mbox{{\raisebox{-0.4ex}{$\stackrel{<}{{\scriptstyle\sim}}
$}}}}
\newcommand{\ph}{\phantom{1}}
\SetEPSFDirectory{./}
\SetRokickiEPSFSpecial
\HideDisplacementBoxes

\begin{document}
\title[The 6C* sample I: radio data]{A sample of 6C radio sources
designed to find objects at redshift $> 4$: I --- the radio data}

\author[Blundell et al.]{Katherine M.\,Blundell$^1$, Steve Rawlings$^1$, 
Stephen A.\,Eales$^{2}$\\ 
\vspace{-1.5mm}\\
{\LARGE Gregory B.\,Taylor$^3$ \& Alistair D.\,Bradley$^{1,4}$}\\
$^1$Astrophysics, Department of Physics, Keble Road, Oxford, OX1 3RH, U.K. \\
$^2$Department of Physics and Astronomy, University of Cardiff, Wales,
U.K. \\
$^3$National Radio Astronomy Observatory, Socorro, NM 87801, U.S.A. \\
$^4$Present address: Superconductivity Research Group, University of Birmingham, B15 2TT, U.K. \\
}
     
\maketitle

\begin{abstract}
We describe the selection of a sample of 34 radio sources from the 6C
survey (Hales, Baldwin \& Warner 1993) from a region of sky covering
$0.133 ~ \rm sr$. The selection criteria for this sample, hereafter
called 6C*, were chosen to optimise the chances of finding radio
galaxies at redshift $z > 4$. Optical follow-up observations have
already led to the discovery of the most distant known radio galaxy at
$z = 4.41$ (Rawlings et al. 1996). We present VLA radio maps and
derive radio spectra for all the 6C* objects.

\end{abstract}


\section{Introduction}

The primary goal of this study was to find radio galaxies with
redshift $> 4$.  Such systems probe the Universe about $10^{9}$ years
after the Big Bang (we assume throughout that ${\em
H}_{\circ}~=~50~{\rm km~s^{-1}~Mpc^{-1}}$, ${\em q_{\circ}}~=~0.5$ and
that the cosmological constant is zero) and may inform on the early
evolutionary history of elliptical galaxies (e.g., Eales \& Rawlings
1996). This study can also be viewed as part of a larger programme
designed to measure the cosmic evolution of the
co-moving density $\rho$ of radio-loud objects (e.g., Dunlop \&
Peacock 1990).

\begin{figure*}
  \ForceWidth{162mm} \BoxedEPSF{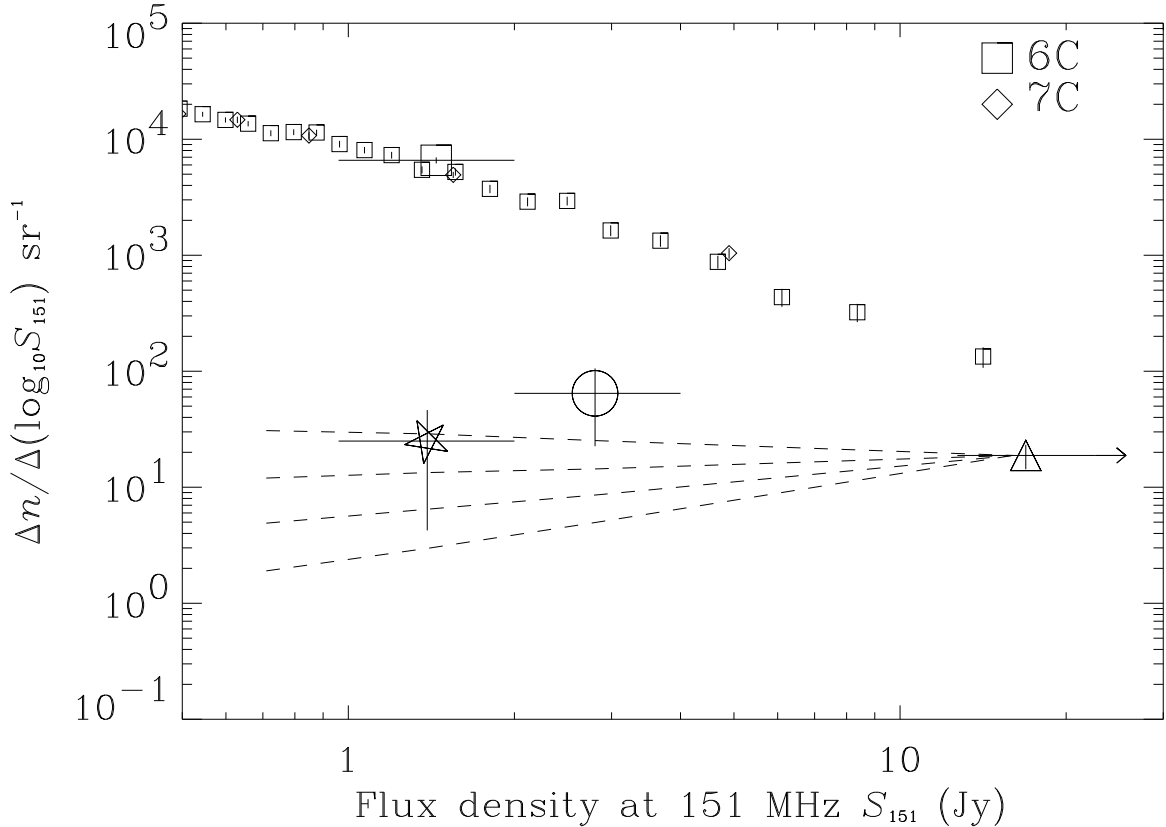}
  {\caption[junk]{\label{fig:scount} Predicted areal density of
  high-redshift radio sources as a function of 151-MHz flux density
  $S_{151}$ and in comparison with the total source count; the source
  count is reproduced from Hales, Baldwin \& Warner (1988) for 6C
  (squares) and from McGilchrist et al. (1990) for 7C (diamonds). The
  largest 6C symbol shows the source count for the 6C* sample
  discussed in this paper.  The lowest 3 symbols represent
  measurements of the number of high radio luminosity ($28.0 <
  \log_{10} L_{151} < 28.7$) sources in three samples: represented by
  a triangle are (i) the 20 radio sources, galaxies plus quasars, with
  $z > 1.3$ in the revised 3CR sample (Laing et al.\ 1983); by an open
  circle the two sources, both galaxies, with $z > 3$ in the 6C sample
  (Eales et al.\ 1997); and by an open 5-pointed star the one radio
  galaxy at $z > 4$ found in a preliminary analysis of the 6C* sample
  (Rawlings et al.\ 1996). The horizontal bars show the $S_{151}$
  range of each sample, and the vertical bars are 68\% confidence
  intervals calculated by adopting a Jeffreys' prior for the the
  probability density function of the areal density and by assuming a
  Poisson likelihood function for the number of high-$L_{151}$ objects in
  each sample (see Sivia 1996); note in the case of 6C* that the true
  areal density may be systematically higher if the $\alpha$ and $\theta$
  filtering has led to the exclusion of some high redshift objects.
  The dashed lines show the results of a crude but, for illustrative
  purposes, adequate model of the high-$z$ evolution of the radio
  luminosity function: we parameterise this by $\rho \propto \rho_{0}
  \times (1 + z)^{- \gamma} \times L_{151}^{- \beta}$, set $\beta = 2$
  (following Dunlop \& Peacock 1990), and fix $\rho$ to reproduce the
  3C data at $z > 1.3$; the dashed lines show (from top to bottom) the
  effects of increasing $\gamma$ from 0 to 3.  }}
\end{figure*}

The considerable difficulties involved in finding $z > 4$ radio
galaxies are illustrated by Fig.~\ref{fig:scount}. Such objects are
absent from the brightest radio source samples (e.g., the revised 3CR
sample: Laing, Riley \& Longair 1983) because of an upper limit to the
low-frequency (151 MHz) radio luminosity $L_{151}$ of radio sources of
$\log_{10} L_{151} < 28.8$ (where we measure $L_{151}$ in units of
$\rm W ~ Hz^{-1} ~ sr^{-1}$). Radio sources at $z > 4$ may be present
in fainter low-frequency selected samples (see Fig.~\ref{fig:scount})
but only amongst a much larger number of less distant radio sources;
the {\it fraction} of radio-luminous $z > 4$ radio galaxies in a
flux-limited sample declines as the flux limit is decreased even if
there is no high-redshift cutoff in $\rho$ (Fig.~\ref{fig:scount}).
At most only about 1 per cent of a complete sample selected at a
151-MHz flux density $S_{151} \approx 1 ~ \rm Jy$ could lie at $z >
4$.  To ensure that optical follow-up is confined to a reasonable
number of objects requires more selective criteria than just a simple
radio flux limit.

We chose additional selection criteria which took into account the
characteristics often seen in very distant radio galaxies namely,
small angular size $\theta$ and steep radio spectral index $\alpha$
[e.g., McCarthy (1993)]. This latter criterion has been used
extensively by a number of groups [e.g., Chambers, Miley \& van
Breugel (1990); R\"{o}ttgering et al.\ (1994); Rhee et al.\ (1996);
Chambers et al. (1997)].  Our chosen selection criteria were $\theta <
15^{\prime\prime}$ and $\alpha > 0.981$. Note that these criteria are
imperfect in two ways: first, they are unlikely to filter out all the
low-redshift objects, and second, they will exclude at least some
high-redshift objects.

\begin{figure*}
\ForceWidth{162mm}
\BoxedEPSF{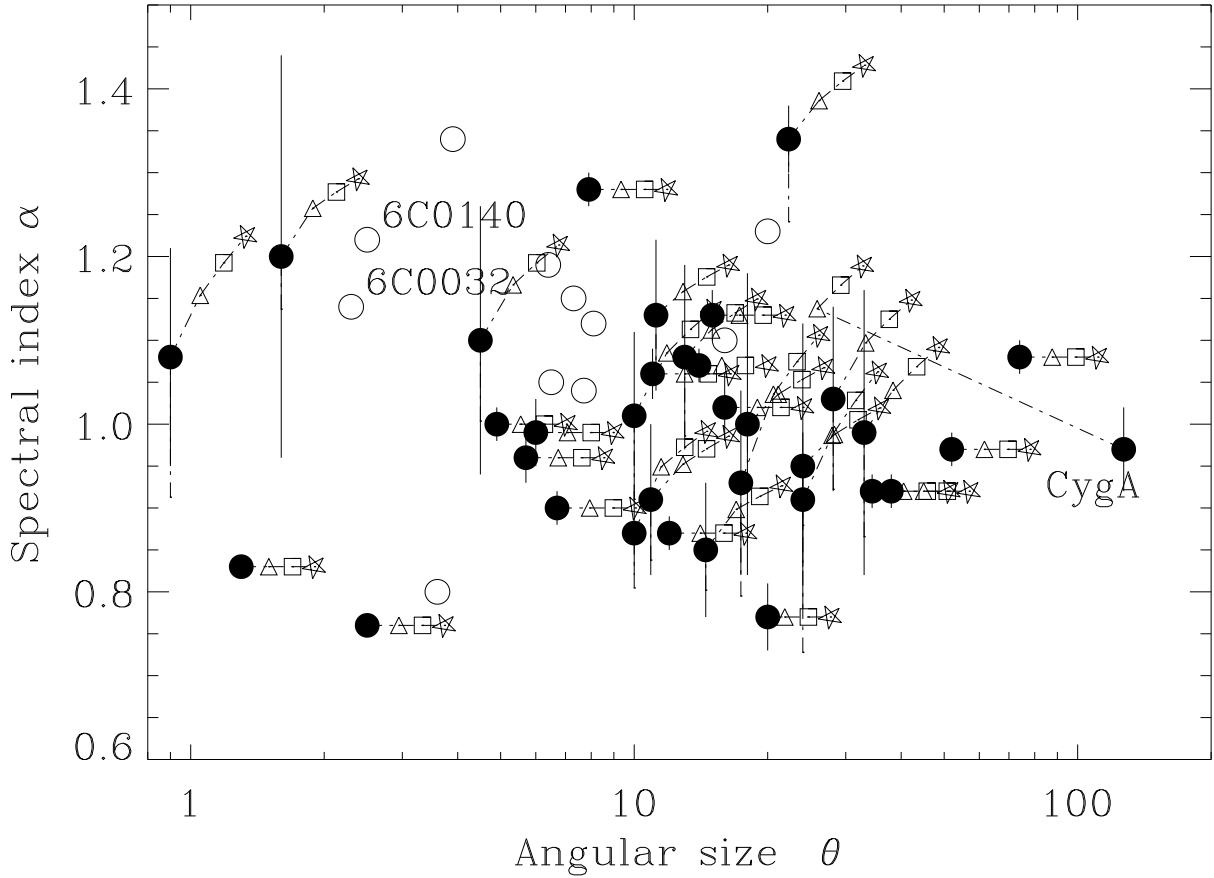}
\caption[ki]{\label{fig:az} 
The radio spectral index $\alpha$ (evaluated at 1 GHz as described
in Section 3.1), angular size
$\theta$ (in arcsec) plane for 3C radio galaxies (filled circles,
Cyg~A marked) and the $z > 3$ radio galaxies known prior to submission
of this paper (open circles, 6C\,0032+412 \& 6C\,0140+326 marked).  For
the 3C sources we have used their integrated radio spectra 
(taken from Laing \& Peacock 1980) and
observed angular sizes to predict the loci of similar radio sources at
$z = 3$ (triangles), $z = 4$ (squares) and $z = 5$ (stars).} 
\end{figure*}

This second imperfection has a more profound impact on the aims of our
project, so to examine the extent to which the spectral index and
angular size cut-offs contribute to this we compared the angular sizes
and spectral indices of the known $z > 3$ objects, with those of the
most radio-luminous 3C sources (see Fig.\ ~\ref{fig:az}). Many, but
not all, high-$z$ radio galaxies have concave radio spectra like
Cyg~A, so the {\it k}-correction means that an $\alpha > 1$ criterion
should exclude only a minority of the $z > 3$ objects. For the
illustrative example depicted in Fig.\ ~\ref{fig:az}, 35\% of 3C
objects redshifted to $z = 3$ or $z = 4$ would be excluded by our
chosen spectral index criterion, while 29\% would be excluded by the
same criterion if they were redshifted to $z = 5$. Note, also that
there are significant selection effects in the positioning of the
known $z > 3$ radio galaxies in Fig.\ \ref{fig:az}: most of these
objects (represented by the open circles) have been selected as
candidate high-$z$ radio galaxies on the basis of their steep spectra,
and in some cases account has been taken of their small angular size.

\begin{figure*}
  \ForceWidth{162mm} \BoxedEPSF{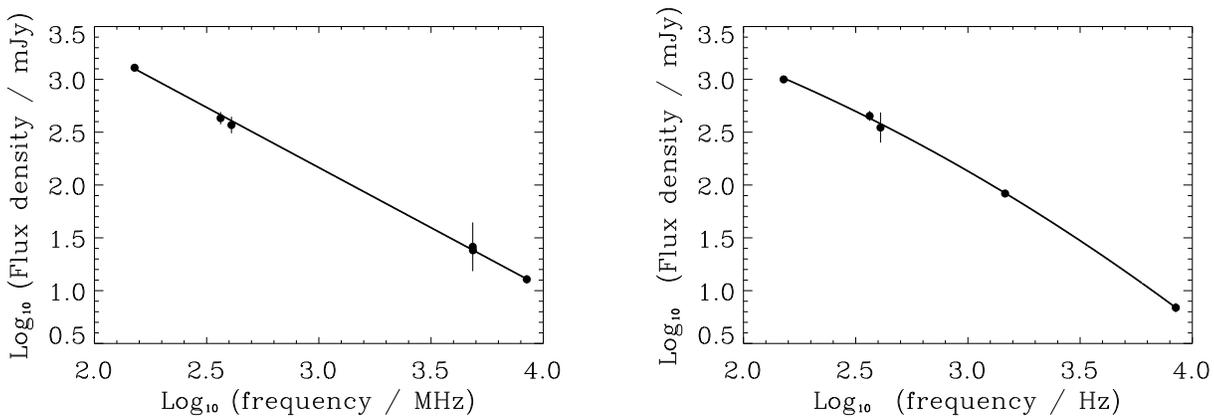}
  {\caption[junk]{\label{fig:specfit} These plots show the flux
  densities against frequency for two high redshift radio galaxies:
  {\em left:} 6C\,0032+412 at $z = 3.67$ and {\em right:} 6C\,0140+326 at
  $z = 4.41$. In the case of 6C\,0032+412 the fit shown is a straight
  line while for 6C\,0140+326 a curvature is required as described
  in the text.
}}
\end{figure*}

We considered, but rejected, the possibility of using spectral
curvature as an additional radio selection criterion [e.g., Lacy et
al.\ (1994)]: although this might have improved the efficiency of our
search for high-$z$ galaxies, it would certainly have been a further
cause of incompleteness. A clear demonstration of this is provided by
the comparison in Fig.~\ref{fig:specfit} of the radio spectra of the
two highest redshift radio galaxies found to date in the 6C* sample:
these objects, 6C\,0140+326 with $z = 4.41$ and 6C\,0032+412 with $z =
3.67$, have similarly steep values of $\alpha$ (and appear very close
to one another in Fig.\ \ref{fig:az}) but in one case (6C\,0140+326)
there is obvious spectral curvature, whereas in the other a single
straight power law provides an excellent fit over the full range of
the available data.  We quantify these considerations via the spectral
fitting described in Section 3.1.  Spectral curvature, although
common, is thus not a ubiquitous feature of the radio spectra of $z
\sim 4$ radio galaxies despite the high rest-frame frequencies probed
(up to 40 GHz in the case of 6C\,0032+412).  This is perhaps not
entirely surprising: the `leaky resevoir' model for radio hotspots
(e.g. Eales, Alexander \& Duncan 1989) predicts that spectral
curvature arises because of breaks at the extreme ends of the spectrum
($\ltsim\ 0.1 \, {\rm GHz}$ and $\gtsim\ 10\, {\rm GHz}$; see also
Chambers et al. 1990). The break values in a given $z \sim 4$ source
can perfectly plausibly fall outside the rest-frame frequency range
spanned by the common radio surveys, namely $\approx$ 0.75--25\, GHz;
note also that there is empirical evidence for an upper break
frequency $\gtsim\ 400 \, \rm GHz$ in the hotspots of Cyg A (Eales et
al.\ 1989).

For the example shown in Fig.\ \ref{fig:az}, a naive calculation would
imply that 50\% of the 3C objects would be excluded by our angular
size cut if they were redshifted to $z = 3$, 44\% if redshifted to $z
= 4$ and 32\% if redshifted to $z = 5$. However, this simple
calculation ignores any negative evolution of linear size with
$z$. Statistically speaking there is good evidence for just such an
evolutionary trend (e.g., Neeser et al.\ 1995; Blundell et al., in
prep.), but an intrinsic spread in $\theta$ means that at least some
$\theta > 15 ~ \rm arcsec$ sources are already known at $z > 3$ (Fig.\
~\ref{fig:az}).

The 6C* sample of 34 radio sources has already led to the detection of
one radio galaxy (6C\,0140+326) with $z = 4.41$ (Rawlings et al.\ 1996),
and one (6C\,0032+412) with $z = 3.67$ (Rawlings et al., in prep).  The
former was not the first radio galaxy to be found at $z > 4$ --- this
was 8C\,1435+635 at $z = 4.25$ (Lacy et al.\ 1994) --- but it is
currently the most distant radio galaxy known. In this paper
(hereafter Paper I) and a companion paper (Paper II) we will present
full details of the 6C* sample.  In Paper I we describe the sample
selection criteria, present VLA radio maps of the objects in the
sample, and tabulate various parameters of the radio sources including
spectral information derived from a combination of survey and VLA
data.  In Paper II (Rawlings et al., in prep) we will describe optical
spectroscopy and near-infrared imaging of these sources.  B1950.0
co-ordinates are used, and the convention for spectral index
($\alpha$) is that $S_{\nu} \propto \nu^{-\alpha}$, where $S_{\nu}$ is
the flux density at frequency $\nu$.

\section{Sample definition}

\begin{figure*}
  \ForceWidth{162mm}
  \BoxedEPSF{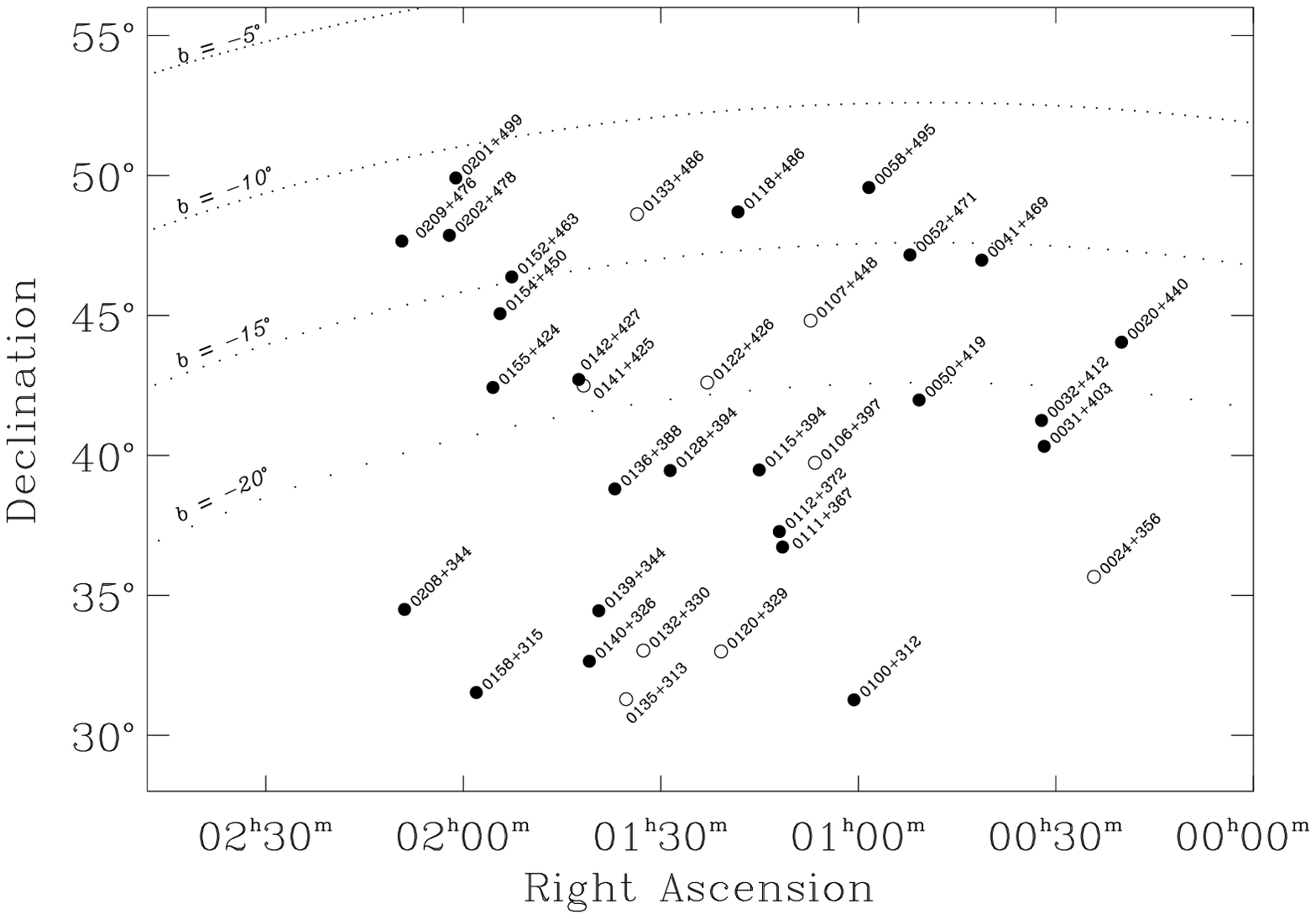}
{\caption[junk]{\label{fig:skycov} The sky locations of the members of the
6C* sample; we have discriminated between objects in the
preliminary version of the sample (filled symbols) and those only in
the final version (open symbols).  Lines of constant galactic latitude
($b$) are plotted as dotted lines. }}
\end{figure*}

This project commenced with criteria based on a preliminary version of
part VI of the 6C survey of radio sources at 151\,MHz (kindly supplied
by S. Hales) which includes the continuous zone of Right Ascension
between $22^{\rm h}35^{\rm m}$ and $09^{\rm h}05^{\rm m}$, passing
through $0^{\rm h}$, and the range of Declination between $+30^\circ$
and $+51^\circ$. A patch of sky $\sim 0.1$ sr was chosen as the
minimum size necessary to allow a reasonable chance of finding a $z >
4$ radio galaxy (see Fig.~\ref{fig:scount}), and the choice of
location of this patch was determined by two considerations: first,
avoidance of the galactic plane (see Fig.~\ref{fig:skycov}); and
second, a time allocation on an optical telescope suited to
observations near $1^{\rm h} 30^{\rm m}$ R.A.\ at northern
declinations.

In this patch of sky all sources with $S_{\rm 151}$ less than 1.00 Jy
(according to the preliminary flux scales of the 6C catalogue) or
greater than 2.00 Jy were rejected. Spectral indices of the remaining
objects were found by obtaining positional coincidences (within
120\arc) of sources in the 6C-VI catalogue and those in the 87GB
catalogue of radio sources, the latter resulting from a survey at
4.85\,GHz (Gregory \& Condon 1991).  Any object with a spectral index
between 151\,MHz and 4.85\,GHz (i.e., $\rm \alpha^{4.85 G}_{151M}$)
which was less than 1.0 was then rejected.  Sources which met the flux
density criterion for the 6C-VI survey but which were not detected in
the 87GB survey were retained in the sample since such sources (if
real) would inevitably have spectra steeper than this threshold.

The sources which were remaining were then
cross-correlated\footnote{The searches were made for {\em all} matches
in the Texas and in the 87GB samples within a $120^{\prime\prime}$
search radius. There were 3 cases of more than one match to a 6C
object being found, each of these arising from a double entry in the
Texas catalogue. These 3 examples were in each case 2 objects with
fluxes differing by not more than 50 mJy (i.e., $< 2\sigma$) and
separated by not more than 5 arcsec. We used the nearer of the two
matches to the 6C position in each case, which also happened to be in
each case the match with the lower flux density. For details on where
this occurred, please see `Notes on Individual Sources' --- Section
4.} with those in the Texas Survey at 365\,MHz (Douglas et al.\
1996). Sources with entries in both the 6C-VI and the Texas surveys
whose positions were coincident within 120\arc\ were only retained in
the sample if their angular size as listed in the Texas survey
($\theta_{\rm Texas}$) was less than 15\arc. This set of sources is
henceforth denoted (6C-T-C).  Those 6C sources with entries in the
Texas survey but not in the 87GB survey were classed as
6C-T-$\overline{\rm C}$.

We were concerned to ensure we would not exclude an object whose size
was $\ltsim\ 15^{\prime\prime}$ and whose two hotspots had
sufficiently asymmetric spectra that the hotspot which dominated the
overall source flux density at the frequency of one survey was
different from the hotspot dominating at the frequency of a different
survey. The worst case positional uncertainty (combining the
uncertainties in quadrature) is $25^{\prime\prime}$. We experimented
using somewhat over a $3\sigma$ tolerance for positional uncertainty
of $90^{\prime\prime}$ and found there was no difference if a very
conservative search radius of $120^{\prime\prime}$ was employed.  At
the flux levels of these surveys there is a minimal risk of excluding
an instrinsically steep spectrum source by spuriously matching it with
an inverted-spectrum neighbour in the higher frequency survey since
the areal density of these objects in these surveys is so low.

With these preliminary (sub-)samples established, an extensive series
of multi-wavelength observations began. Radio maps were obtained with
the VLA, enabling accurate positioning and orientation of slits for
spectroscopy.

We reviewed the selection criteria and sample membership once the
final version of the 6C catalogue became available in this region
(Hales et al.\ 1993).  All of the final catalogue analysis and
manipulation was carried out using the Starlink Cursa Project software
written by Clive Davenhall.  It was found that all objects for which
optical and infra-red follow-up had occurred were selected by imposing
the flux limits 0.96\,Jy $\leq S_{151} \leq$ 2.00\,Jy, and a radio
spectral index limit $\rm \alpha^{4.85G}_{151M} \geq 0.981$.

In this final version of our 6C* sample we included those 6C souces
with no entry in the Texas catalogue (within 120\arc) nor in the 87GB
survey (within 120\arc) and denoted these as 6C-$\overline{\rm
T}$-$\overline{\rm C}$ since it was plausible that such sources have
very steep spectra.\footnote{Three objects in this subset were found
to be not appropriate members of the sample after inspection of the 6C
radio maps and consideration of other data --- see Section 6.} We also
retained those 6C sources in the 87GB survey but not in the Texas
survey and named this sub-set as 6C-$\overline{\rm T}$-C.

The final selection criteria may thus be summarised as follows:
\begin{itemize}
\item $00^{\rm h}20^{\rm m} \leq$ R.A. $\leq 02^{\rm h}10^{\rm m}$ 

\item $30^\circ \leq$ Dec. $\leq 51^\circ$

\item 0.96\,Jy $\leq S_{151} \leq$ 2.00\,Jy

\item $\rm \alpha^{4.85G}_{151M} \geq 0.981$, where calculable.

\item $\theta_{\rm Texas} < 15^{\prime\prime}$, where available.

\end{itemize}

The final 6C* sample comprises 34 objects. Of these 9 were missing
from our original version of the sample (marked by asterisks in Table
~\ref{tab:flux} and see also Fig.~\ref{fig:skycov}).  The only
systematic difference between the original and final versions of the
sample lies in the inclusion of the 4 sources in the 6C-$\overline{\rm
T}$-$\overline{\rm C}$ sub-sample. The remaining 5 objects were not
part of the original sample because of changes in the finalised 6C
survey, and because of inadvertant omission in a process which was not
fully automated.

\begin{figure*}
  \ForceWidth{162mm} \BoxedEPSF{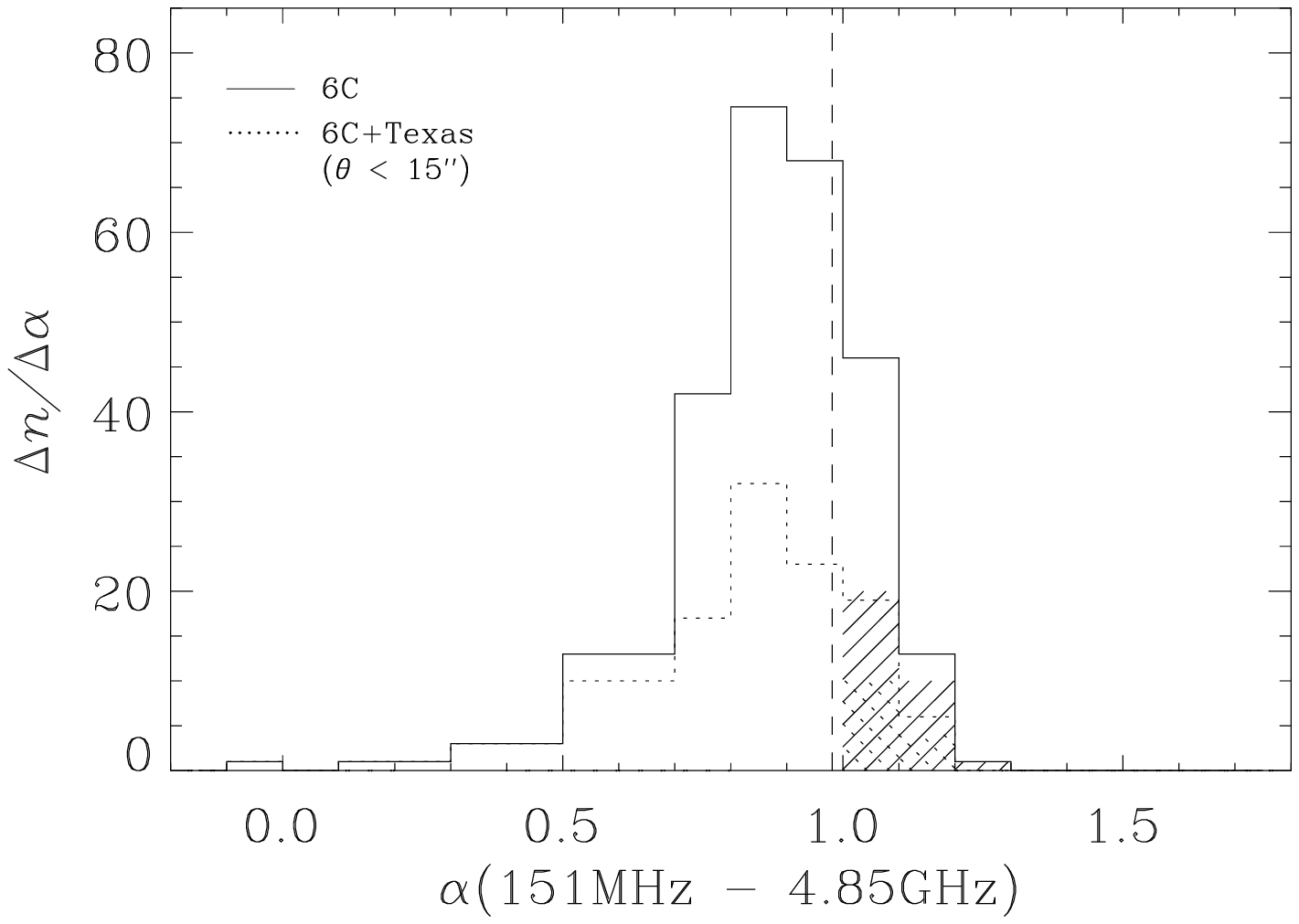}
  \parbox{162mm}{\caption[junk]{\label{fig:spixhist} The solid line
  delineates the histogram of spectral indices between 151 MHz (from
  the 6C survey) and 4.85 GHz (from the 87GB survey) for those 279 6C
  objects satisfying the flux limits and sky area boundaries described
  in Section 2. Of these 279 objects, 248 have matches in the 87GB
  catalogue and their calculated spectral indices are used.  The
  remaining 31 of these objects did not have matches (within 120\arc)
  and so lower limits to their spectral indices were derived taking
  the upper limit for their flux density at 4.85 GHz to be 25 mJy (the
  flux limit to which the 87GB survey is complete). The vertical
  dashed line is plotted at the spectral index limit of 0.981.  The
  hatched lines show the contributions from the lower limits to the
  spectral indices of objects without matches in the 87GB survey. The
  dotted line shows the histogram for those 112 objects in 6C which
  had a match in the Texas catalogue and whose angular size was quoted
  in this survey as being $< 15^{\prime\prime}$; for these objects
  which have matches in the 87GB catalogue, their calculated spectral
  indices between 151 MHz and 4.85 GHz are used, while for those 14
  objects which do not have matches in the 87GB catalogue lower limits
  for spectral indices were derived assuming the same upper limit to
  flux density as previously, namely 25 mJy. The dotted hashed lines
  indicate the contribution from these 14 objects. }}
\end{figure*}

The effects of the $\theta$ and $\alpha$ selection criteria on the
original flux-limited sample are illustrated in
Fig.~\ref{fig:spixhist}: about 12 per cent of the original
flux-limited sample were retained once these filtering criteria were
applied. Only one of the objects matched in the Texas survey and
retained in the final sample was deemed by this survey to have a size
other than unresolved. This object is 6C\,0133+486 whose angular size,
deemed by the Texas survey to be 10\arc, turned out to be 10.9\arc.
One of the multiply-matched objects (6C\,0031+403) had one match with
a purported angular size of 23\arc\ but a nearer match deemed to be
unresolved. We measured a size of $\approx 20$\arc\ for this object.

The sky area of the 6C* survey is 0.133 sr.  The radio flux density
measurements of the sample members obtained from the various radio
surveys are compiled in Table~\ref{tab:flux}. One final complication
concerns the slightly discrepant 151 MHz source count in this region
of the 6C survey (Fig.~\ref{fig:scount}) but this effect is at the
level where it is plausibly a statistical fluctuation.

\scriptsize
\begin{table*}
\begin{center}
\begin{tabular}{l|r|l|r|r|r|r|c}
\hline\hline
\mc{1}{c|}{Source}  & \mc{1}{c|}{$\rm \alpha_{151M}^{4.85G}$} &\mc{1}{c|}{$\alpha_{1000}$} & \mc{1}{c|}{6C} & \mc{1}{c|}{Texas} & \mc{1}{c|}{B2/3} & \mc{1}{c|}{WB} & \mc{1}{c|}{87GB} \\
\mc{1}{c|}{      }  & \mc{1}{c|}{      } & \mc{1}{c|}{      }         & \mc{1}{c|}{151 MHz} & \mc{1}{c|}{365 MHz}  & \mc{1}{c|}{408 MHz} & \mc{1}{c|}{1400 MHz} & \mc{1}{c|}{4850 MHz}\\ 
\hline
\mc{8}{c}{ }\\[-0.25cm]
\mc{8}{c}{6C-T-C} \\
\hline
6C\,0020+440       & $1.13 \pm 0.10$ & $1.13 \pm 0.03$      & $2000 \pm 45$ & $ 697 \pm 23$    & $700 \pm 36$ & $178 \pm 30$ & $40 \pm 6$  \\
6C\,0031+403       & $0.99 \pm 0.13$ & $0.95 \pm 0.03$      &  $960 \pm 45$ & $ 432 \pm 28$    & $390 \pm 30$ & $138 \pm 30$ & $31 \pm 6$  \\
6C\,0032+412       & $1.13 \pm 0.15$ & $1.19 \pm 0.03$      & $1290 \pm 45$ & $ 430 \pm 25$    & $370 \pm 29$ &  -           & $26 \pm 6$  \\
6C\,0050+419       & $1.01 \pm 0.13$ & $1.02 \pm 0.01$      & $1000 \pm 45$ & $ 396 \pm 19$    & $370 \pm 29$ &  -           & $30 \pm 6$  \\
6C\,0052+471       & $1.00 \pm 0.10$ & $1.00 \pm 0.03$      & $1310 \pm 45$ & $ 561 \pm 22$    &  -           & $100 \pm 30$ & $41 \pm 6$  \\
6C\,0058+495       & $1.02 \pm 0.12$ & $1.00 \pm 0.02$      &  $970 \pm 45$ & $ 420 \pm 29$    &  -           &  -           & $28 \pm 5$  \\ 
6C\,0111+367       & $1.00 \pm 0.12$ & $0.96 \pm 0.04$      & $1110 \pm 45$ & $ 511 \pm 21$    & $390 \pm 50$ &  -           & $34 \pm 6$  \\
6C\,0128+394       & $0.99 \pm 0.09$ & $1.06 \pm 0.41^{c}$  & $1940 \pm 45$ & \hspace{-1.5mm}$1111 \pm 20$ & $920 \pm 40$ & $261 \pm 30$ & $62 \pm 9$  \\
6C\,0133+486$^{*}$ & $1.06 \pm 0.09$ & $1.08 \pm 0.02$      & $1890 \pm 45$ & $ 685 \pm 36$    &  -           & $169 \pm 30$ & $48 \pm 7$  \\
6C\,0139+344       & $1.05 \pm 0.14$ & $1.09 \pm 0.16$      & $1100 \pm 45$ & $ 589 \pm 22$    & $570 \pm 60$ &  -           & $29 \pm 6$  \\
6C\,0152+463       & $1.02 \pm 0.10$ & $1.01 \pm 0.04$      & $1290 \pm 45$ & $ 493 \pm 20$    & $530 \pm 33$ & $145 \pm 30$ & $38 \pm 6$  \\
6C\,0154+450       & $1.02 \pm 0.12$ & $1.08 \pm 0.07$      & $1150 \pm 45$ & $ 357 \pm 25$    & $380 \pm 30$ &  -           & $34 \pm 6$  \\
6C\,0155+424       & $1.06 \pm 0.10$ & $1.06 \pm 0.01$      & $1510 \pm 45$ & $ 593 \pm 20$    & $520 \pm 33$ & $143 \pm 30$ & $38 \pm 6$  \\
6C\,0158+315       & $1.03 \pm 0.12$ & $0.88 \pm 0.07$      & $1510 \pm 45$ & $ 808 \pm 30$    & $710 \pm 80$ & $197 \pm 30$ & $43 \pm 8$  \\
\hline							       			            
\mc{8}{c}{ }\\[-0.25cm]				       			                                           	                
\mc{8}{c}{6C-T-$\overline{\rm C}$} \\			       			                                           	                
\hline							       			                                           	                
6C\,0024+356$^*$   & $>1.09$         & $1.12 \pm 0.20$      & $1090 \pm 45$ & $515 \pm 19$     &  -           & -            & -     \\
6C\,0100+312       & $>1.11$         & $1.68 \pm 0.14\dag$  & $1160 \pm 45$ & $^{\ddag}223 \pm 27$ & $290 \pm 50$ & -            & -     \\
6C\,0107+448$^{*}$ & $>1.07$         & $1.50 \pm 0.07$      & $1040 \pm 45$ & $258 \pm 19$     & $260 \pm 26$ & -            & -     \\
6C\,0112+372       & $>1.07$         & $1.11 \pm 0.08$      & $1030 \pm 45$ & $502 \pm 18$     & $370 \pm 29$ & -            & -     \\
6C\,0115+394       & $>1.05$         & $1.12 \pm 0.05$      &  $960 \pm 45$ & $369 \pm 20$     & $290 \pm 28$ & -            & -     \\
6C\,0118+486       & $>1.06$         & $1.22 \pm 0.20$      &  $980 \pm 45$ & $333 \pm 21$     &  -           & -            & -     \\
6C\,0122+426$^{*}$ & $>1.08$         & $1.09 \pm 0.08$      & $1050 \pm 45$ & $451 \pm 23$     & $460 \pm 31$ & $182 \pm 30$ & -     \\
6C\,0136+388       & $>1.06$         & $1.10 \pm 0.07$      &  $990 \pm 45$ & $478 \pm 20$     & $360 \pm 29$ & -            & -     \\
6C\,0140+326       & $>1.06$         & $1.11 \pm 0.05$      & $1000 \pm 45$ & $451 \pm 21$     & $350 \pm 50$ & -            & -     \\
6C\,0142+427       & $>1.17$         & $1.21 \pm 0.12$      & $1460 \pm 45$ & $448 \pm 20$     & $530 \pm 33$ & -            & -     \\
6C\,0201+499       & $>1.10$         & $1.05 \pm 0.16$      & $1140 \pm 45$ & $450 \pm 21$     & -            & -            & -     \\
6C\,0201+478       & $>1.08$         & $1.22 \pm 0.19$      & $1060 \pm 45$ & $362 \pm 21$     & -            & -            & -     \\
6C\,0208+344       & $>1.05$         & $1.05 \pm 0.08$      &  $970 \pm 19$ & $471 \pm 19$     & -            & -            & -     \\
6C\,0209+476       & $>1.10$         & $1.13 \pm 0.08$      & $1140 \pm 45$ & $526 \pm 19$     & -            & -            & -     \\
\hline							       			                                           	                
\mc{8}{c}{ }\\[-0.25cm]				       			                                           	                
\mc{8}{c}{6C-$\overline{\rm T}$-C} \\			       			                                           	                
\hline							       			                                           	                
6C\,0041+469       & $1.07 \pm 0.12$ & $0.98 \pm 0.06$      & $1530 \pm 45$ & -                & $650 \pm 35$ & -            & $38 \pm 7$ \\
6C\,0141+425$^{*}$ & $1.06 \pm 0.11$ & $0.95 \pm 0.07$      & $1640 \pm 45$ & -                & (see notes)  & $265 \pm 30$ & $41 \pm 7$ \\ 
\hline							       			                            
\mc{8}{c}{ }\\[-0.25cm]				       			                            
\mc{8}{c}{6C-$\overline{\rm T}$-$\overline{\rm C}$}\\	       			                            
\hline
6C\,0106+397$^{*}$  & $>1.05$        & $1.08 \pm 0.30$      &  $960 \pm 45$ & -                & $460 \pm 31$ & -            & -          \\
6C\,0120+329$^{*}$  & $>1.24$        & $>1.24$              & $1870 \pm 45$ & -                &    -         & (see notes)  & -          \\
6C\,0132+330$^{*}$  & $>1.19$        & $>1.19$              & $1560 \pm 45$ & -                &    -         & -            & -          \\
6C\,0135+313$^{*}$  & $>1.12$        & $>1.12$              & $1240 \pm 45$ & -                &    -         & -            & -          \\
\hline\hline
\end{tabular}
\normalsize {\caption[junk]{\label{tab:flux} The survey flux densities
of the 6C* sources in mJy and spectral indicies derived from these.
The first column gives the source name, with an asterisk indicating if
the source had been excluded from the preliminary version of the
sample.  The second column lists the spectral index $\alpha$ derived
from the 6C and 87GB flux densities, i.e., the value used in the
selection criterion for sample membership.  The third column lists the
spectral index at 1000 MHz derived from the fitting procedure
described in the Section 3.1; the fit requiring spectral curvature is
marked with a `c', and values for the curvature $\beta$ for this
object, 6C\,0128+394, is 0.64.  The remaining columns give survey flux
densities, properties of these surveys are: 6C, a 151-MHz survey with
a resolution of $\approx 4 ~ \rm arcmin$ (e.g., Hales et al.\ 1988),
the numbers listed are fitted peak flux densities; Texas, a 365-MHz
survey with a complicated beam meaning that the flux densities of
sources larger than a few arcmin are unreliable (Douglas et al.\
1996): B2.1, B2.3 and B3, the Bologna 408-MHz B2.1 (Colla et al.\
1970), B2.3 (Colla et al.\ 1974) and B3 (Ficarra, Grueff \& Tomassetti
1985) surveys with resolutions of $\ltsim\ 10 ~ \rm arcmin$ and
$\ltsim\ 5 ~ \rm arcmin$ for the B3 survey (B3 is used except in the
cases of 6C\,0100+312, 6C\,0140+326 and 6C\,0158+315 [B2.1] and in the
cases of 6C\,0111+367, 6C\,0139+344 [B2.3]); WB, data compiled by
White \& Becker (1992) from the 1400-MHz survey of Condon \& Broderick
(1986) which had a resolution of $\approx 12 ~ \rm arcmin$; 87GB, a
4850-MHz survey with a resolution of $\approx 3.5 ~ \rm arcmin$
(Gregory \& Condon 1991). The $\dag$ symbol indicates that the
spectral index is unreliable because the flux density marked with a
$\ddag$ is possibly underestimated because of the large angular size
of the radio source.  The errors in 6C fluxes were obtained following
the assumption in Hales et al.\ (1993; 6C-VI) that the errors in this
survey are 50\% worse than in Baldwin et al.\ (1985; 6C-I) --- thus we
take 45 mJy as the typical error in a flux density measurement from
6C. Errors on the Texas, B2.1, B2.3, B3, WB and 87GB flux density
measurements were either taken from the catalogues, or estimated from
the noise statistics of the survey.

}}

\end{center}
\end{table*}

\normalsize

\section{Radio observations}

Radio observations of the sources were made using the VLA in various
configurations, details of which are tabulated in
Table~\ref{tab:radobs}.  In each case, the bandwidth was 50\,MHz
(except for the 1.5 GHz observations in 1995/1996 where a bandwidth of
25\,MHz was used).  The primary flux calibrators used were either
3C\,286 or 3C\,48, and the flux scale used in each case was the then most
recently determined for the VLA.  We calibrated and reduced the data
using standard procedures in \aip, including self-calibration for
phase only, where there was sufficient signal-to-noise.

\begin{figure*}
\begin{center}
\setlength{\unitlength}{1mm}
\begin{picture}(150,220)
\put(0,140){\includegraphics{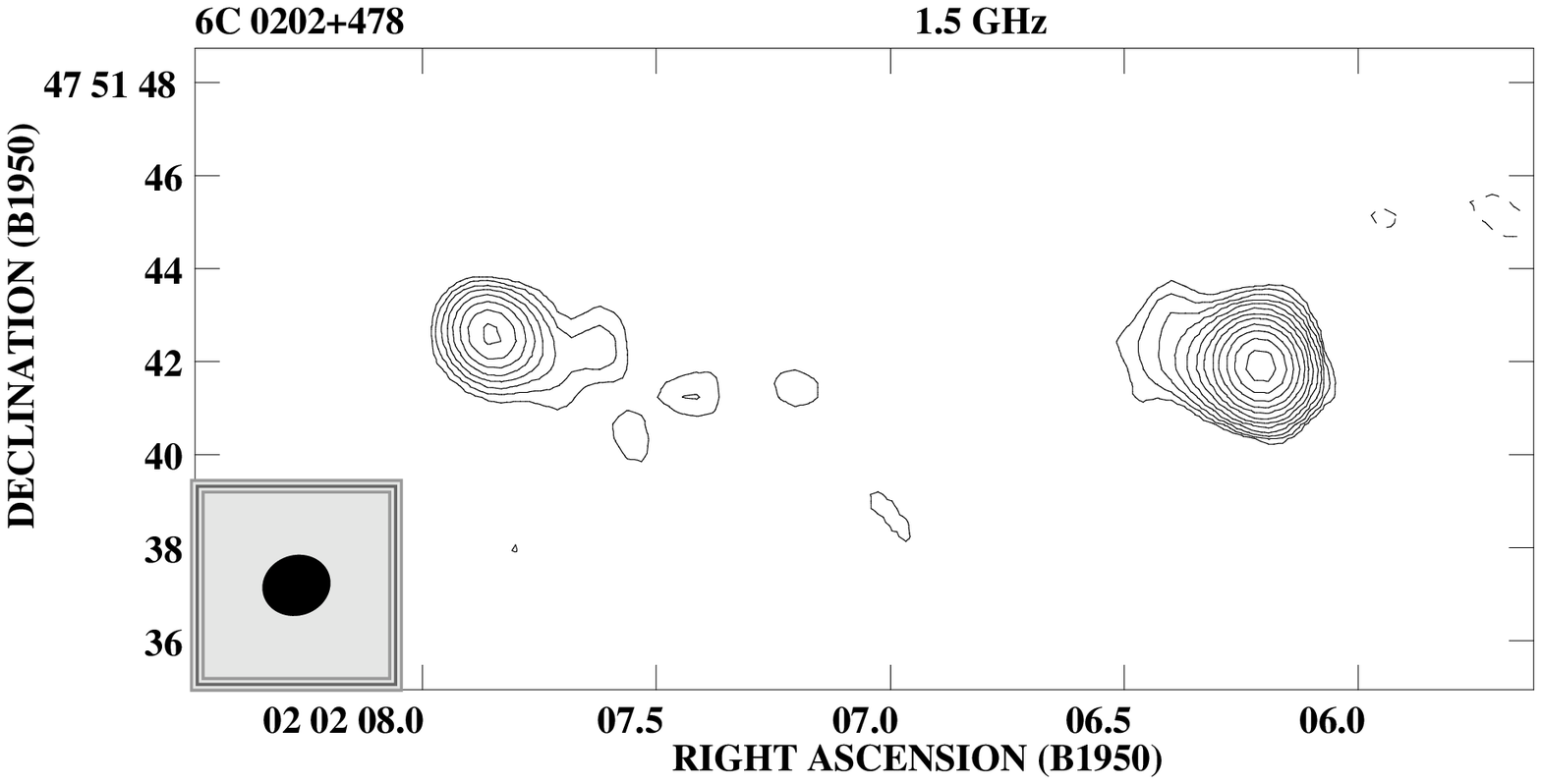}}
\put(75,125){\includegraphics{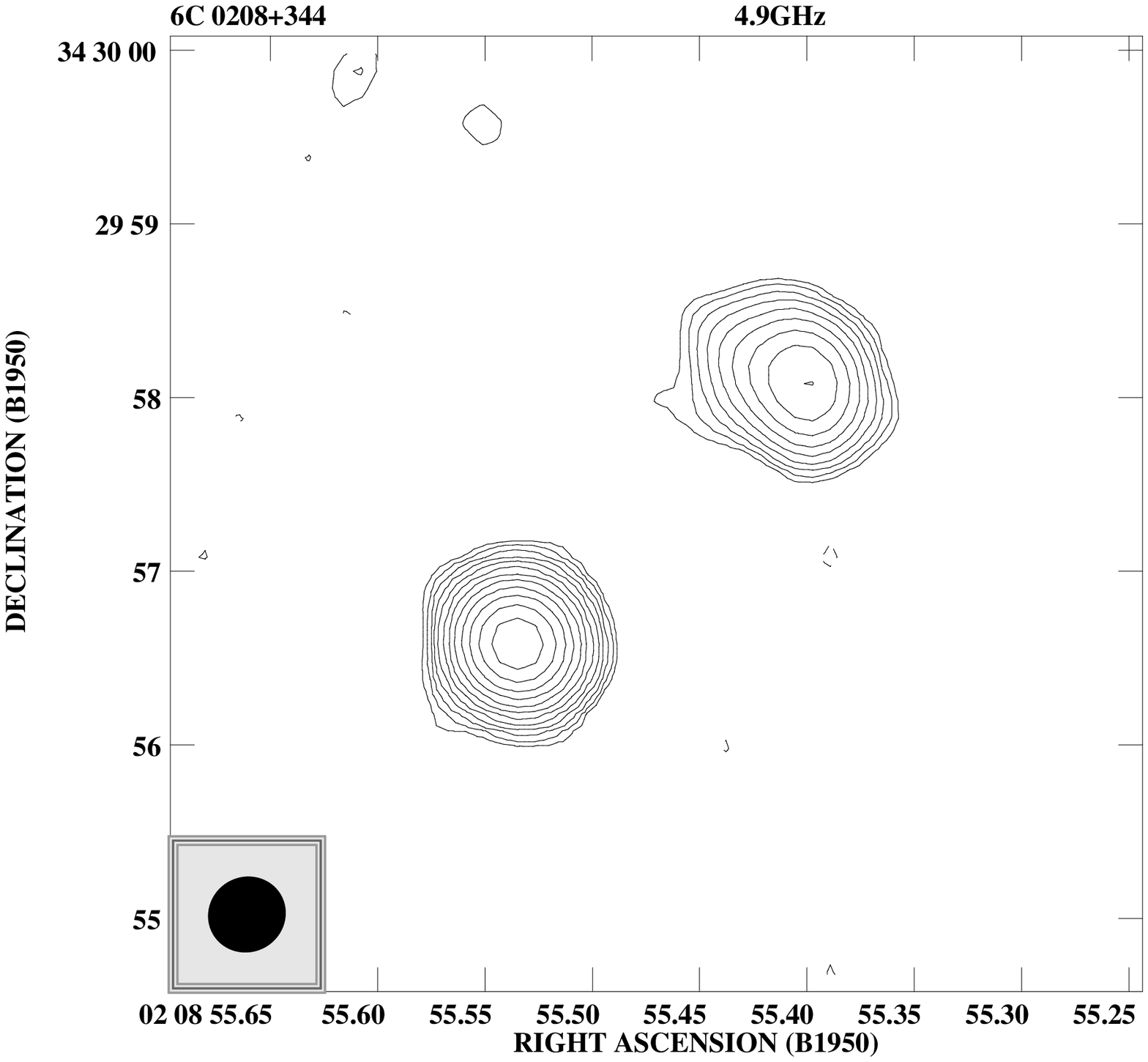}}
\put(-2,70){\includegraphics{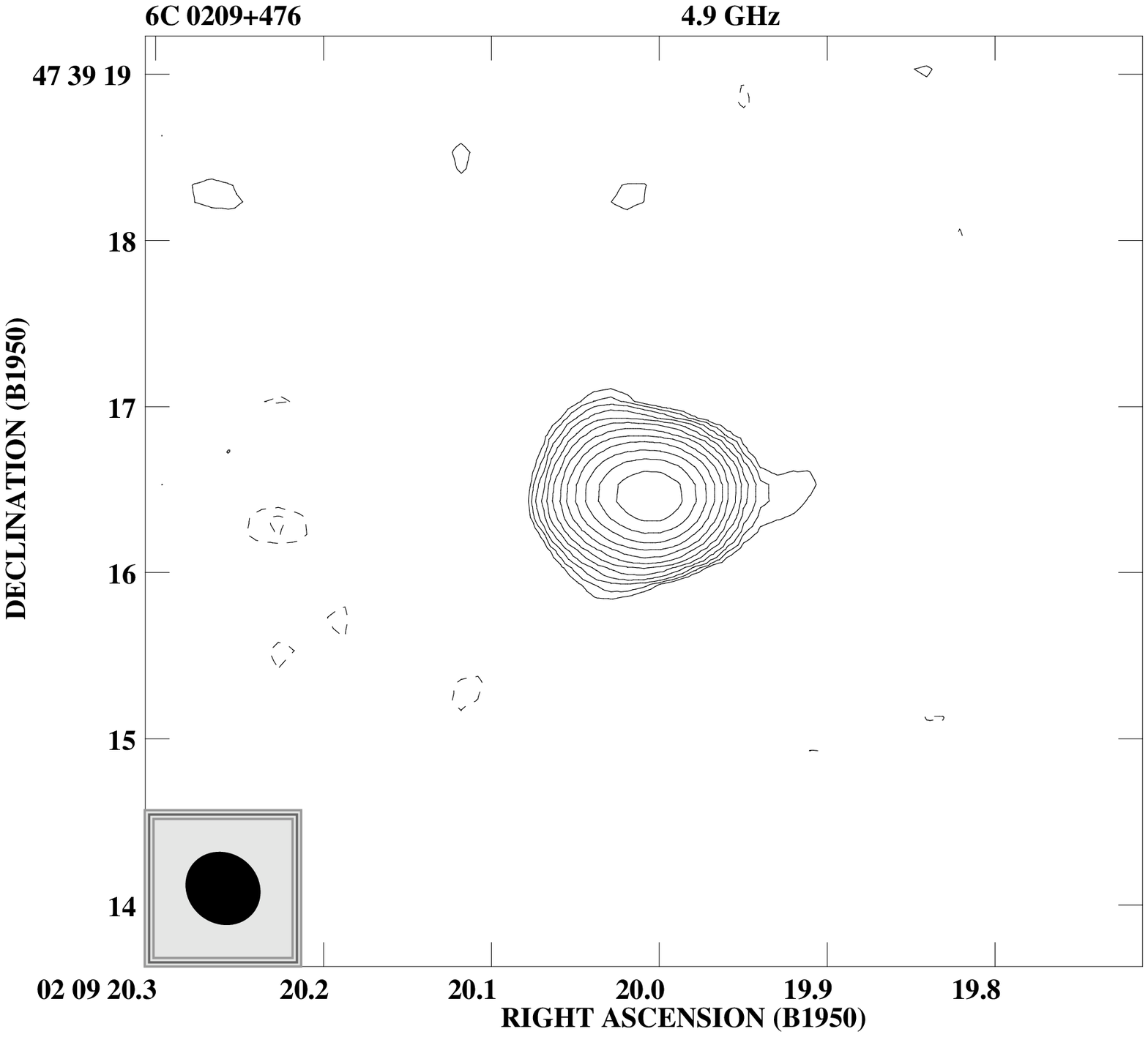}}
\end{picture}
\end{center}
{\caption[junk]{\label{fig:rad_6} Contour maps of the radio sources.}}
\end{figure*}

We present maps in Figures~\ref{fig:rad_1}~---~\ref{fig:rad_6};
lowest contour levels, beam sizes and other relevant data are
tabulated in Table~\ref{tab:fignotes}. For each map presented,
adjacent contour levels differ from one another by a factor of
$\sqrt{2}$ in flux density. The exact observing frequencies used were:
L-band 1.4649 GHz, C-band 4.8601 GHz and X-band 8.4399 GHz.

\scriptsize
\begin{table*}
\begin{center}
\begin{tabular}{l|ll|c|r|r|l|c|l|c}
\hline\hline
\mc{1}{c|}{Source} & \mc{1}{c}{R.A.} & \mc{1}{c|}{Dec.} & \mc{1}{c|}{Position} & \mc{1}{c|}{Shape} & \mc{1}{c|}{Angular} & \mc{1}{c|}{VLA-L} 
& \mc{1}{c|}{VLA-C} & \mc{1}{c|}{VLA-X} & \mc{1}{c}{Source} \\
\mc{1}{c|}{ } & \mc{1}{c}{B1950.0 } & \mc{1}{c|}{B1950.0 } & \mc{1}{c|}{flag} & \mc{1}{c|}{ } & \mc{1}{c|}{size } & \mc{1}{c|}{1.5 GHz } & \mc{1}{c|}{4.9 GHz } 
& \mc{1}{c|}{8.4 GHz } & \mc{1}{c}{ }   \\
\hline
\hline
6C\,0020+440           & 00 20 01.71 & +44 02 32.3 & h  & c  & 10.0        &{\bf L1-A}   &            & X1-C         & 6C\,0020+440 \\
6C\,0024+356           & 00 24 12.89 & +35 39 47.7 & h  & c  &  9.6        &             &            & {\bf X3-BnA} & 6C\,0024+356 \\
6C\,0031+403           & 00 31 46.79 & +40 19 26.3 & c  & dt &$\approx 20$ &{\bf L1-A}   &            & X1-C         & 6C\,0031+403 \\
6C\,0032+412           & 00 32 10.73 & +41 15 00.2 & c  & t  &  2.3        &             & {\bf C1-A} & X1-C         & 6C\,0032+412 \\
6C\,0041+469           & 00 41 16.04 & +46 58 27.2 & c? & t? &  3.8        &             & {\bf C1-A} & X1-C         & 6C\,0041+469 \\
6C\,0050+419           & 00 50 46.52 & +41 58 45.3 & p  & a  & $<0.5$      &             & {\bf C1-A} & X1-C         & 6C\,0050+419 \\
& & & & & & & & & \\			             
6C\,0052+471           & 00 52 09.59 & +47 09 46.2 & h  & c  &  5.6        &{\bf L1-A}   &            & X1-C         & 6C\,0052+471 \\
6C\,0058+495           & 00 58 24.75 & +49 34 04.3 & h  & c  &  3.2        &             & {\bf C1-A} & X1-C         & 6C\,0058+495 \\ 
6C\,0100+312           & 01 00 39.71 & +31 16 06.5 & c  & t  & 89.7        &{\bf L1-A}   &            & X1-C         & 6C\,0100+312 \\
6C\,0106+397           & 01 06 35.31 & +39 44 02.8 & h  & c  &  3.0        &{\bf L2-A}   &            &              & 6C\,0106+397 \\
6C\,0107+448           & 01 07 15.42 & +44 48 44.5 & c? & dt?&  5.8        &             &            & {\bf X2-BnA} & 6C\,0107+448 \\
6C\,0111+367           & 01 11 30.99 & +36 43 32.8 & p  & a  &  1.2        &             & {\bf C1-A} & X1-C         & 6C\,0111+367 \\
& & & & & & & & & \\			             
6C\,0112+372           & 01 12 00.33 & +37 16 41.4 & c  & t? &  3.6        &             & {\bf C1-A} & X1-C         & 6C\,0112+372 \\
6C\,0115+394           & 01 15 03.47 & +39 28 45.9 & c  & t  &  7.7        &             &            & {\bf X3-BnA} & 6C\,0115+394 \\
6C\,0118+486           & 01 18 16.49 & +48 41 58.0 & c? & t? & 16.1        &{\bf L1-A}   &            & X1-C         & 6C\,0118+486 \\
6C\,0120+329$^{\ddag}$ & 01 20 50.7  & +32 59 45   & g  & dt & 130.0       &{\bf L2-A}   &            &              & 6C\,0120+329 \\
6C\,0122+426           & 01 22 56.80 & +42 36 16.0 & h  & c  &  0.9        &             &            & {\bf X2-BnA} & 6C\,0122+426 \\
6C\,0128+394           & 01 28 34.67 & +39 27 32.0 & h  & c  &  2.6        &             &            & {\bf X3-BnA} & 6C\,0128+394 \\
& & & & & & & & & \\			             
6C\,0132+330           & 01 32 39.22 & +33 01 42.0 & p  & u  &  $< 8$      &             &            & {\bf X4-DnC} & 6C\,0132+330 \\
6C\,0133+486           & 01 33 36.58 & +48 37 04.0 & h  & c  & 10.9        &             &            & {\bf X3-BnA} & 6C\,0133+486 \\
6C\,0135+313           & 01 35 16.13 & +31 17 27.3 & p  & a  &  1.8        &{\bf L2-A}   &            &              & 6C\,0135+313 \\
6C\,0136+388           & 01 36 59.09 & +38 48 09.2 & c  & dt &  4.2        &             & {\bf C1-A} & X1-C         & 6C\,0136+388 \\
6C\,0139+344           & 01 39 25.87 & +34 27 02.5 & c? & t? &  6.5        &             & {\bf C1-A} & X1-C         & 6C\,0139+344 \\
6C\,0140+326           & 01 40 51.53 & +32 38 45.8 & h  & c  &  2.5        &{\bf L1-A}   &            & X2-BnA       & 6C\,0140+326 \\
& & & & & & & & & \\			             
6C\,0141+425           & 01 41 42.00 & +42 29 15.5 & h  & d? & 113?        &             &            & {\bf X4-DnC} & 6C\,0141+415 \\
6C\,0142+427           & 01 42 28.13 & +42 42 41.6 & c  & t  & 15.2        &{\bf L1-A}   &            & X1-C         & 6C\,0142+427 \\
6C\,0152+463           & 01 52 38.22 & +46 22 30.4 & c? & dt?&  5.7        &L3-BnA       & {\bf C1-A} & X1-C         & 6C\,0152+463 \\
6C\,0154+450           & 01 54 24.85 & +45 03 46.4 & c  & t  & 21.6        &             &            & {\bf X1-C}   & 6C\,0154+450 \\
6C\,0155+424           & 01 55 29.74 & +42 25 38.0 & p  & a  &  4.0        &{\bf L3-BnA} &            & X1-C         & 6C\,0155+424 \\
6C\,0158+315           & 01 58 01.53 & +31 31 47.0 & p  & a  &  1?         &             & {\bf C1-A} & X1-C         & 6C\,0158+315 \\
& & & & & & & & & \\
6C\,0201+499           & 02 01 08.41 & +49 54 38.1 & h  & c  &  0.9        &             & {\bf C1-A} & X1-C         & 6C\,0201+499 \\
6C\,0202+478           & 02 02 07.03 & +47 51 42.2 & h  & c  & 16.5        & {\bf L1-A}  &            & X1-C         & 6C\,0202+478 \\
6C\,0208+344           & 02 08 55.47 & +34 29 57.3 & h  & c  &  2.2        &             & {\bf C1-A} & X1-C         & 6C\,0208+344 \\
6C\,0209+476           & 02 09 20.01 & +47 39 16.5 & p  & u  &  $<0.5$     &             & {\bf C1-A} & X1-C         & 6C\,0209+476 \\
\hline\hline
\end{tabular}
\end{center}
{\caption[Table of observations]{\label{tab:radobs} Table summarising
our snapshot observations with the VLA.  The positions of the objects
are indicated in columns 2 \& 3; the positions were measured from the
radio core if present (this is denoted `c' in column 4), or the peak
of the source if unresolved or amorphous (use of these methods are
denoted `p' in column 4), or in the case of a classical double by
taking the mid-point of the hotspot peaks (use of this method is
denoted `h' in column 4). The structures of each object as inferred
from inspection of the VLA maps are indicated in column 5 as follows:
u: unresolved, c: classical double, t: classical triple, a:
amorphous/partially resolved; a `d' preceding `c' or `t' indicates
significant distortion in the structure. Largest angular sizes (or
separation of the hotspots for the classical doubles) are indicated in
column 5; the units are arcseconds. $^{\ddag}$For this source the
position and structure are quoted from de Ruiter et al.\ (1986) ---
the `g' in column 4 indicates that the position given is that of the
galaxy identification.  Observations we made using the VLA at 1.5, 4.9
and 8.4 GHz are denoted by the letters L, C, and X respectively
followed by a number referring to the date on which the observations
were made. Data presented as maps in this paper are indicated by bold
type.
The codes are as follows:
X1-C:  1990 November 8, C-array.
C1-A:  1991 August 4, A-array.
L1-A:  1991 August 2, A-array.
L2-A:  1995 July 29, A-array.
X2-BnA:  1995 September 25, BnA-array.
X3-BnA:  1995 September 27, BnA-array.
L3-BnA:  1995 September 27, BnA-array.
L4-DnC:  1996 May 21, DnC-array.
X4-DnC:  1996 May 21, DnC-array.
}}
\end{table*}
\normalsize

\begin{table*}
\begin{center}
\begin{tabular}{l|l|r|r|r|r|r|r}
\hline\hline
\mc{1}{c|}{{\bf Source}} &\mc{1}{c|}{{\bf Beam}} & \mc{1}{c|}{{\bf Peak}} & \mc{1}{c|}{{\bf Lowest }} 
& \mc{1}{c|}{VLA L} & \mc{1}{c|}{VLA C} & \mc{1}{c|}{VLA X} &
\mc{1}{c|}{$\alpha_{\rm VLA}$} \\
\mc{1}{c|}{{\bf name}} &\mc{1}{c|}{{\sc fwhm}} & \mc{1}{c|}{{\bf
flux}} & \mc{1}{c|}{{\bf contour}} 
& \mc{1}{c|}{1.5 GHz} & \mc{1}{c|}{4.9 GHz} & \mc{1}{c|}{8.4 GHz} & \\
\hline\hline
6C\,0020+440  & 1.40$^2$                                & 99.58 & 1.00 & $166.4 \pm 3.3$\ph  &                    & $21.0 \pm 0.7$\ph   & $1.12 \pm 0.02$\ph \\
6C\,0024+356  & 0.71 $\times$ 0.59 p.a.\ 112$^{\circ}$  &  2.12 & 0.25 &                     &                    & $18.1 \pm 0.6$\ph   & $1.04 \pm 0.02$\ph \\
6C\,0031+403  & 1.40$^2$                                & 16.31 & 0.88 & $>134.8 \pm 4.7$\ph &                    & $48.8 \pm 1.0^{*}$  & $0.87 \pm 0.03$\ph \\
6C\,0032+412  & 0.50$^2$                                & 13.27 & 0.35 &  $79.9 \pm 2.4\dag$ & $24.1 \pm 0.4$\ph  & $12.8 \pm 0.5$\ph   & $1.14 \pm 0.01$\ph \\
6C\,0041+469  & 0.45$^2$                                &  9.00 & 0.60 &                     & $35.2 \pm 1.5$\ph  & $17.3 \pm 0.6$\ph   & $1.11 \pm 0.02^{c}$\\
6C\,0050+419  & 0.50$^2$                                &  7.53 & 0.50 &                     & $19.7 \pm 0.8$\ph  & $14.6 \pm 0.6$\ph   & $1.09 \pm 0.03$\ph \\ 
& & & & & & & \\
6C\,0052+471  & 1.45$^2$                                & 48.97 & 1.20 & $145.1 \pm 5.5$\ph  &                    & $24.1 \pm 0.8$\ph   & $0.99 \pm 0.01$\ph  \\
6C\,0058+495  & 0.45$^2$                                & 18.98 & 0.55 &                     & $36.3 \pm 1.1$\ph  & $19.2 \pm 0.7$\ph   & $0.96 \pm 0.02$\ph \\
6C\,0100+312  & 1.25$^2$                                & 14.22 & 0.40 & $>49.6 \pm 0.7\ddag$&                    & $>16.6 \pm 1.1^{*}$ & $1.39 \pm 0.20$\ph \\
6C\,0106+397  & 1.86 $\times$ 1.41 p.a.\ 55$^{\circ}$   & 81.85 & 1.00 & $145.9 \pm 1.5$\ph  &                    &                     & $1.10 \pm 0.16^{c}$ \\
6C\,0107+448  & 1.12 $\times$ 0.97 p.a.\ 2$^{\circ}$    &  1.34 & 0.20 &                     &                    &  $>4.5 \pm 0.4$\ph  & $1.35 \pm 0.03$\ph \\
6C\,0111+367  & 0.40$^2$                                & 14.70 & 0.50 &                     & $31.8 \pm 1.1$\ph  & $20.3 \pm 0.7$\ph   & $1.02 \pm 0.02$\ph \\
& & & & & & & \\
6C\,0112+372  & 0.45$^2$                                &  4.05 & 0.50 &                     & $14.2 \pm 1.1$\ph  &  $7.8 \pm 0.4^{*}$  & $1.21 \pm 0.58^{c}$ \\ 
6C\,0115+394  & 0.72 $\times$ 0.60 p.a.\ $-87^{\circ}$  &  2.00 & 0.25 &                     &                    &  $>8.7 \pm 0.5^{*}$ & $1.17 \pm 0.01$\ph \\ 
6C\,0118+486  & 1.50$^2$                                & 24.94 & 0.80 & $70.1 \pm 3.1$\ph   &                    & $12.4 \pm 1.1^{*}$  & $1.09 \pm 0.02$\ph \\
6C\,0122+426  & 0.71 $\times$ 0.60 p.a.\ 59$^{\circ}$   &  8.02 & 0.25 &                     &                    & $12.1 \pm 0.3$\ph   & $1.09 \pm 0.53^{c}$\\
6C\,0128+394  & 0.75 $\times$ 0.57 p.a.\ 51$^{\circ}$   & 19.85 & 0.30 &                     &                    & $43.5 \pm 0.5$\ph   & $0.93 \pm 0.27^{c}$\\
6C\,0132+330  & 7.55 $\times$ 6.12 p.a.\ 71$^{\circ}$   &  8.80 & 0.30 &                     &                    &  $9.3 \pm 0.3$\ph   & $1.27 \pm 0.03$\ph \\
& & & & & & & \\
6C\,0133+486  & 0.78 $\times$ 0.57 p.a.\ 50$^{\circ}$   &  3.69 & 0.30 &                     &                    & $>6.1 \pm 0.5\ddag$ & $^{d}$ \\
6C\,0135+313  & 1.63 $\times$ 1.19 p.a.\ 86$^{\circ}$   & 43.07 & 0.40 &  $83.4 \pm 0.6$\ph  &                    &                     & $1.19 \pm 0.04$\ph \\
6C\,0136+388  & 0.45$^2$                                & 32.03 & 0.50 &                     & $23.4 \pm 1.1$\ph  &  $9.4 \pm 0.6$\ph   & $1.12 \pm 0.38^{c}$\\
6C\,0139+344  & 0.50$^2$                                & 11.55 & 0.40 &                     & $>34.4 \pm 1.6$\ph & $17.4 \pm 0.8$\ph   & $1.02 \pm 0.21^{c}$\\
6C\,0140+326  & 1.30$^2$                                & 41.76 & 0.80 &  $83.1 \pm 1.1$\ph  & $17.1 \pm 0.5\dag$ &  $6.9 \pm 0.3$\ph   & $1.22 \pm 0.12^{c}$\\
6C\,0141+425  & 7.21 $\times$ 5.89 p.a.\ 85$^{\circ}$   &  0.81 & 0.40 &                     &                    & $>3.4 \pm 0.6^{*}$  & $^{d}$ \\
& & & & & & & \\
6C\,0142+427  & 1.45$^2$                                & 25.58 & 1.00 & $117.9 \pm 2.3$\ph  &                    & $20.7 \pm 1.0$\ph   & $1.07 \pm 0.03$\ph \\
6C\,0152+463  & 0.50$^2$                                & 14.43 & 0.50 & $144.7 \pm 0.4$\ph  & $43.4 \pm 0.9$\ph  & $22.5 \pm 0.6$\ph   & $0.99 \pm 0.02$\ph \\
6C\,0154+450  & 2.44 $\times$ 2.07 p.a.\ 42$^{\circ}$   &  8.36 & 0.40 &                     &                    & $20.0 \pm 1.0$\ph   & $1.00 \pm 0.03$\ph \\
6C\,0155+424  & 4.19 $\times$ 1.56 p.a.\ 87$^{\circ}$   & 51.71 & 0.30 & $163.5 \pm 0.5$\ph  &                    & $27.5 \pm 0.7$\ph   & $0.99 \pm 0.02$\ph \\
6C\,0158+315  & 0.50$^2$                                & 38.93 & 0.30 &                     & $49.8 \pm 0.6$\ph  & $28.2 \pm 0.6$\ph   & $1.00 \pm 0.02$\ph \\
6C\,0201+499  & 0.47 $\times$ 0.42 p.a.\ $-36^{\circ}$  & 21.19 & 0.30 &                     & $25.1 \pm 0.5$\ph  & $10.8 \pm 0.6$\ph   & $1.10 \pm 0.01$\ph \\
& & & & & & & \\
6C\,0202+478  & 1.29 $\times$ 1.13 p.a.\ $18^{\circ}$   & 43.22 & 0.80 &  $85.0 \pm 2.2$\ph  &                    &  $7.7 \pm 1.0$\ph   & $1.10 \pm 0.02$\ph \\
6C\,0208+344  & 0.45 $\times$ 0.43 p.a.\ $30^{\circ}$   & 18.34 & 0.30 &                     & $28.8 \pm 0.6$\ph  & $13.7 \pm 0.8$\ph   & $1.02 \pm 0.03^{c}$\\
6C\,0209+276  & 0.47 $\times$ 0.42 p.a.\ $-38^{\circ}$  & 18.13 & 0.30 &                     & $25.8 \pm 0.4$\ph  & $12.1 \pm 0.6$\ph   & $1.12 \pm 0.03^{c}$\\
\hline\hline
\end{tabular}
\end{center}
{\caption[junk]{\label{tab:fignotes} 
The columns in this table are from left to right: source name, {\sc
fwhm} (in arcseconds) of the synthesized beam of the radio maps
(Figures~\ref{fig:rad_1}~---~\ref{fig:rad_6}) and position angle of
its major axis if non-circular, peak flux density (in mJy/beam),
lowest contour (in mJy/beam) on the maps shown, integrated flux
densities measured from all our maps given in mJy.  The right-most
column contains the slope of $\log {\rm (flux)}$ -- $\log {\rm
(frequency)}$ data evaluated at 1 GHz (see Section 3.1 for details).
The $\dag$ symbol means that the flux density is from observations in
the literature at a nearby frequency: MERLIN 1658 MHz data in the case
of 6C\,0032+412 (Rawlings et al., in prep.); VLA 4885 MHz data in the
case of 6C\,0140+426 (Rawlings et al. 1996). The $\ddag$ symbol means
that the integrated flux should be considered a lower limit since
spectral fitting implies that some of the source flux has not been
sampled by the shortest VLA baselines used for the observation, while
a $>$ implies that on the basis of angular size considerations, there
is a possibility that the flux has been undersampled.  An asterisk
means that a flat-spectrum core contributes significantly to the total
flux at this frequency. A `c' means that spectral curvature is
required by the flux data: the values of $\beta$ (see Section 3.1 for
definition) are 0.33 (6C\,0041+469), 0.88 (6C\,0106+397), 0.47
(6C\,0112+372), 0.33 (6C\,0122+426), 0.40 (6C\,0128+394), 0.44
(6C\,0136+388), 0.45 (6C\,0139+344), 0.35 (6C\,0140+326), 0.35
(6C\,0208+344) and 0.38 (6C\,0209+476). A `d' means no useful
additional spectral information is available as a result of the VLA
imaging. 
}}
\end{table*}

\subsection{Spectral fitting}

With regard to radio spectral index $\alpha$ the sample was selected
on the basis of a two frequency (151 MHz and 4850 MHz) criterion,
namely, $\alpha_{151M}^{4.85G} > 0.981$.  More sophisticated fitting
of the radio spectra was made possible by consideration of data from
many radio surveys (Table~\ref{tab:flux}), and furthermore in
combination with data from our VLA mapping.  We fitted the flux
density data using a Bayesian polynomial regression analysis which
assessed the posterior probability density function ({\sc pdf}) for
the required order of a polynomial fit (Gull 1988; Sivia 1996); we
chose $x = \log_{10} (\nu/\rm MHz)$ as the dependent variable and
expanded the independent variable $y = \log_{10} S_{\nu} =
\sum_{r=0}^{N} a_{r} x^{r}$.  In most cases the peak in the {\sc pdf}
led us to prefer a first-order polynomial, and the slope of the fit
($- a_{1}$) evaluated to give $\alpha_{1000}$ for the survey data or
$\alpha_{\rm VLA}$ for the full radio dataset.  In some cases the {\sc
pdf} led us to prefer a second-order fit, implying significant
curvature in the radio spectrum: in these cases we calculated the
curvature $\beta = -2 a_{2}$, and $\alpha_{1000}$ (or $\alpha_{\rm VLA}$)
as the slope evaluated at 1 GHz, namely $- a_{1} + 6 \beta$. In no
case was a higher order fit preferred.  In the cases of significant
spectral curvature there are often fairly large errors in the values
of $\alpha_{1000}$ and $\alpha_{\rm VLA}$.

\normalsize

\section{Notes on individual sources}

\begin{figure*}
\begin{center}
\setlength{\unitlength}{1mm}
\begin{picture}(150,220)
\put(0,145){\includegraphics{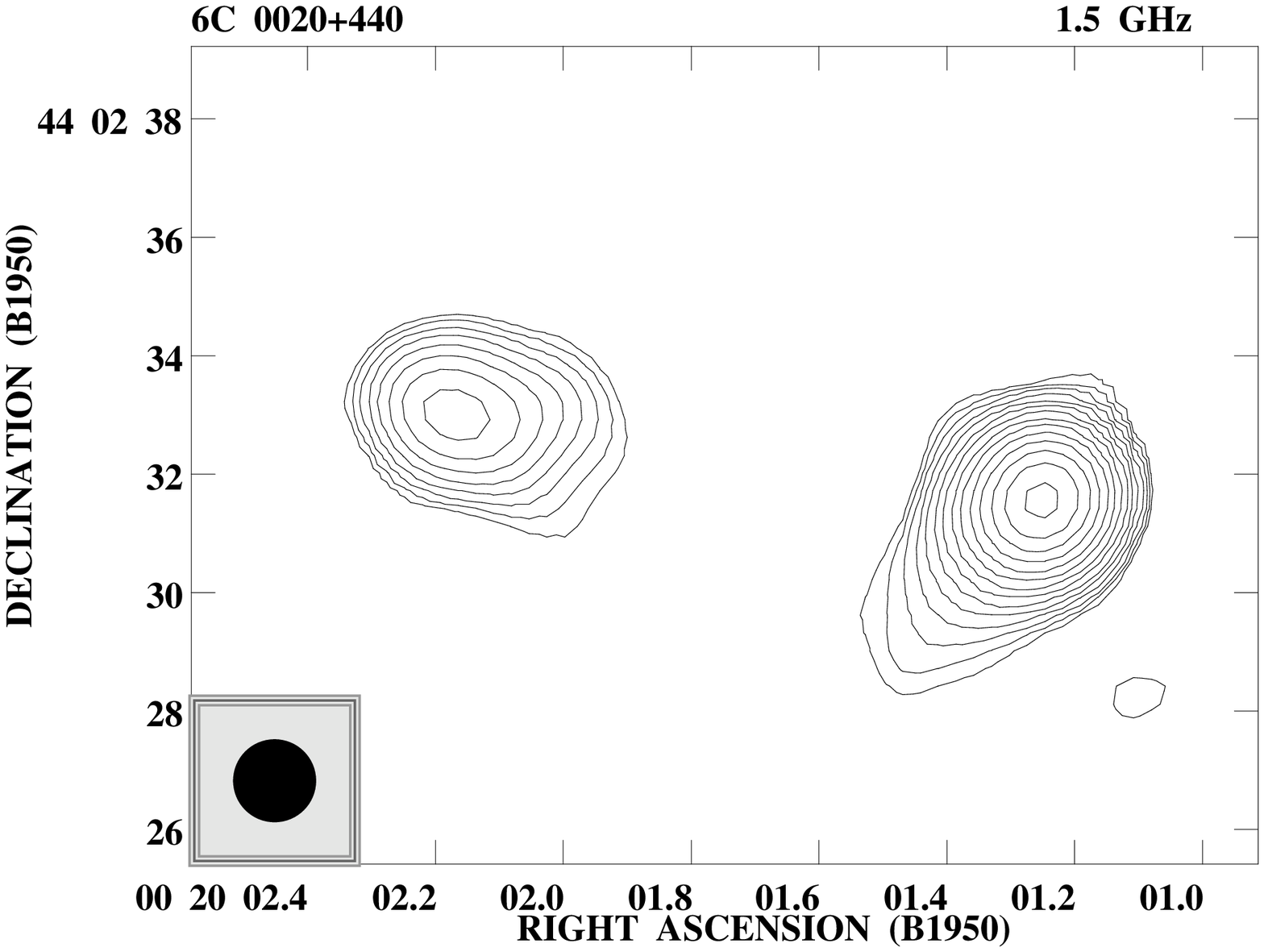}}
\put(75,145){\includegraphics{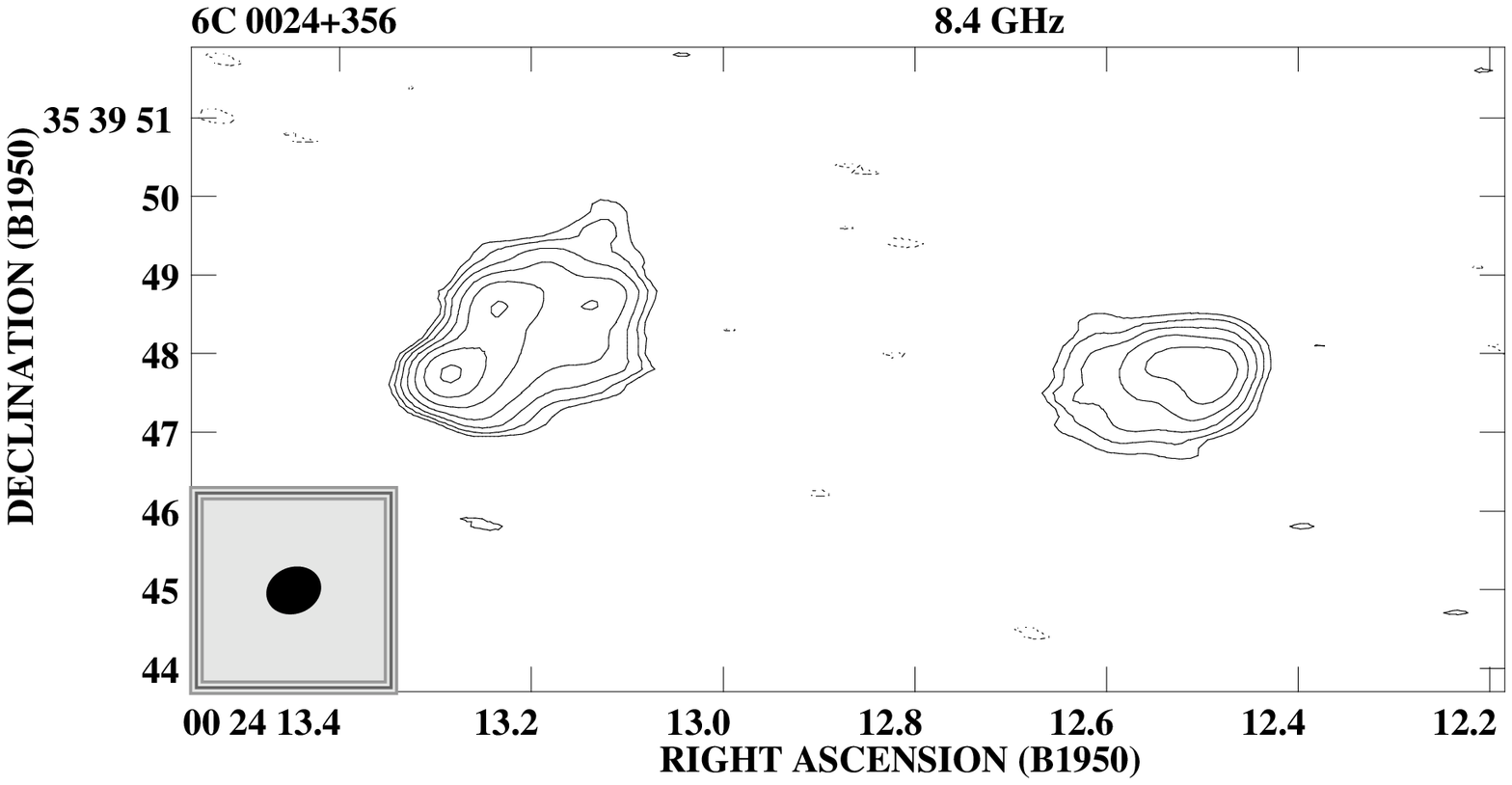}}
\put(0,70){\includegraphics{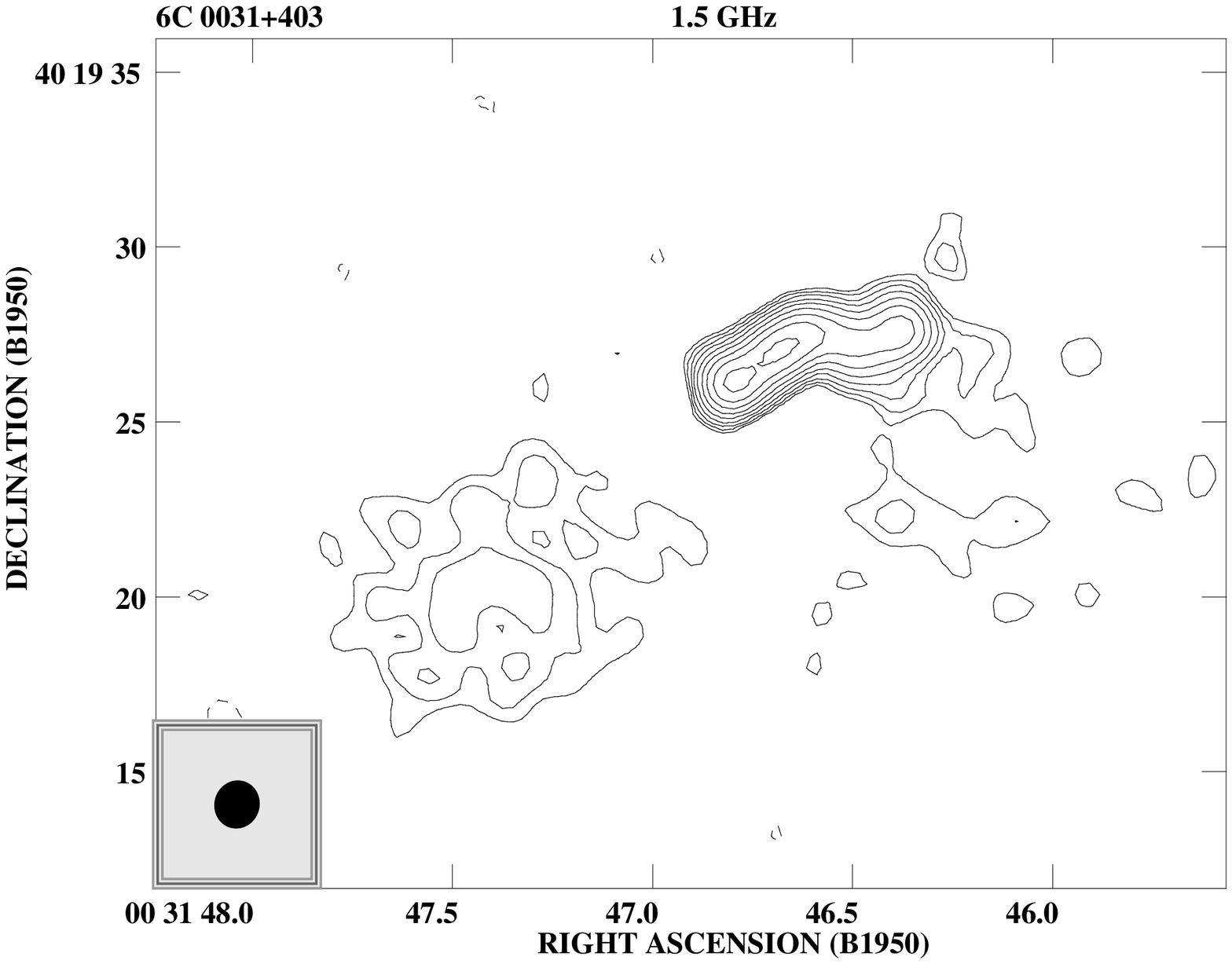}}
\put(75,80){\includegraphics{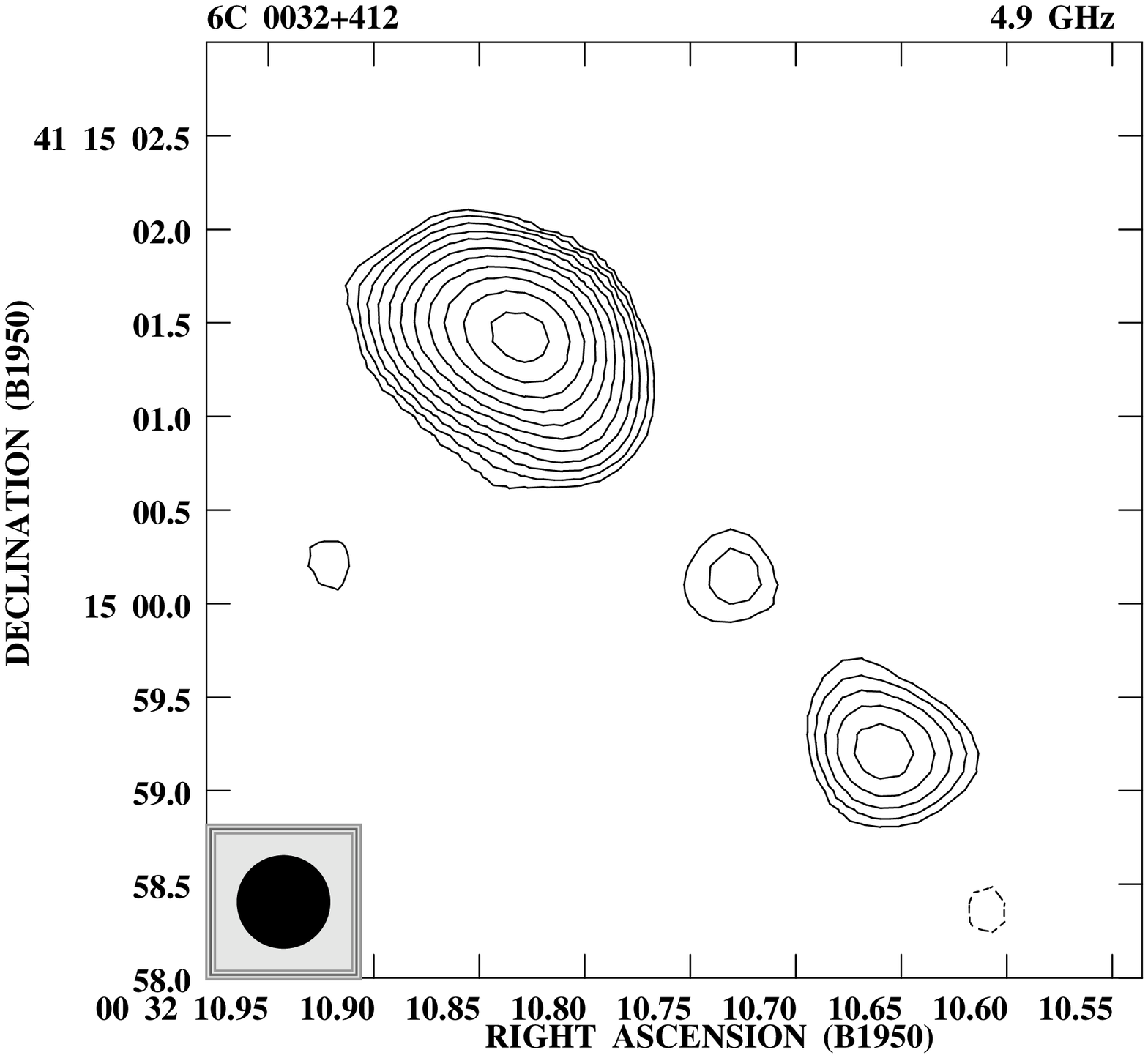}}
\put(0,-5){\includegraphics{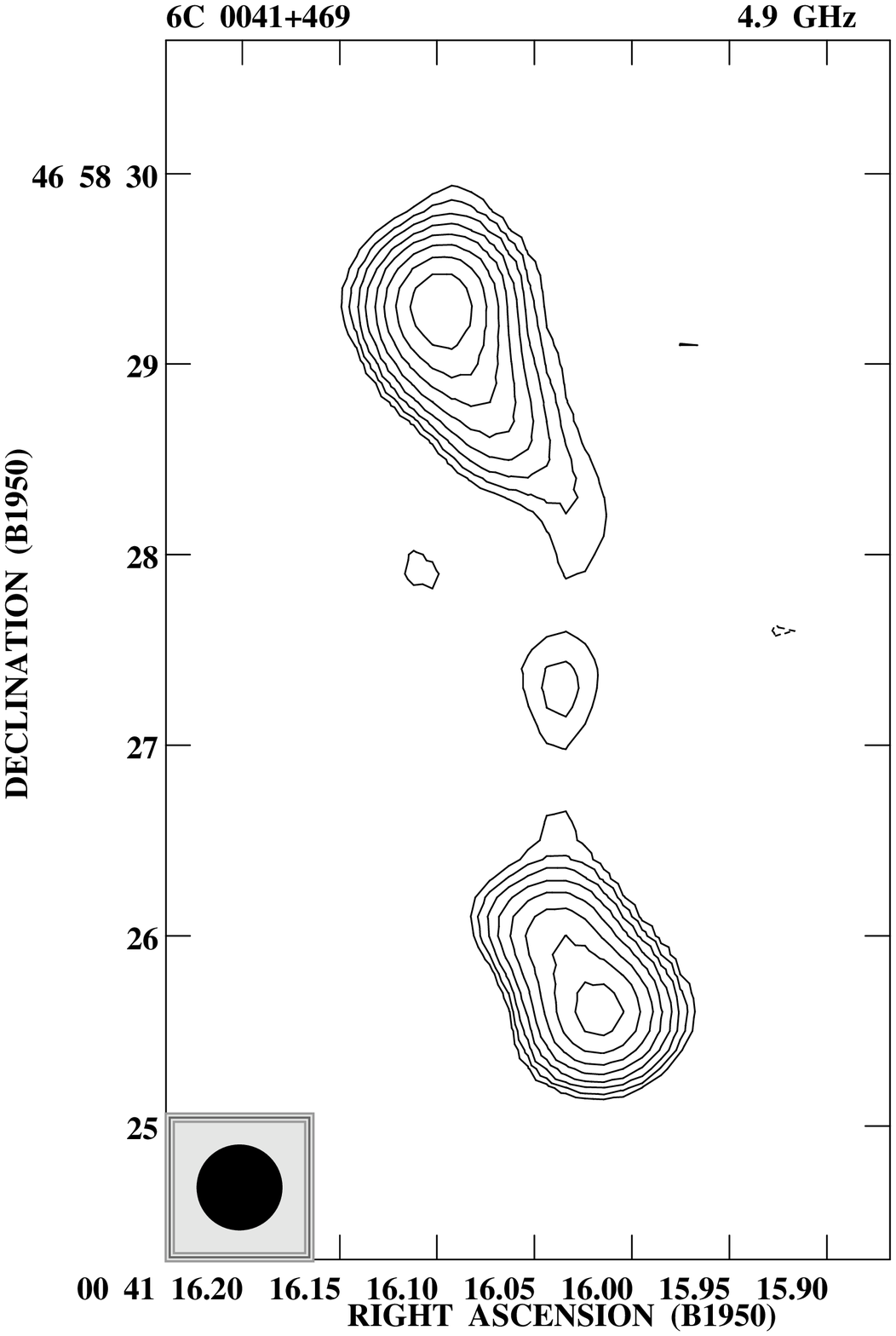}}
\put(80,-5){\includegraphics{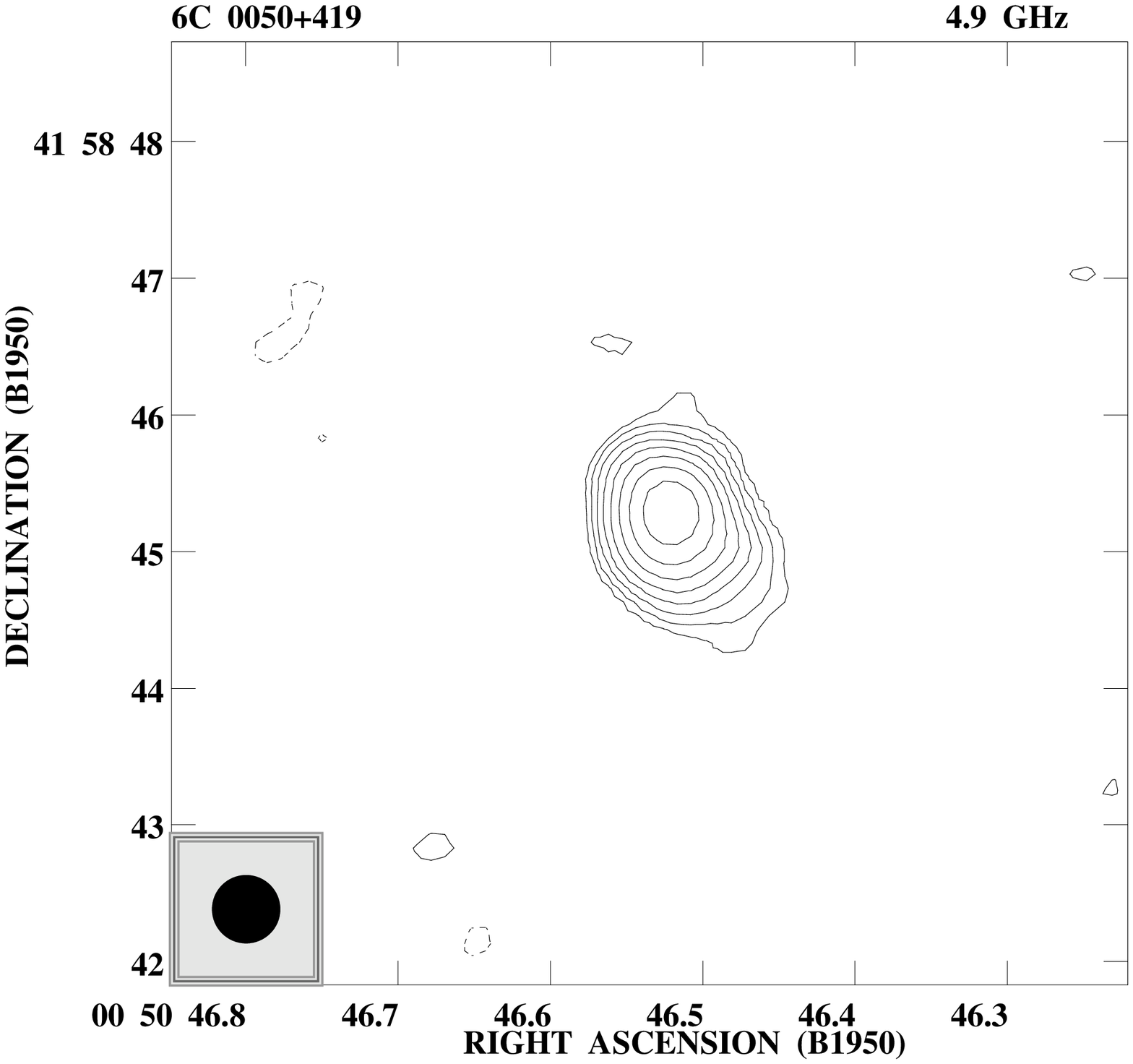}}
\end{picture}
\end{center}
{\caption[junk]{\label{fig:rad_1} Contour maps of the radio sources.}}
\end{figure*}

\begin{figure*}
\begin{center}
\setlength{\unitlength}{1mm}
\begin{picture}(150,220)
\put(0,150){\includegraphics{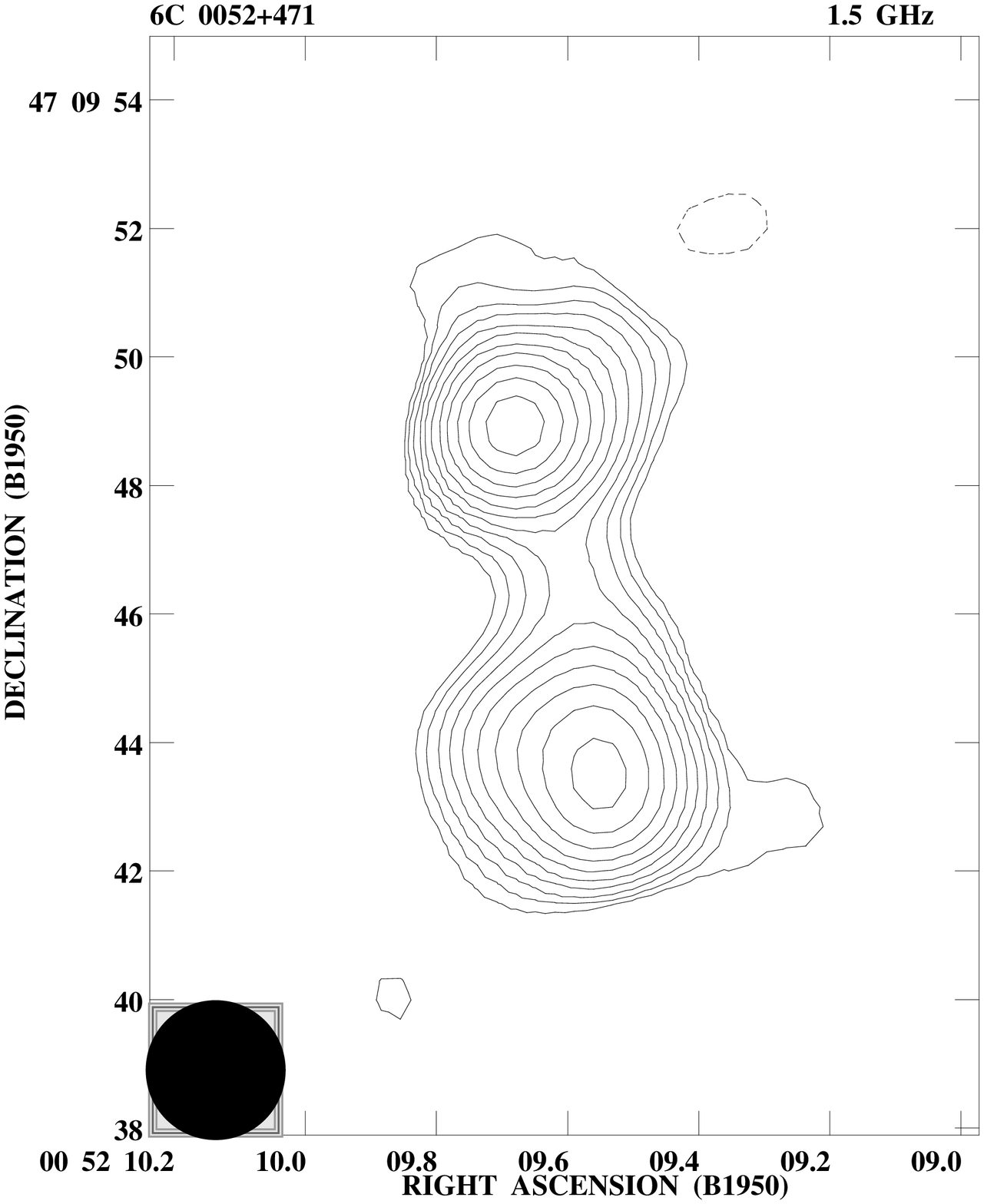}}
\put(75,140){\includegraphics{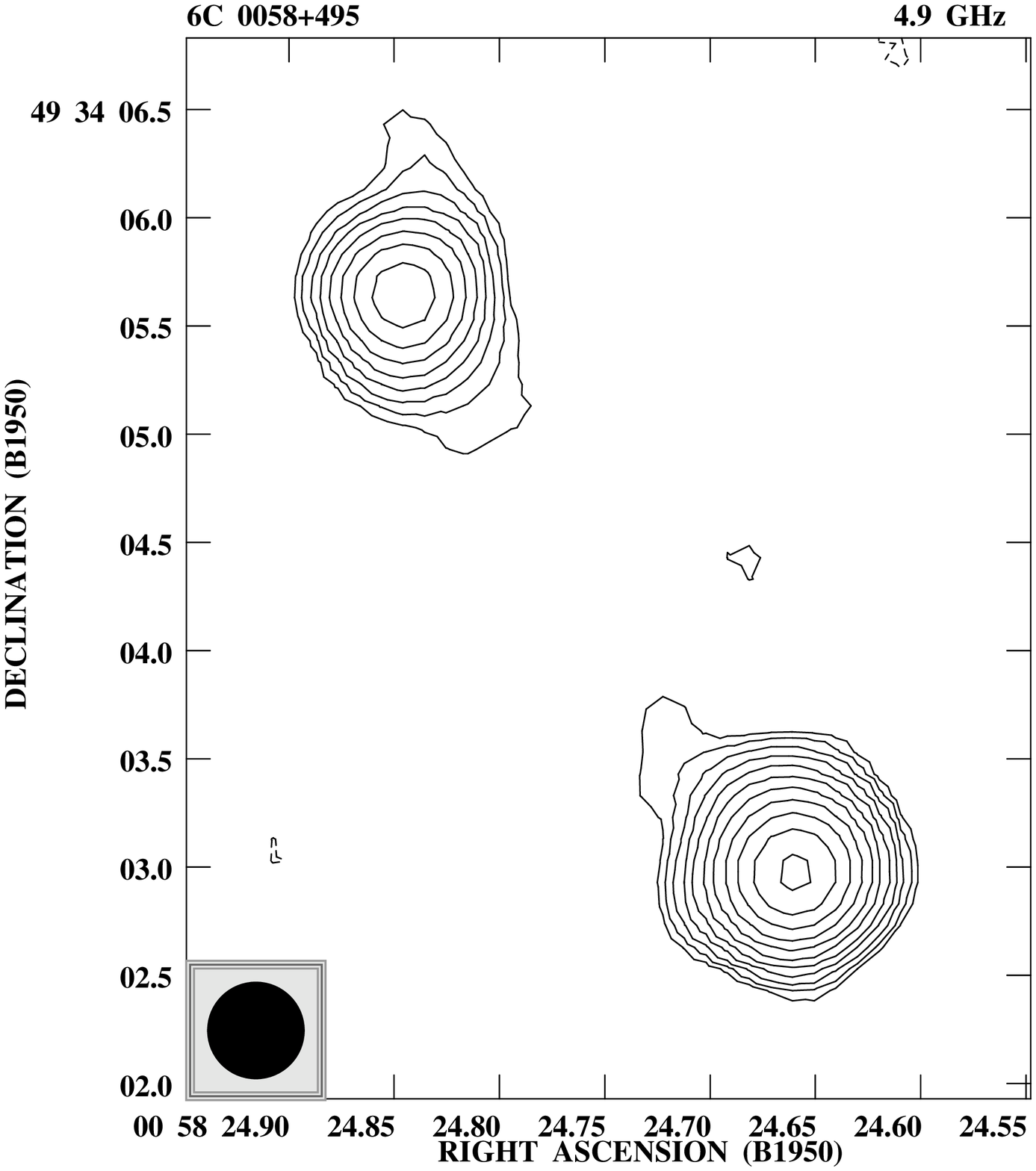}}
\put(0,70){\includegraphics{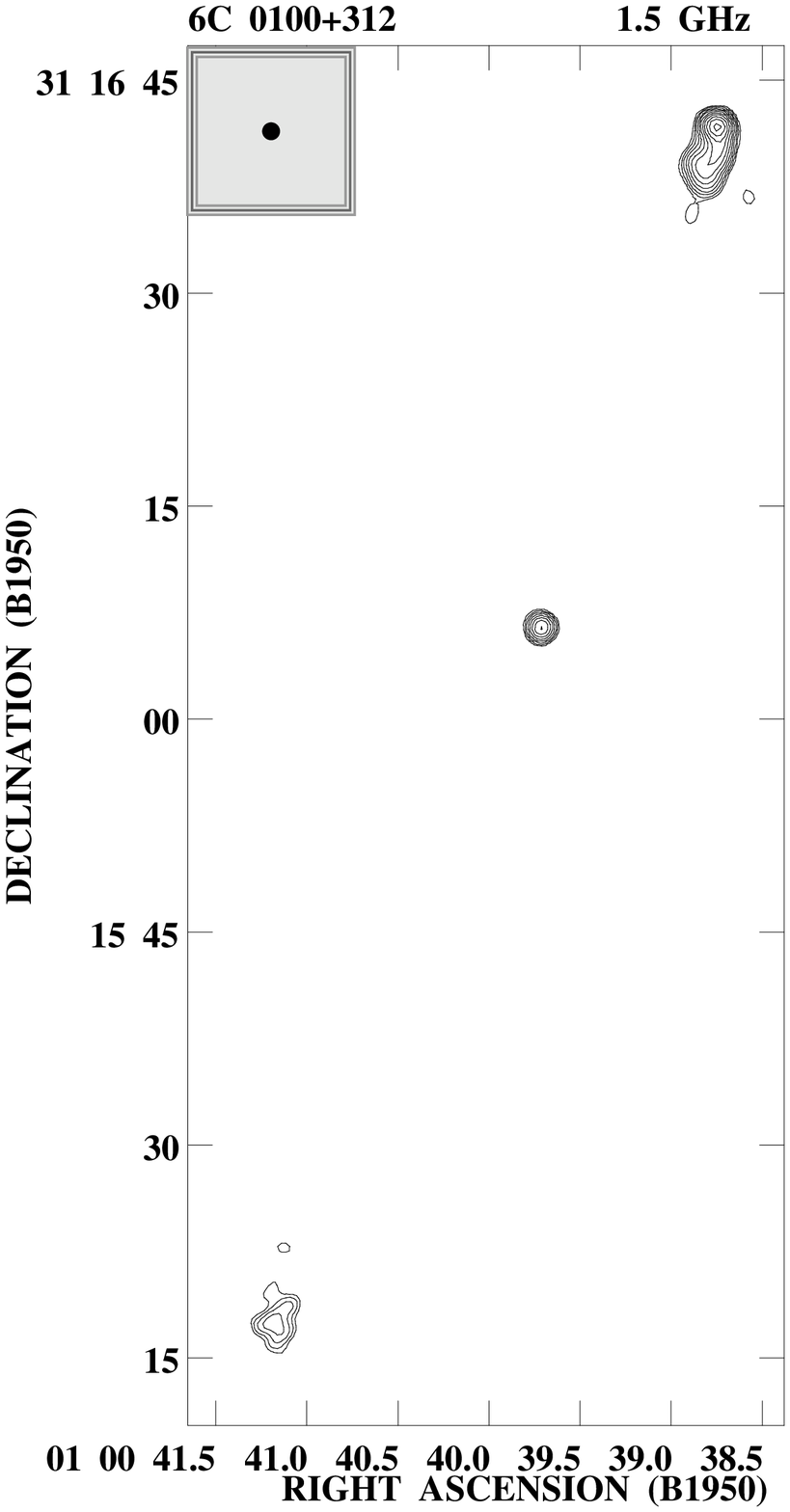}}
\put(75,60){\includegraphics{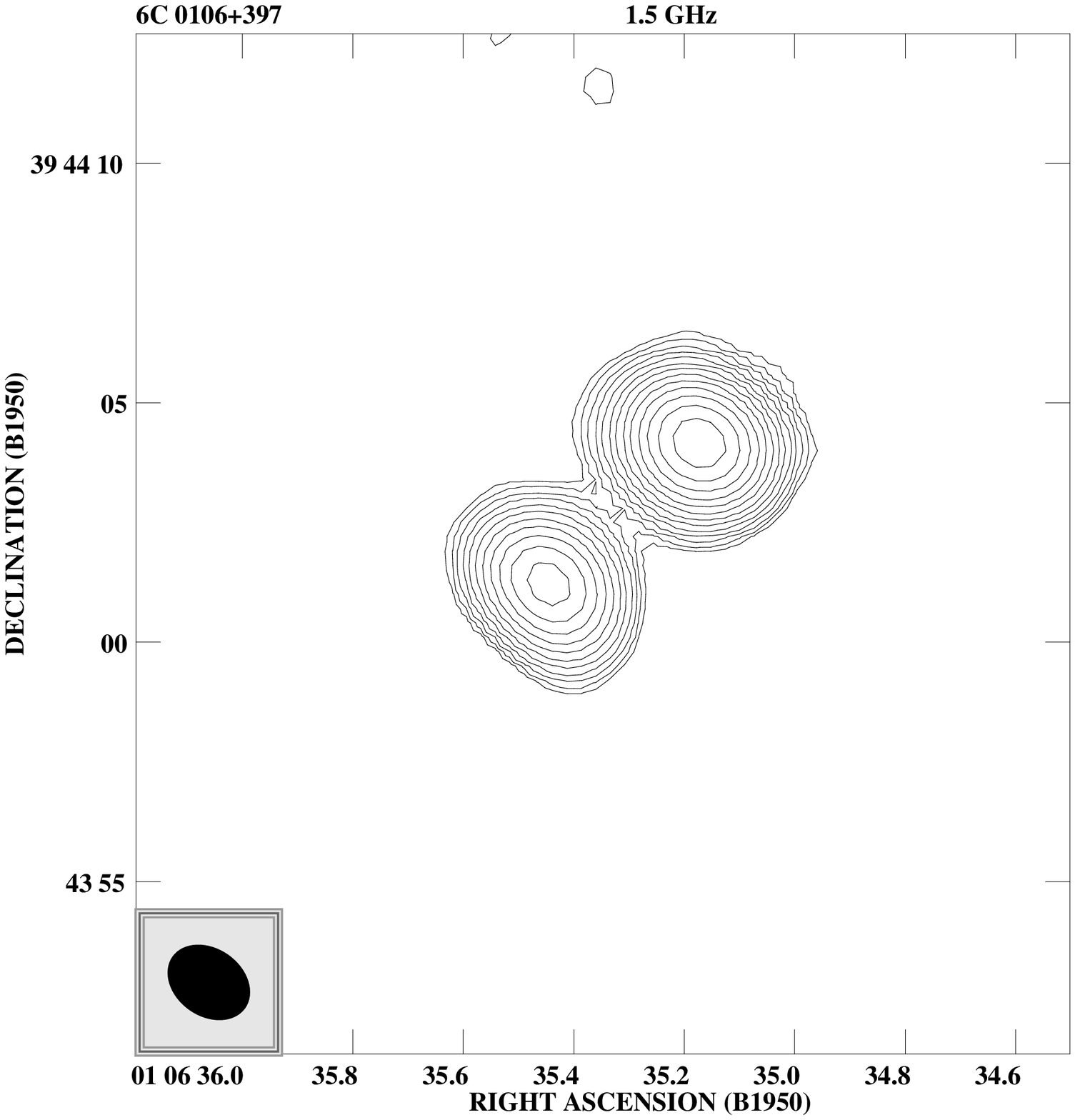}}
\put(0,-5){\includegraphics{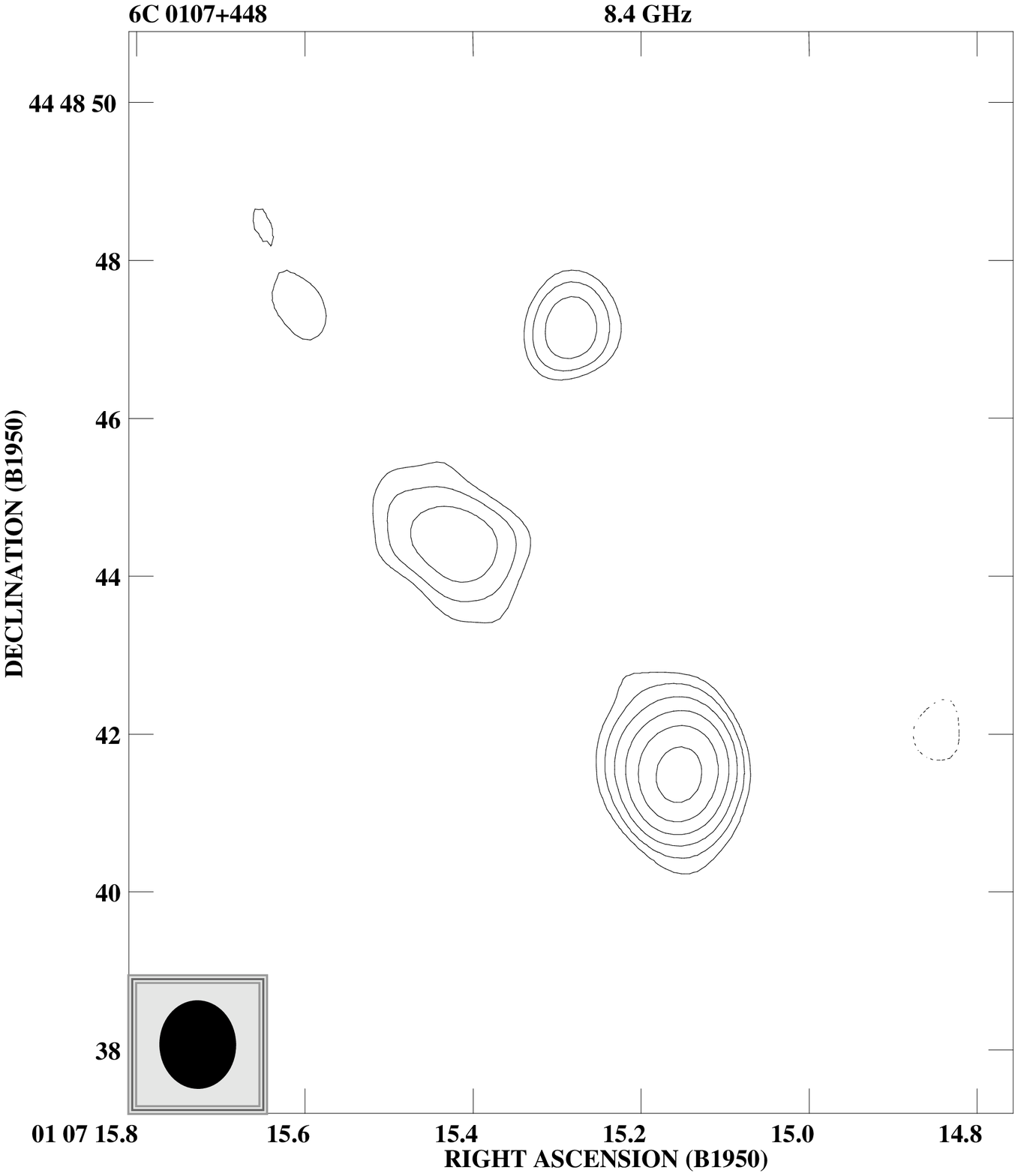}}
\put(80,-5){\includegraphics{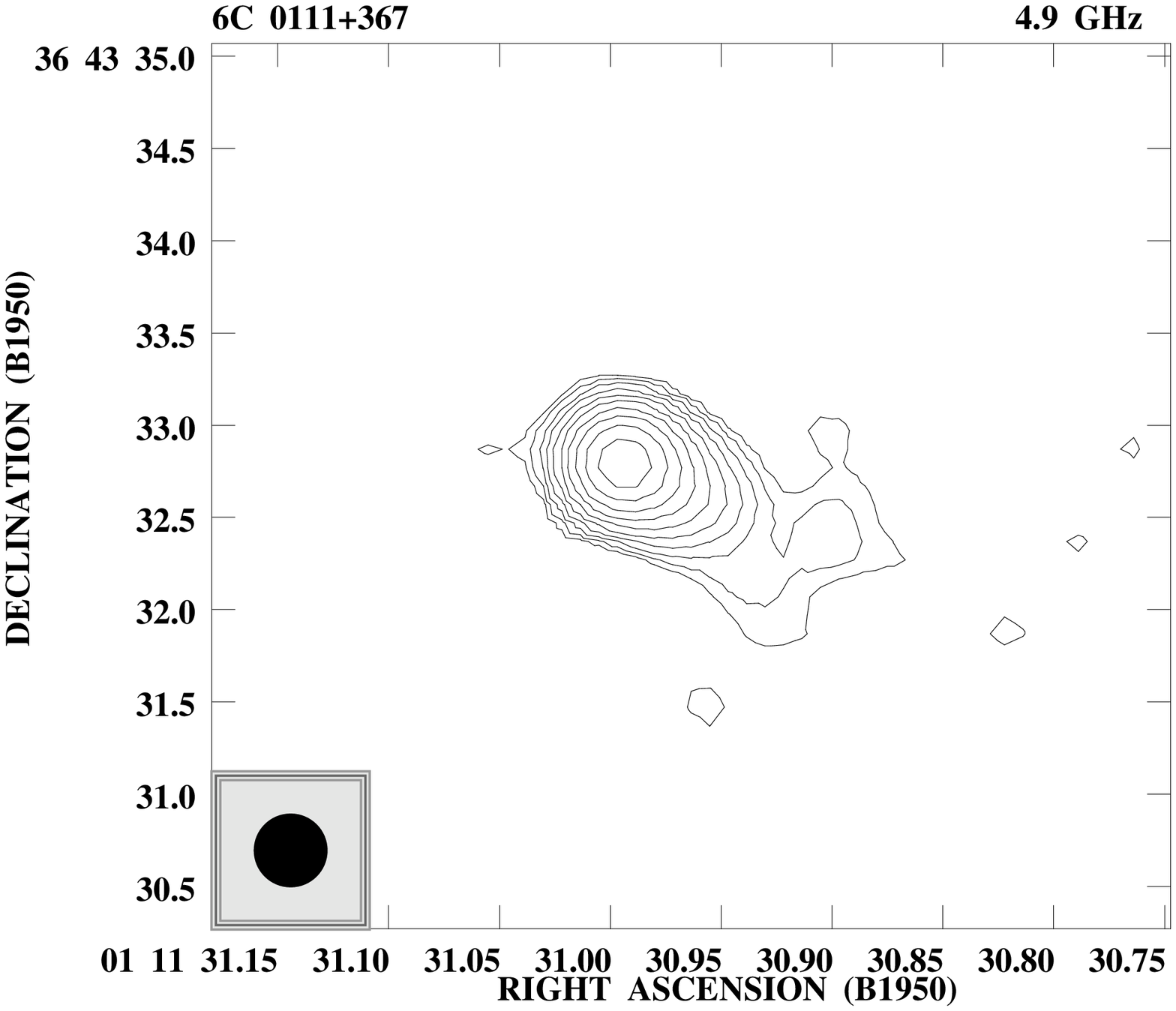}}
\end{picture}
\end{center}
{\caption[junk]{\label{fig:rad_2} Contour maps of the radio sources.}}
\end{figure*}

\begin{figure*}
\begin{center}
\setlength{\unitlength}{1mm}
\begin{picture}(150,220)
\put(0,145){\includegraphics{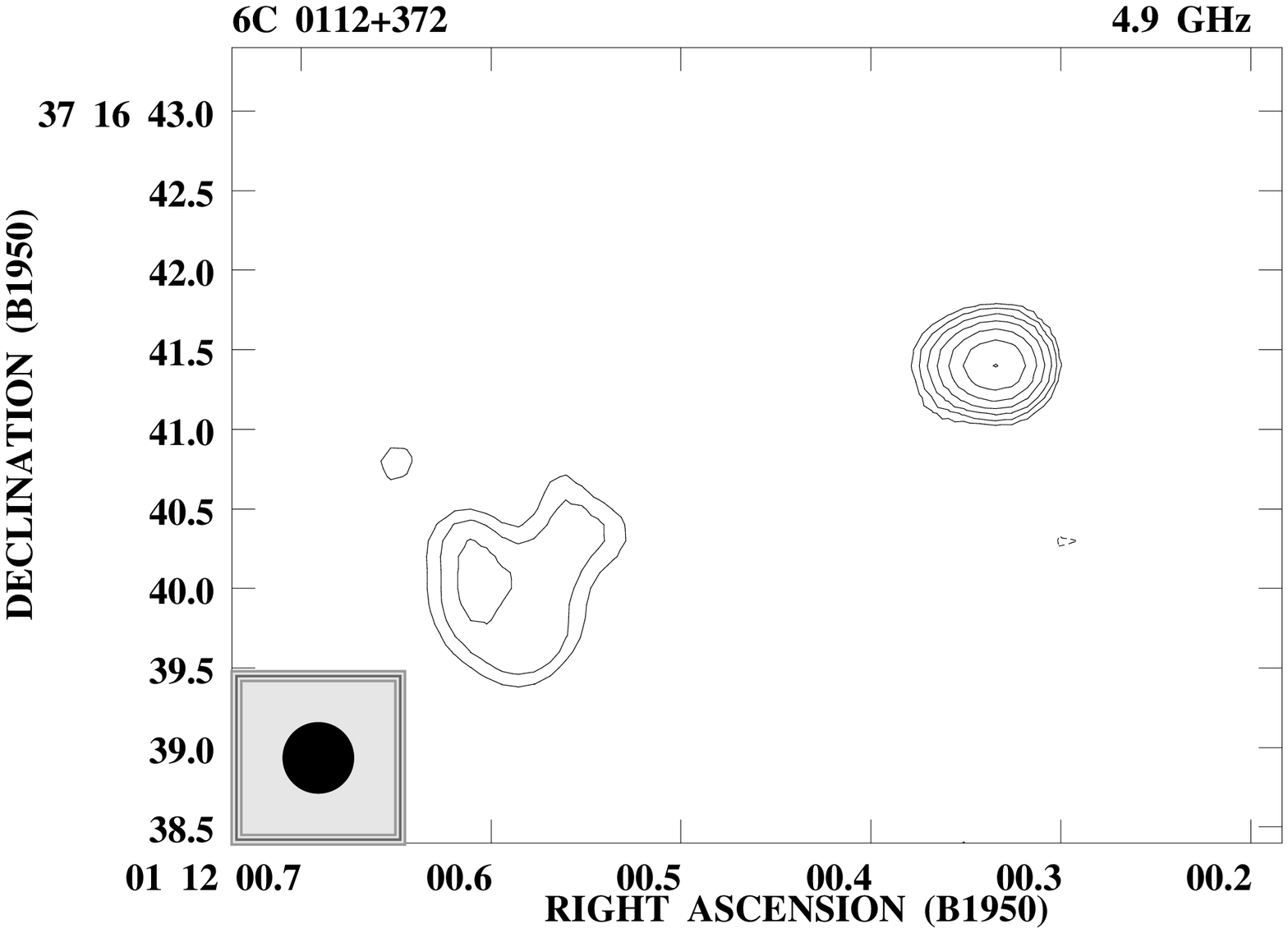}}
\put(75,140){\includegraphics{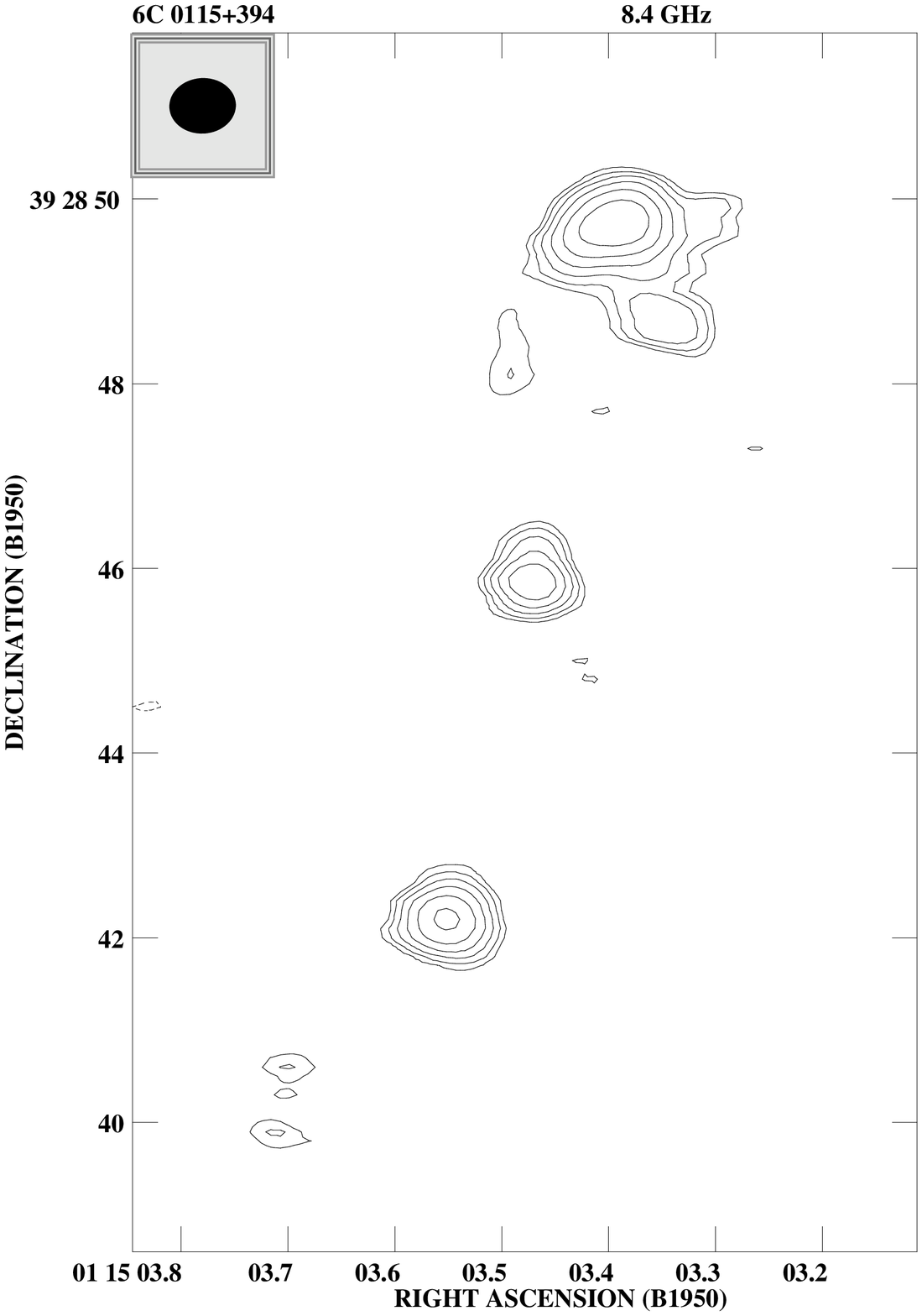}}
\put(0,75){\includegraphics{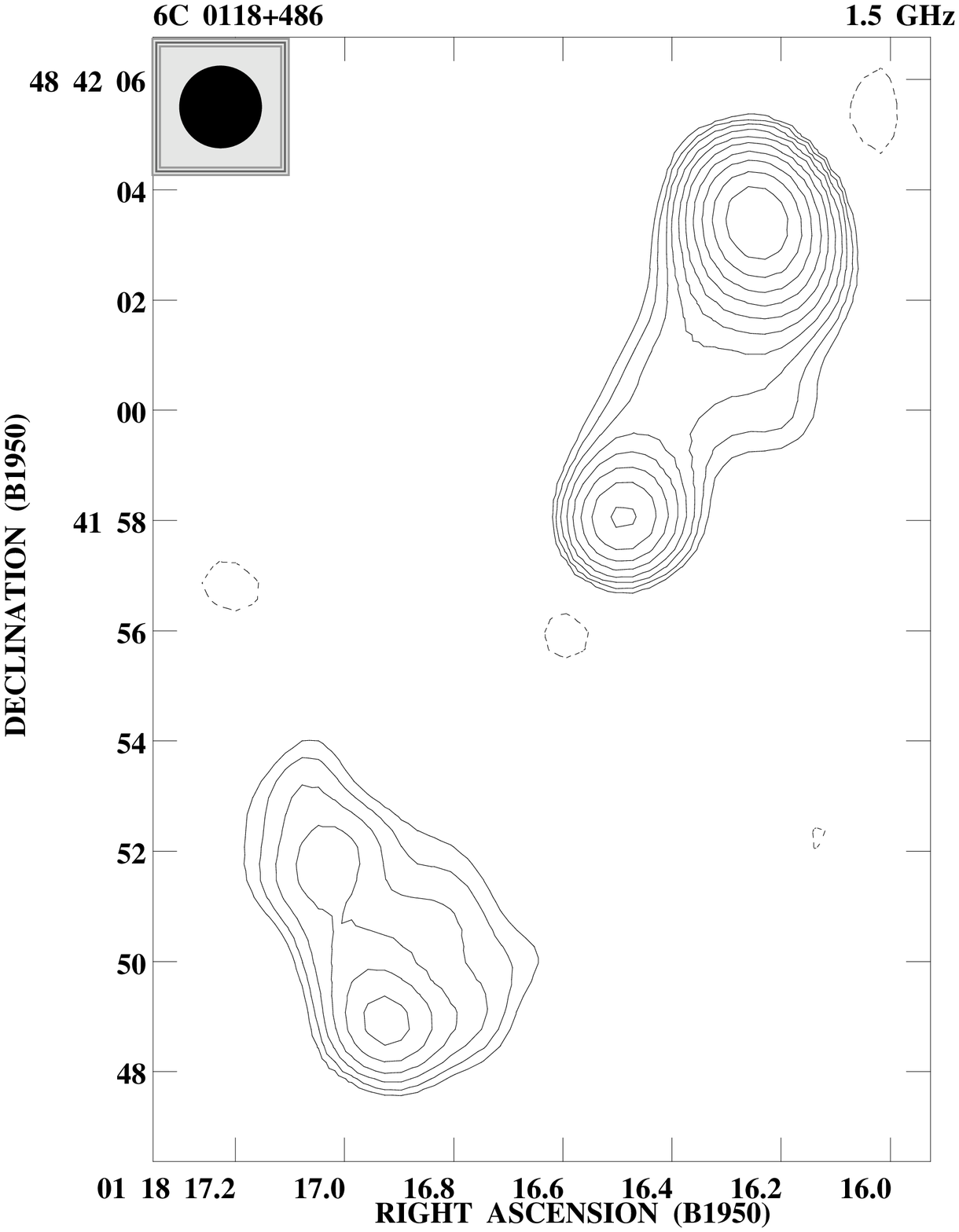}}
\put(75,62){\includegraphics{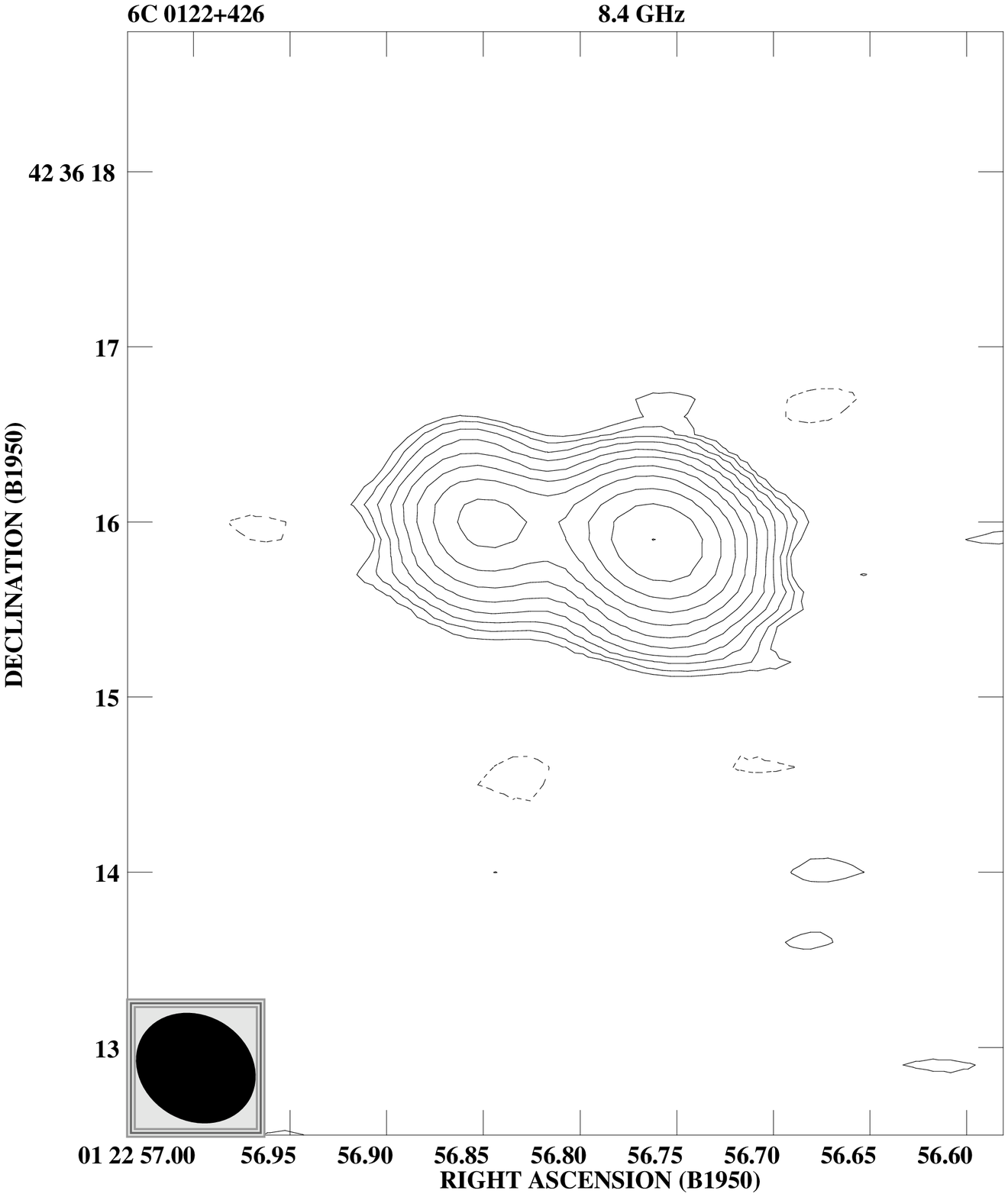}}
\put(0,-5){\includegraphics{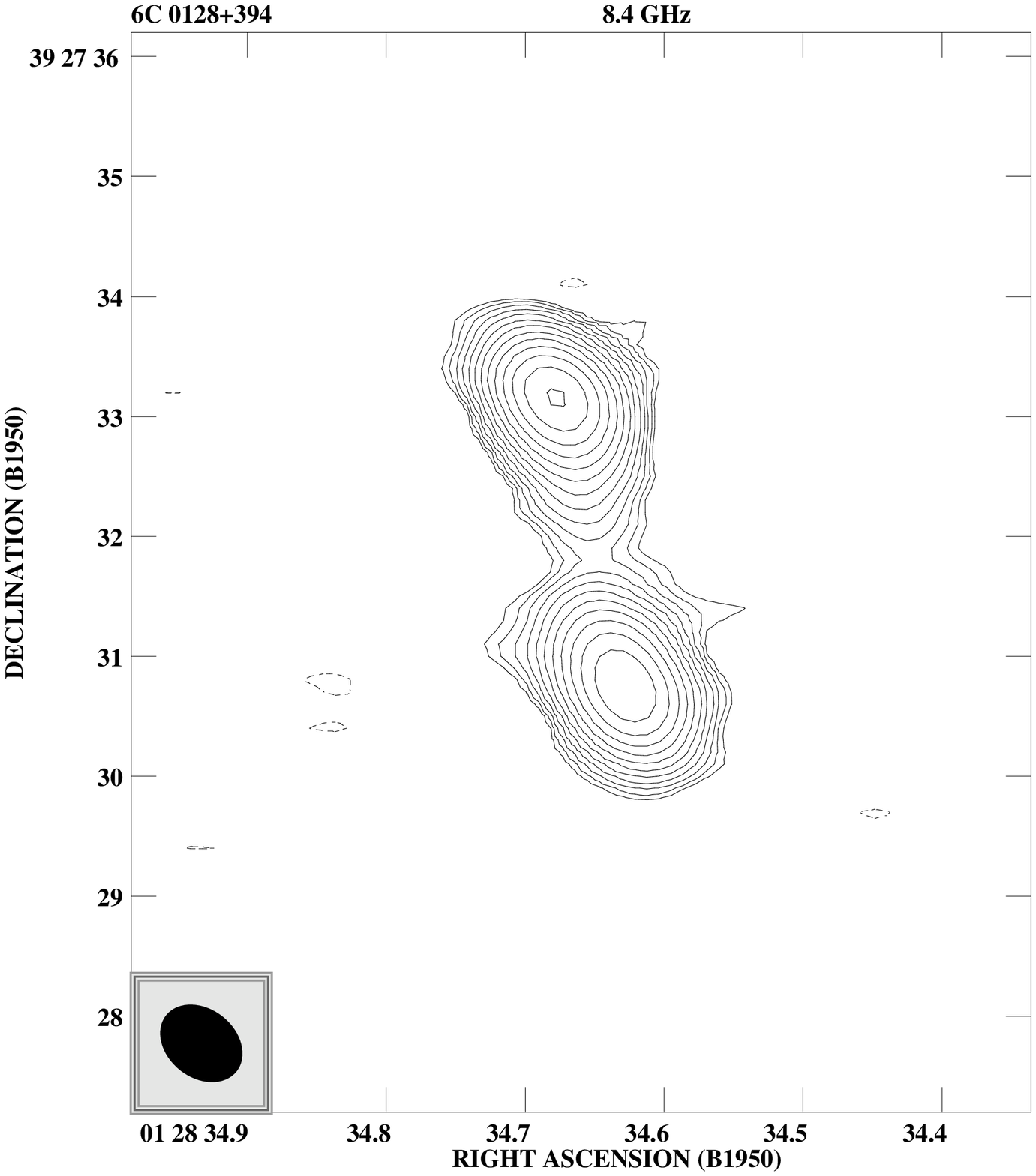}}
\put(80,-10){\includegraphics{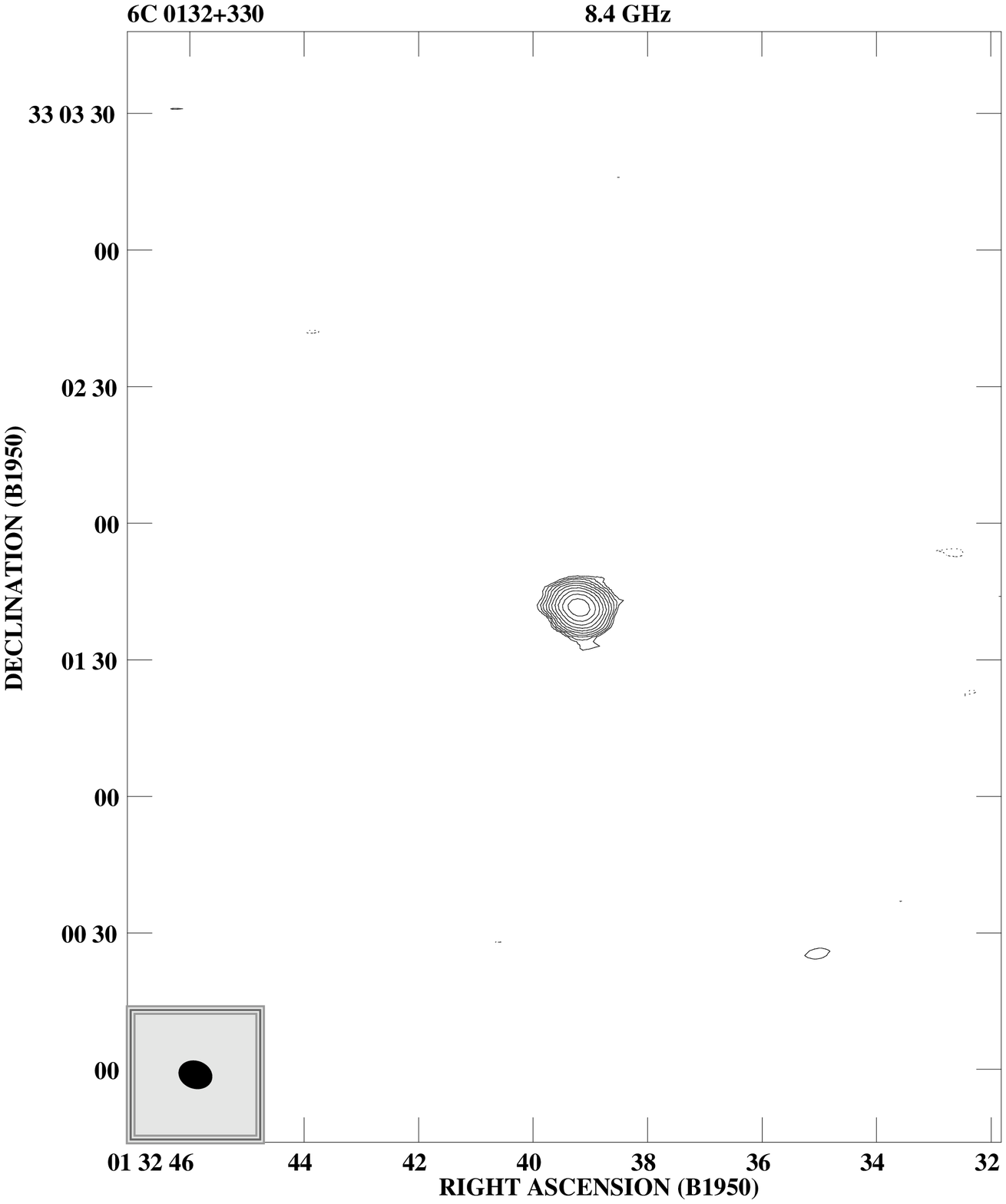}}
\end{picture}
\end{center}
{\caption[junk]{\label{fig:rad_3} Contour maps of the radio sources.}}
\end{figure*}

\begin{description}

\item[{\bf 6C\,0020+440}] Comparison of radio maps at 1.5 and 8.4 GHz show
that both lobes have steep spectra ($\alpha$ \gtsim\ 1.0).

\item[{\bf 6C\,0024+356}] We only have a map of this object at 8.4 GHz but
both components have the resolved appearance of lobes and/or hotspots
and there is no evidence of a core between them.

\item[{\bf 6C\,0031+403}] The core position in Table~\ref{tab:radobs} is from
the 8.4 GHz map, where it is clearly the dominant component, implying a
spectral index $\sim 0$. The diffuse lobes to the south-east and west
of the core are also detected at 8.4 GHz.  This object was found to
have 2 matches in the Texas catalogue, which are separated from one
another by 4.7 arcseconds and fluxes which are 432 and 480 mJy.

\item[{\bf 6C\,0032+412}] The core position in Table~\ref{tab:radobs}
is from the 4.9 GHz map shown in Fig.~\ref{fig:rad_1}; the core has a
spectral index $\sim 0$ (Rawlings et al.\, in prep.).

\item[{\bf 6C\,0041+469}] The core position in Table~\ref{tab:radobs}
is of the weak unresolved component in the 4.9 GHz map shown in
Fig.~\ref{fig:rad_1} although the resolution of the 8.4 GHz map is too low
to allow measurement of the spectral index.

\item[{\bf 6C\,0050+419}] The position given in Table~\ref{tab:radobs}
is the peak of the component seen in Fig.~\ref{fig:rad_1} which appears
to be marginally resolved approximately along p.a.\
$15^{\circ}$. There is some evidence for another component in
the 8.4 GHz map at 00 50 46.12 +41 58 37.52. The structure of this radio
source and the search position for any optical 
identification are therefore uncertain.

\item[{\bf 6C\,0052+471}] Both components shown in the 1.5 GHz map in
Fig.~\ref{fig:rad_2} have steep spectra ($\alpha$ \gtsim\ 1.0).

\item[{\bf 6C\,0058+495}] Both components shown in the 4.9 GHz map in
Fig.~\ref{fig:rad_2} have steep spectra ($\alpha \sim 1.0$). This
object was found to have 2 matches in the Texas catalogue, which are
separated from one another by 2.2 arcseconds and fluxes which are 420
and 444 mJy.

\item[{\bf 6C\,0100+312}] Note that the Texas flux density is likely to be
an underestimate of the true value because of the size of
source.  The peak brightness of the core on the map shown in
Fig.~\ref{fig:rad_2} is $4.7 \pm 0.1$ mJy/beam. The position in
Table~\ref{tab:radobs} is measured from the core seen in the 1.5 GHz
map in Fig.~\ref{fig:rad_2}; the core has a spectral index of $\sim 0$.

\item[{\bf 6C\,0106+397}] Vigotti et al.\ (1989) find this object to be 
unresolved at 1.4 GHz with a flux density of 151 mJy. The radio
spectrum of 6C\,0106+397 between 151 and 1400 MHz requires extreme ($\beta
= 0.88$) curvature at higher frequencies to be consistent with an
assumed 25mJy limit from the 87GB survey.

\item[{\bf 6C\,0107+448}] The position given in Table~\ref{tab:radobs}
is of the eastern-most of the three brightest components shown in
the map in Fig.~\ref{fig:rad_2}. We have no spectral information to
determine whether this feature is in fact a core.

\item[{\bf 6C\,0111+367}] Our 4.9 GHz map shows only the hotspot-like 
feature (with $\alpha \sim 1$) seen in Fig.~\ref{fig:rad_2}; however,
there are two additional very weak features on our 8.4 GHz map at 01 11
29.8 +36 43 27.5 ($2.3 \sigma$) and 01 11 27.8 +36 43 19 ($3.6
\sigma$) which together with the feature seen on the 4.9 GHz map are
roughly co-linear.

\item[{\bf 6C\,0112+372}] The position given in Table~\ref{tab:radobs}
is the position of the brightest component on the 4.9 GHz map in
Fig.~\ref{fig:rad_3}. This has a spectral index $\sim 0$. An eastern
lobe is detected both at 4.9 and 8.4 GHz but no corresponding emission
to the west is seen in either band.

\item[{\bf 6C\,0115+394}]  We have assumed that the component whose
position is given in Table~\ref{tab:radobs} is the core of a triple
source although we have no measure of its spectral index. Vigotti et
al.\ (1989) find this object to be unresolved at 1.4 GHz with a flux
density of 88 mJy.

\item[{\bf 6C\,0118+486}]  The position given in Table~\ref{tab:radobs}
is that of the middle component on the 1.5 GHz map in
Fig.~\ref{fig:rad_3}. The spectral index of this feature is $\sim 0.5$.
Both the northern and southern lobes are detected at 8.4 GHz and have
spectral indices \gtsim\ 1.0.

\item[{\bf 6C\,0120+329}] This is the only object in the sample for which
we do not present a map because a better map is in de Ruiter et al.\
(1986).  This source is associated with NGC507 and has a redshift of
0.0164.

\item[{\bf 6C\,0122+426}] From our single frequency data we are unable to
ascertain whether the core is either of the 
features seen in Fig.~\ref{fig:rad_3}.
The 6C map is marginally resolved in the direction of a second B3
match at 01 23 02.3 42 40 06 (whose B3 flux density 130 mJy) but, with
a separation of 4$^\prime$ from our VLA position, this is unlikely to
contribute significantly to the peak 6C flux density.

\item[{\bf 6C\,0128+394}] We only have a map of this object at 8.4 GHz but
both components have the resolved appearance of lobes and/or hotspots
and there is no evidence of a core between them. 

\item[{\bf 6C\,0132+330}] The 6C flux density of this source is highly
uncertain because of its proximity to 3C\,48.

\item[{\bf 6C\,0133+486}] We only have a map of this object at 8.4 GHz but
the main features seen have the resolved appearance of lobes and/or hotspots.

\item[{\bf 6C\,0135+313}] Our 1.5 GHz map only partially resolves this
source along p.a.\ 30$^{\circ}$.

\item[{\bf 6C\,0136+388}] The brightest feature on the 4.9 GHz map in 
Fig.~\ref{fig:rad_3} has a spectral index $\sim 0$ and we take this to
be the core. The lobe to the south-east is also seen at 8.4 GHz but at
neither frequency is there evidence of a lobe to the north-west.

\item[{\bf 6C\,0139+344}] Both of the two main features on the 4.9 GHz map
have spectral indices \gtsim\ 1; the core position given is for the
weak component between the main components on the 4.9 GHz map. Our
8.4 GHz map has insufficient resolution to confirm whether this is the
core.

\item[{\bf 6C\,0140+326}] This $z = 4.41$ object is discussed in detail in
Rawlings et al.\ (1996). 

\item[{\bf 6C\,0141+425}] There are two sources in B3 at 01 41 46.5 42 29 43
with 408-MHz flux density 390 mJy and at 01 41 37.5 42 28 48.0, with
408-MHz flux density 410 mJy which are within the 6C beam. It is
therefore plausible that this object is a 113\arc\ double. Our 8.4 GHz
data in Fig.~\ref{fig:rad_4} detects only emission associated with the
former B3 object and this emission appears to be resolved, consistent
with it not being detected by the Texas survey. The low resolutions of
the WB and 87GB surveys (listed in the caption to
Table~\ref{tab:flux}) mean that it is not possible to resolve emission
separately from each of the components found in B3.  A further
discussion of these and other data on this object will be found in
Paper II.

\item[{\bf 6C\,0142+427}] The core position given in
Table~\ref{tab:radobs} is that of the feature closest to the centre of
the map presented in Fig.~\ref{fig:rad_4} which also has the flattest
spectrum ($\alpha \sim 0$).

\item[{\bf 6C\,0152+463}] The position given is that of the most
easterly component on the 4.9 GHz map which is shown in
Fig.~\ref{fig:rad_5}. Our 1.5 and 8.4 GHz maps have insufficient
resolution to confirm the identity of this feature as the core.

\item[{\bf 6C\,0154+450}] The position given in Table~\ref{tab:radobs}
is that of the middle component shown in the map in
Fig.~\ref{fig:rad_5}. We have no data at a second frequency.

\item[{\bf 6C\,0155+424}] This source is somewhat resolved along p.a.\ 
$80^{\circ}$.

\item[{\bf 6C\,0158+315}] Our 4.9 GHz data resolves this source,
although it is unclear whether this is a hotspot of a larger source or
a discrete object.  This object was found to have 2 matches in the
Texas catalogue, which are separated from one another by 1.3\arc\ and
flux densities which are 834 and 808 mJy.

\item[{\bf 6C\,0201+499}] This source is most likely to be a small double.

\item[{\bf 6C\,0202+478}] Both components have spectral indices
\gtsim\ 1.

\item[{\bf 6C\,0208+344}] Neither of the components seen in the 4.9
GHz map in Fig.~\ref{fig:rad_5} has a spectral index which implies a
core.

\item[{\bf 6C\,0209+276}] This object appears to be marginally
resolved.

\end{description}

\section{Note on sources from 6C-$\overline{\rm T}$-$\overline{\rm C}$
subsequently excluded from this sample}

\begin{description}

\item[{\bf 6C\,0034+375}] Our own map at 1.4 GHz from VLA-DnC shows
four discrete objects at the following positions: a) 00 33 59.05 +37
52 32.0 (flux density 35.4 mJy), b) 00 34 05.82 +37 33 32.0 (flux
density 143.7 mJy), c) 00 34 10.02 37 27 22.0 (flux density 95.8 mJy)
and d) 00 35 04.05 36 42 51.1 (flux density 29.6 mJy).  Objects a, b,
c and d appear to lie on a curved path. It is most likely (from
inspection of the 6C-map at 151 MHz) that only b and c are
related. Objects a, b and c have counterparts in the 87GB survey and
objects a and b have counterparts in the Texas survey. The 6C position
falls mid-way between b and c and thus these counterparts were not
included by the selection criteria which only searched for
counterparts in a $2^{\prime}$ radius. If b and c form part of one
source, then it fails to meet the spectral index selection
criterion. Object c alone does not satisfy the 6C flux criteria and
object b alone does not satisfy the spectral index
criterion. 6C\,0034+375 is thus excluded from the 6C* sample.

\item[{\bf 6C\,0059+506}] There is an object in the 87GB catalogue
which is further than $2^{\prime}$ away from the 6C position, at
00 59 03.4 +50 38 51, with flux $94 \pm 11$ mJy.  From our own VLA
maps we see that this may be identified with the the north hotspot.
The spectral index of this north hotspot derived using the 87GB flux
density and the 6C flux density is too flat for this source to remain
in this sample.

\item[{\bf 6C\,0104+318}] This radio source is probably a mixture of 
background source lying in the tail of 3C31 and 3C31 tail emission
itself (Laycock 1990). It is thus excluded from the 6C* sample as it
is unlikely that it satisfies the 6C criterion for flux density.

\end{description}


\begin{figure*}
\begin{center}
\setlength{\unitlength}{1mm}
\begin{picture}(150,220)
\put(0,140){\includegraphics{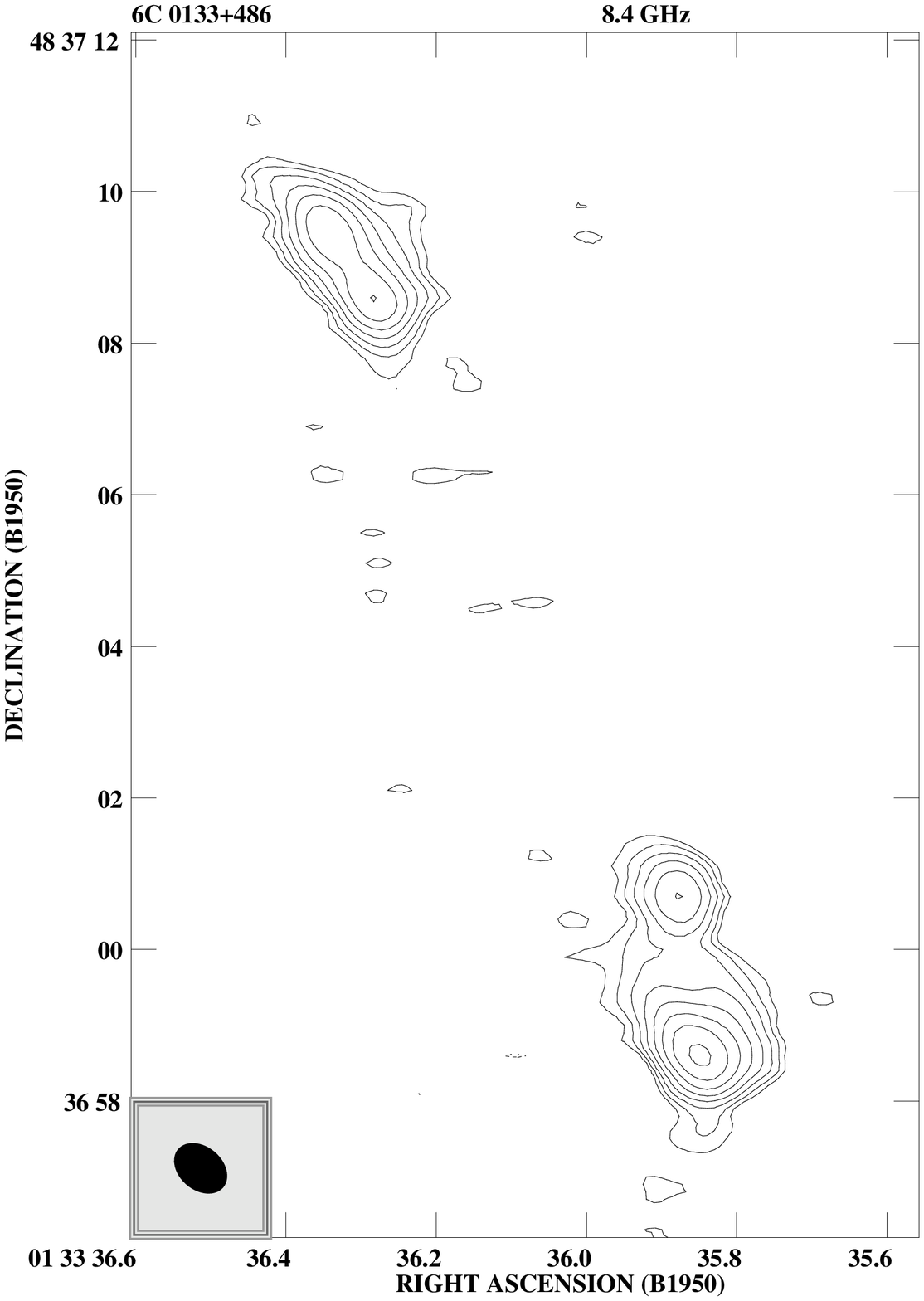}}
\put(75,135){\includegraphics{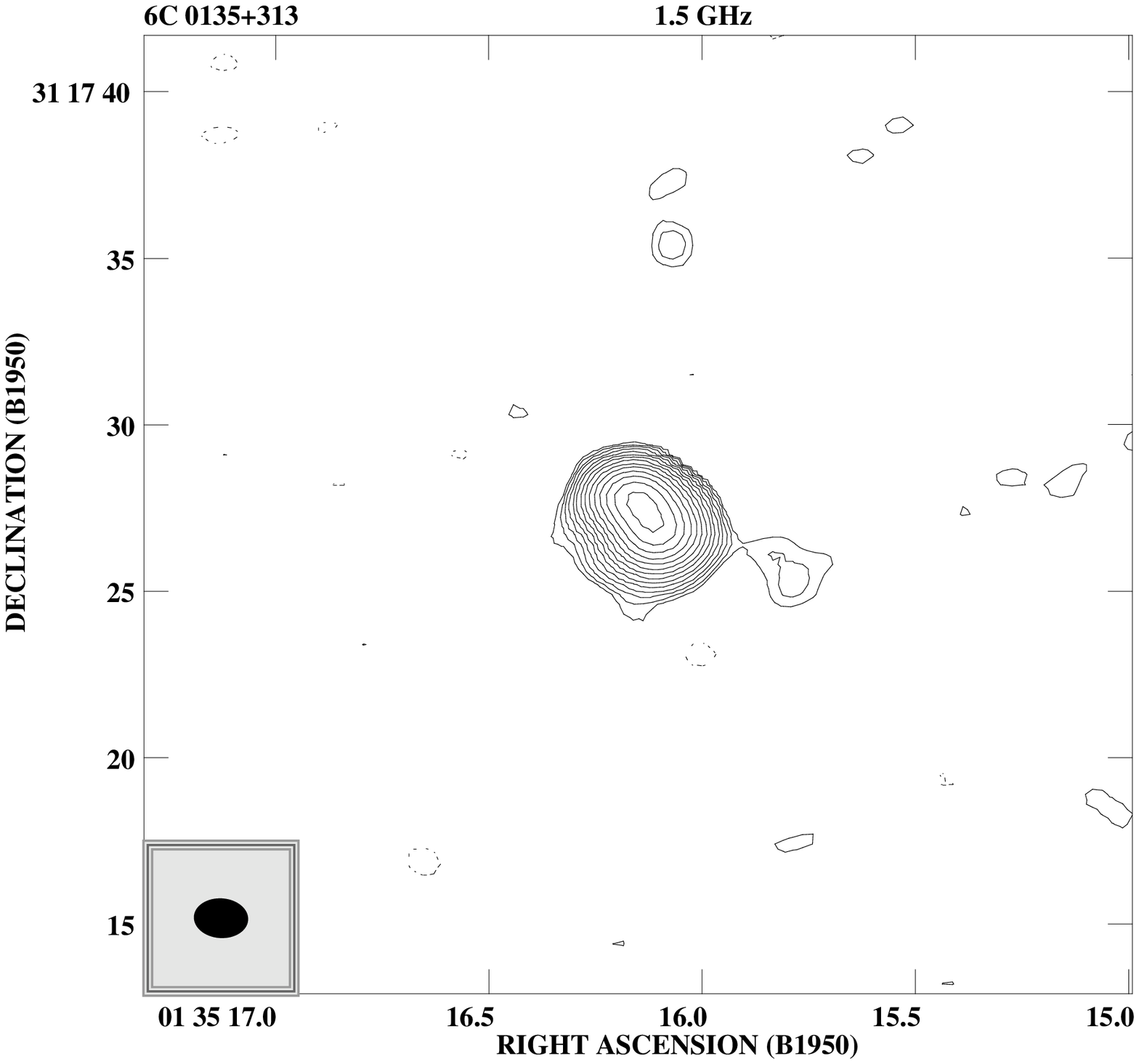}}
\put(0,70){\includegraphics{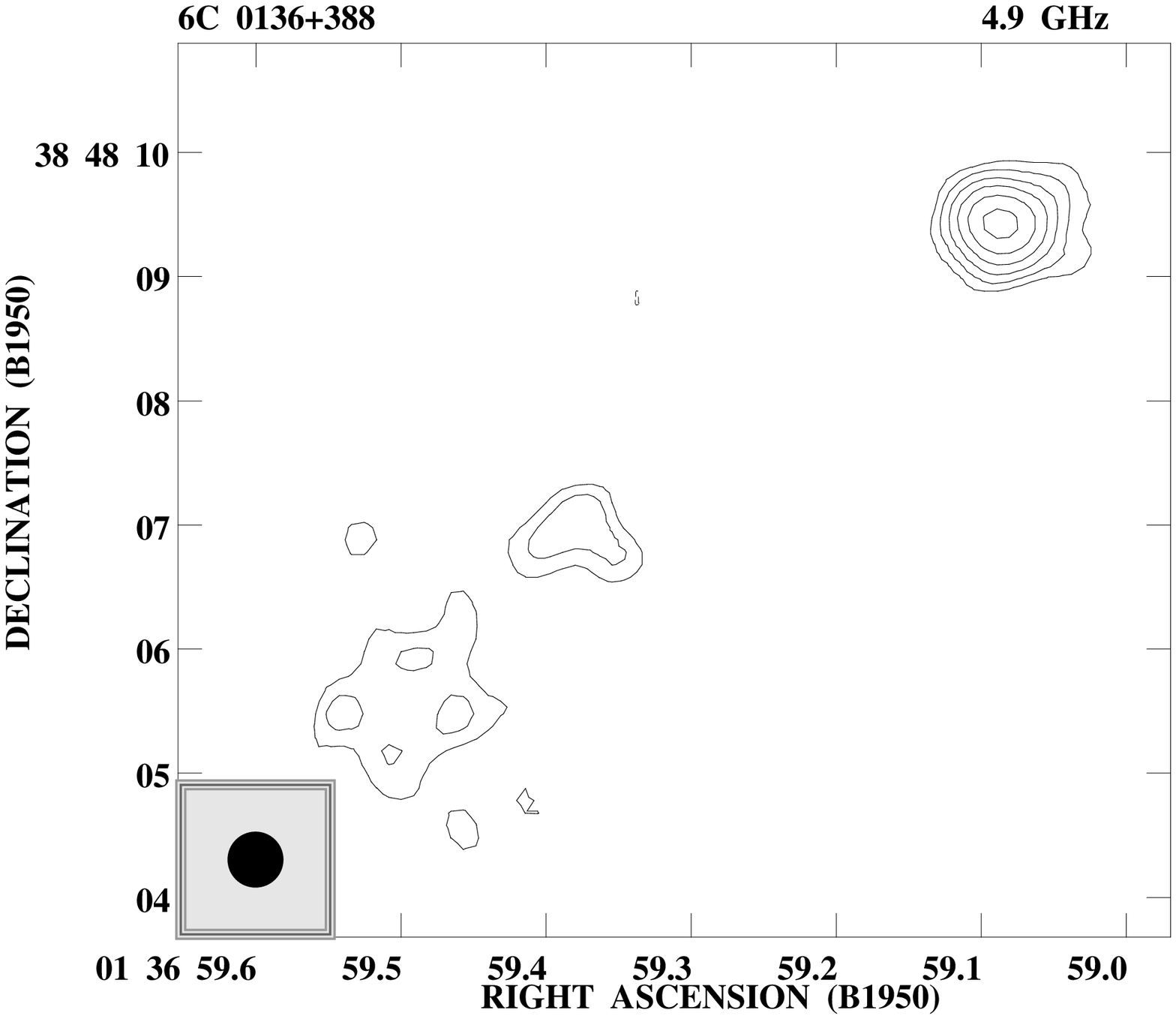}}
\put(75,65){\includegraphics{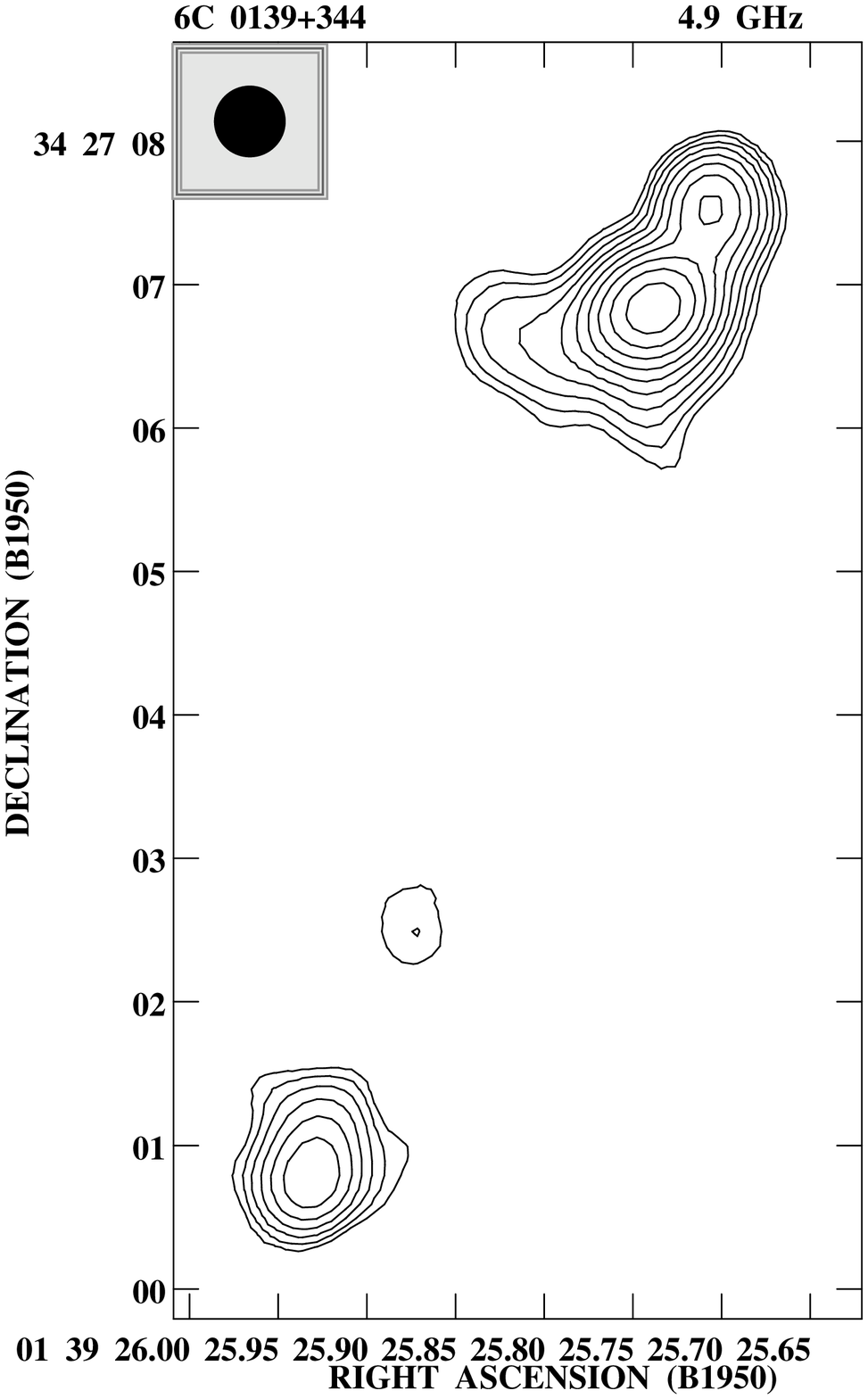}}
\put(0,-5){\includegraphics{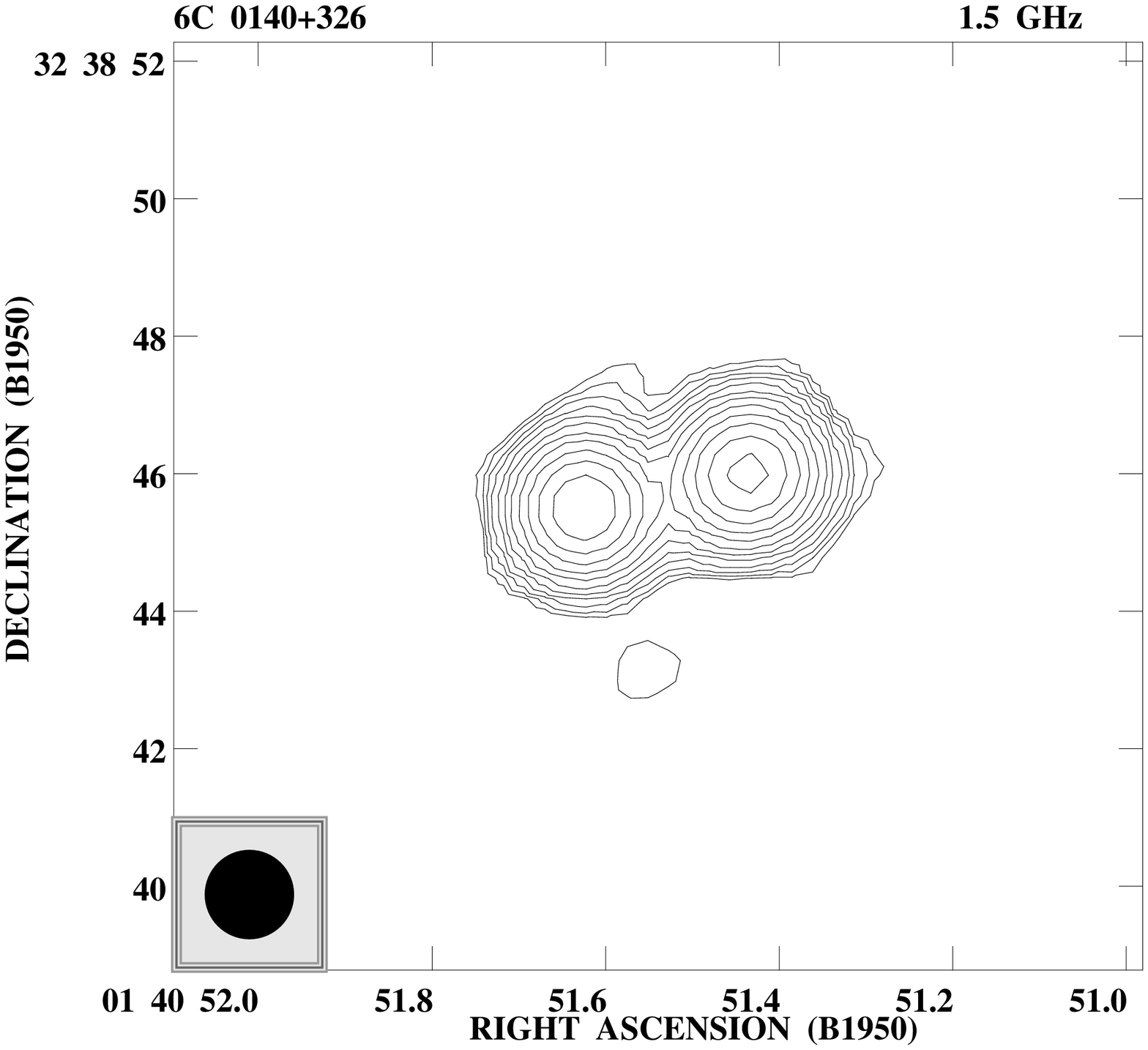}}
\put(75,-5){\includegraphics{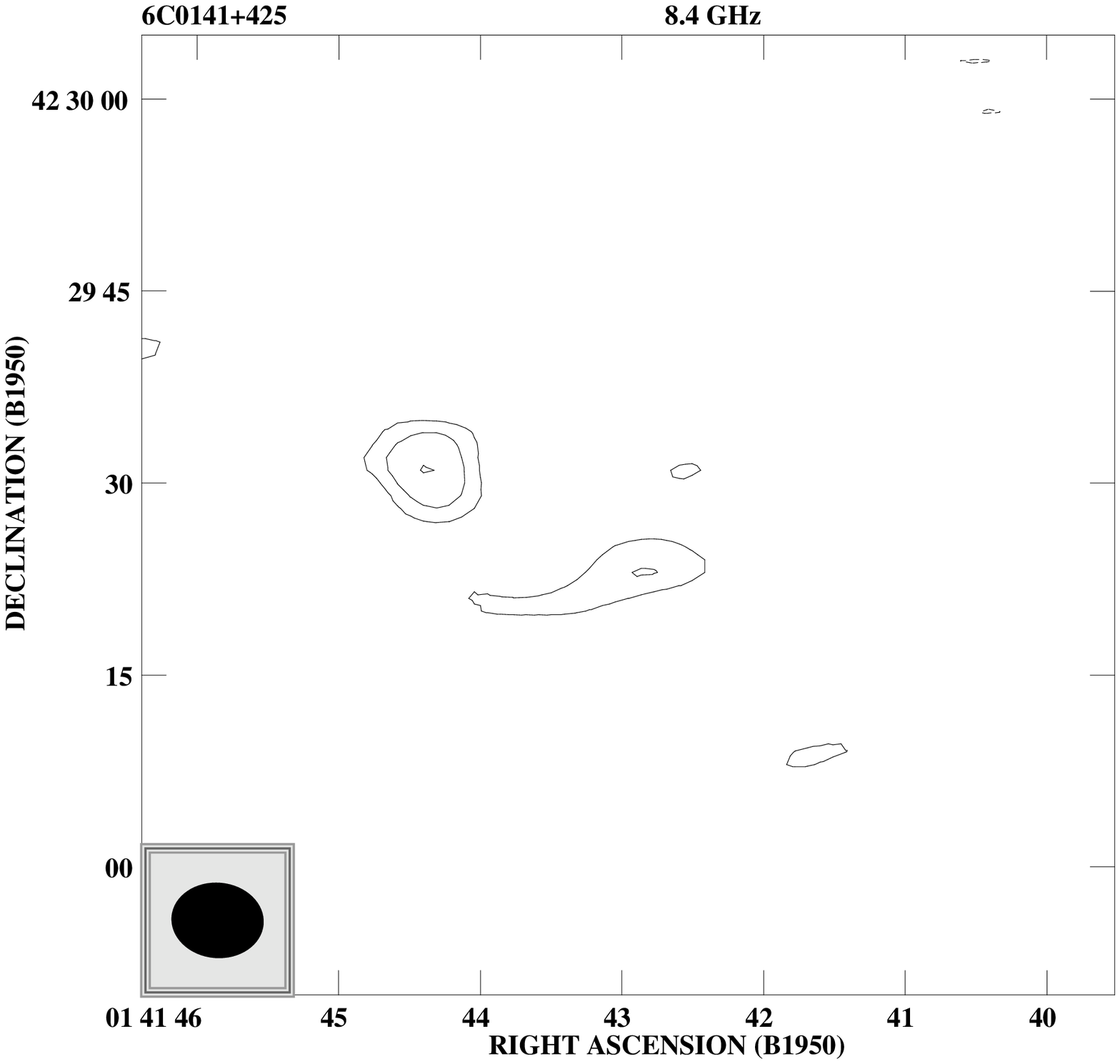}}
\end{picture}
\end{center}
{\caption[junk]{\label{fig:rad_4} Contour maps of the radio sources.}}
\end{figure*}

\begin{figure*}
\begin{center}
\setlength{\unitlength}{1mm}
\begin{picture}(150,220)
\put(0,145){\includegraphics{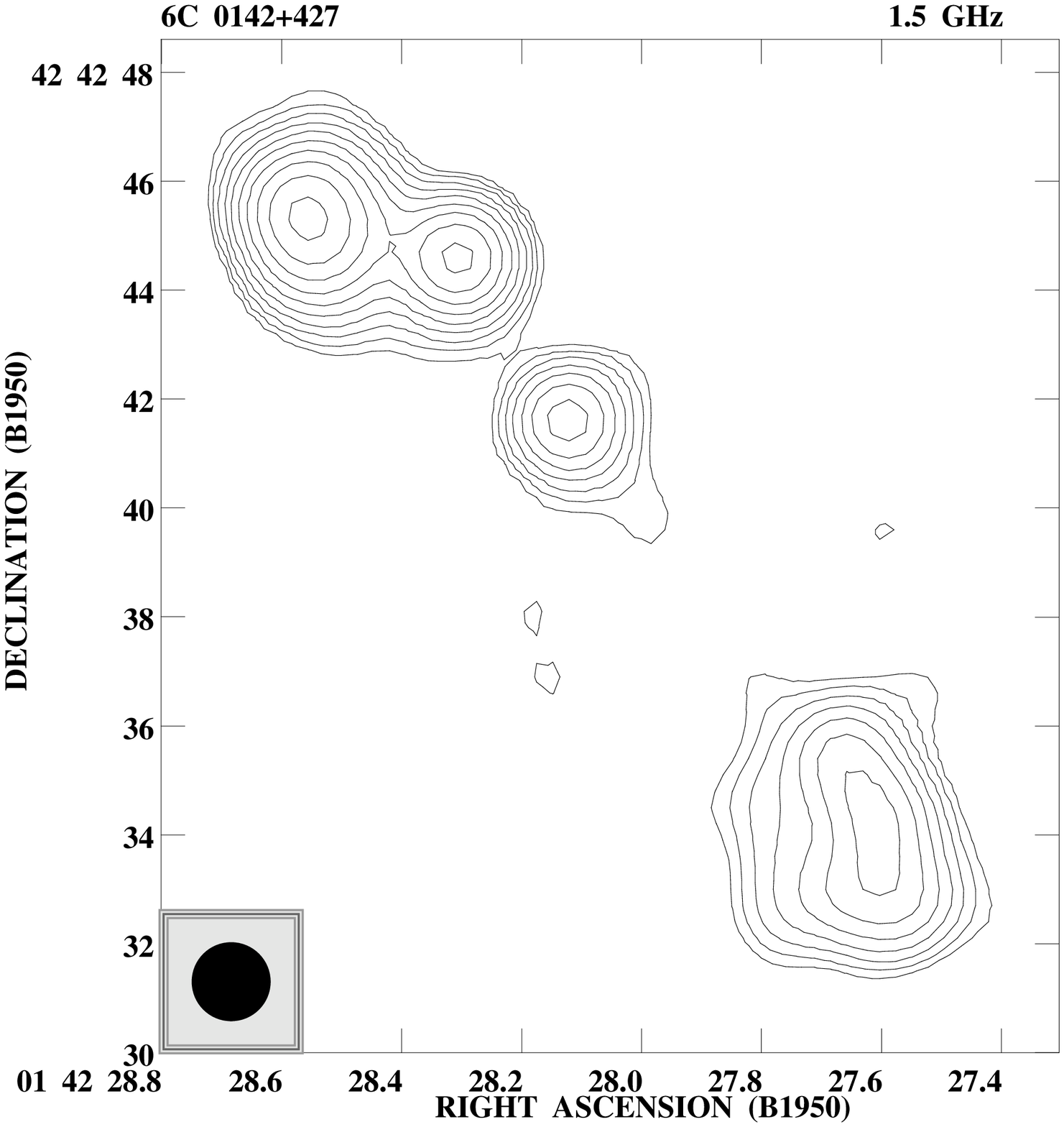}}
\put(75,138){\includegraphics{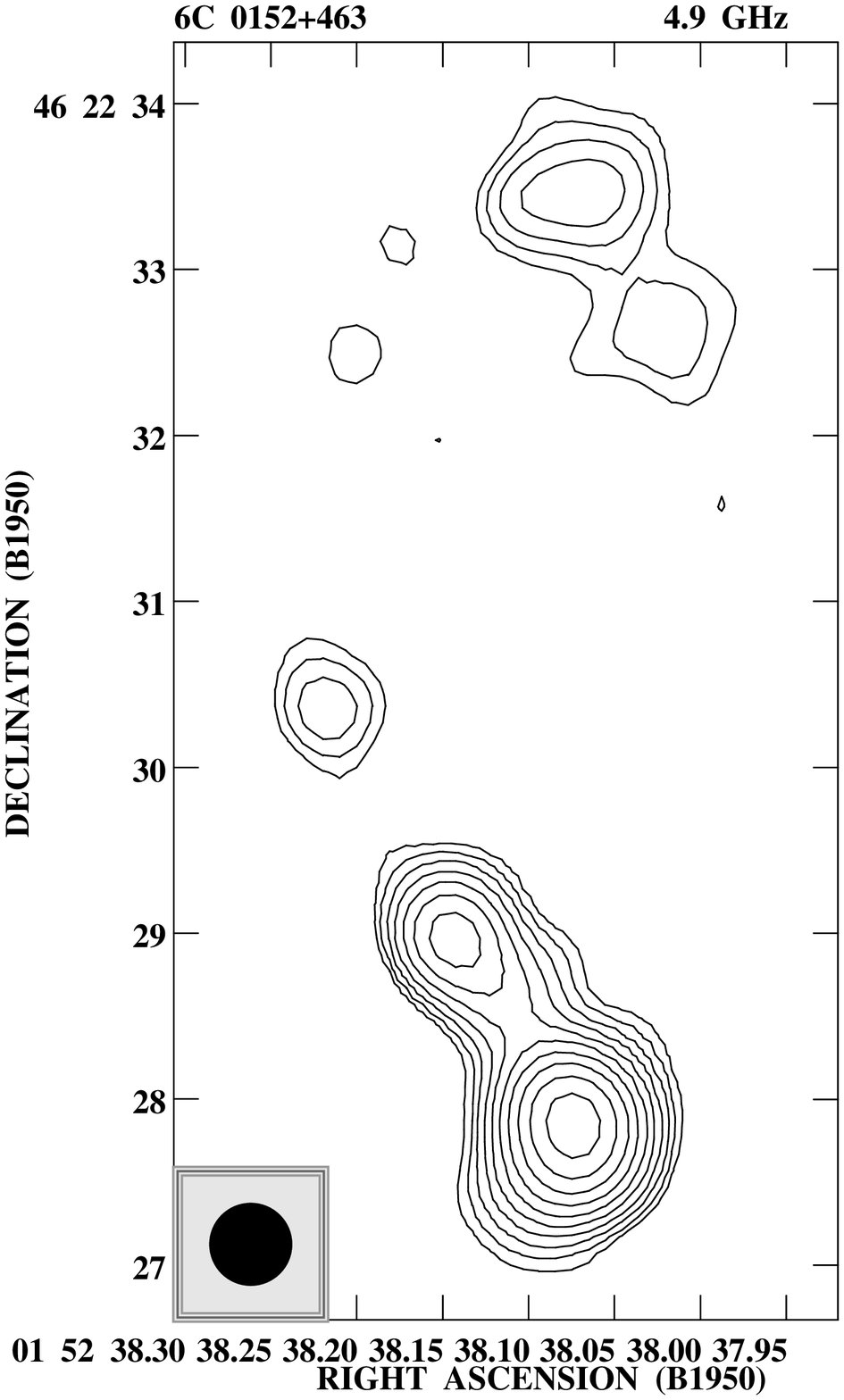}}
\put(0,70){\includegraphics{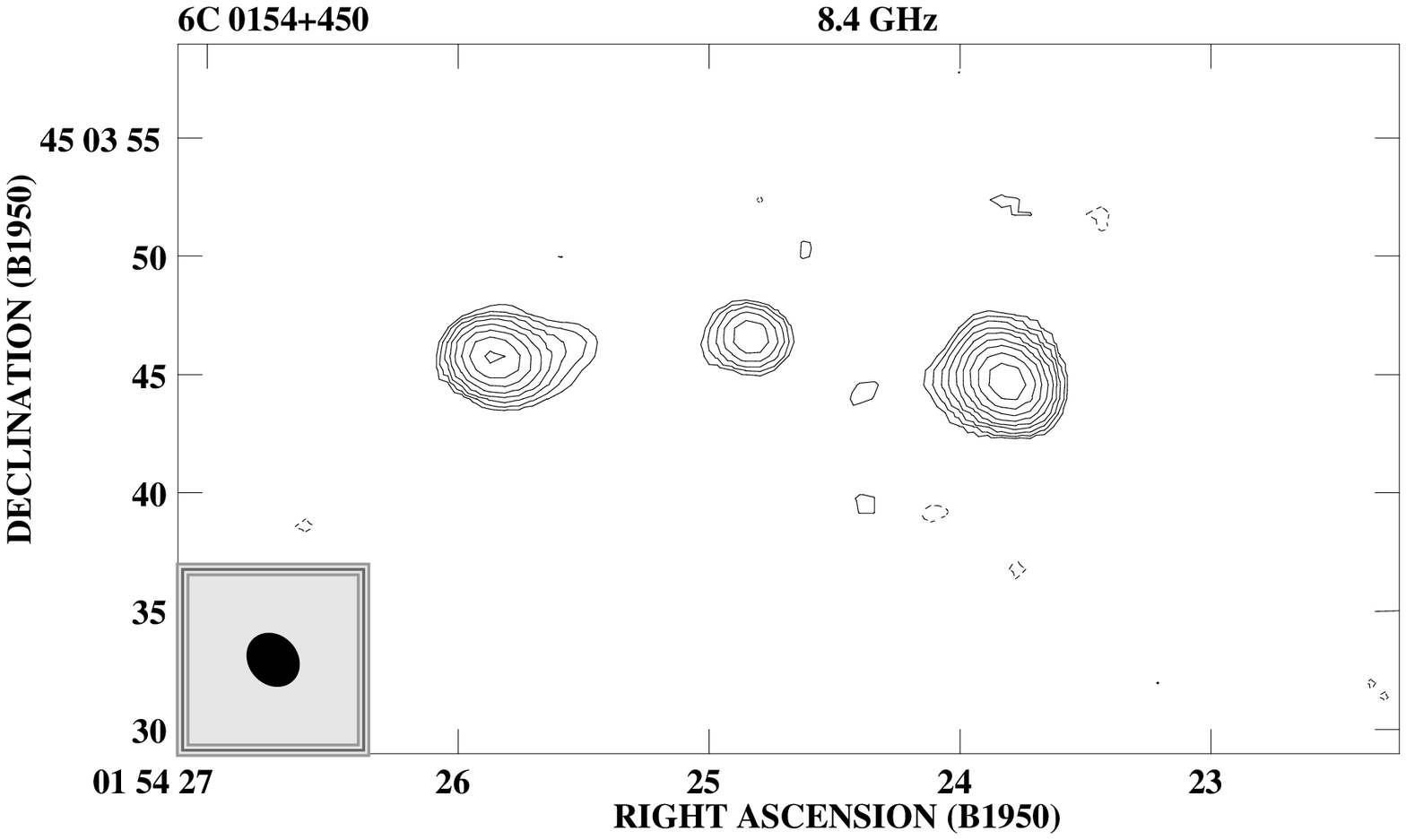}}
\put(75,70){\includegraphics{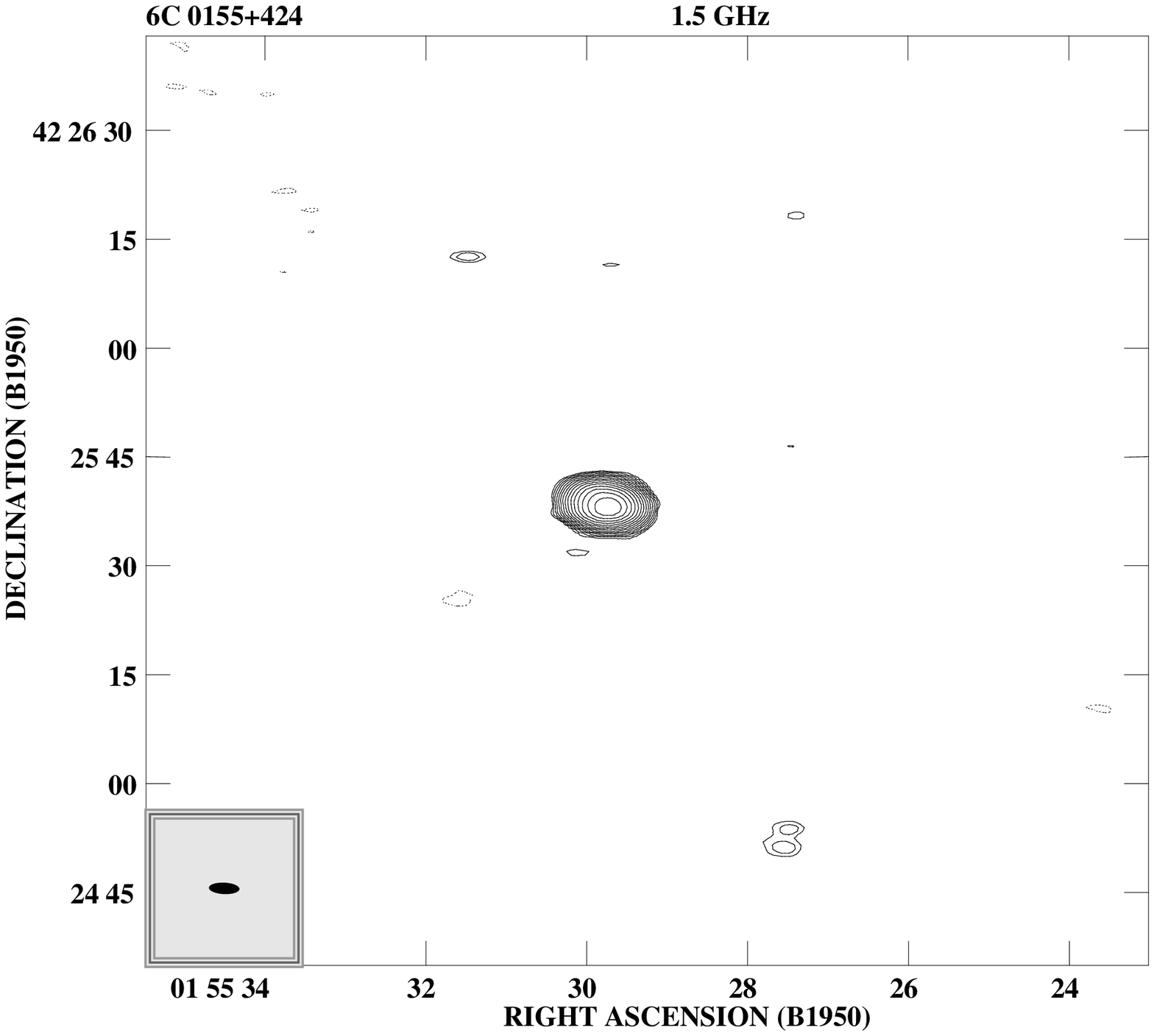}}
\put(0,-5){\includegraphics{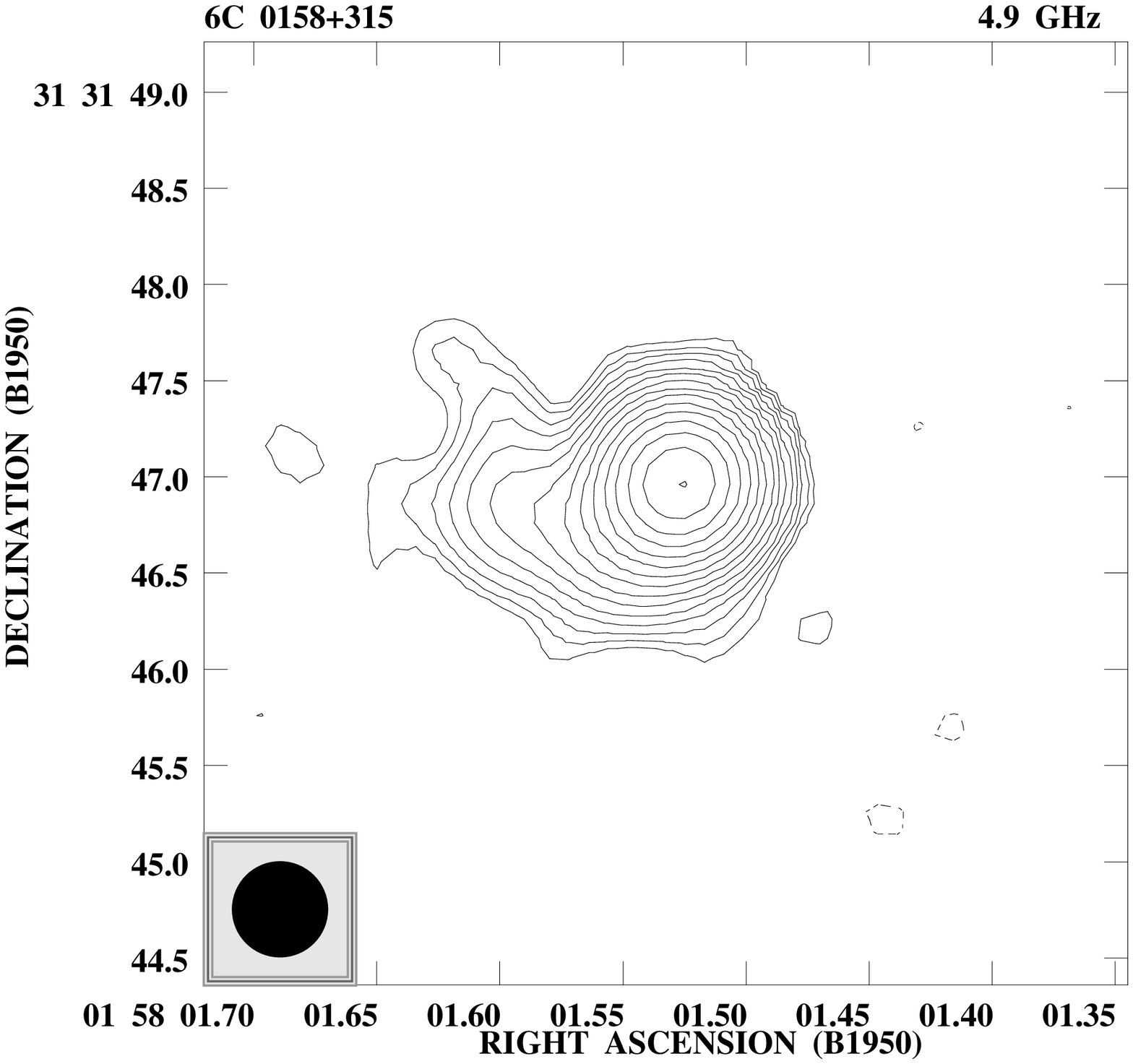}}
\put(80,-5){\includegraphics{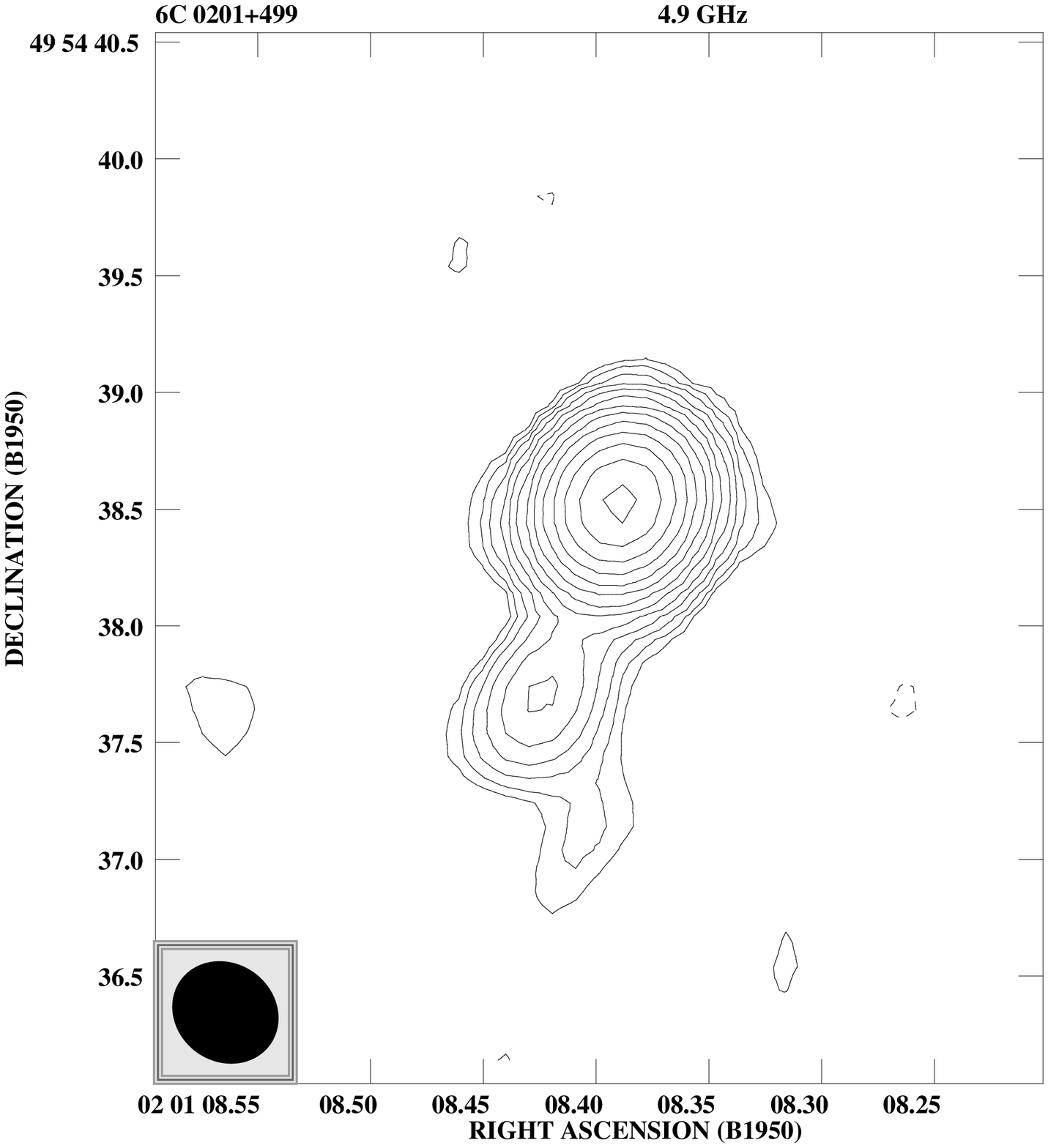}}
\end{picture}
\end{center}
{\caption[junk]{\label{fig:rad_5} Contour maps of the radio sources.}}
\end{figure*}

\section{Concluding remarks}

The search for $z > 4$ radio galaxies requires optical follow-up of
relatively faint ($S_{151} \sim 1$ Jy) surveys covering at least $\sim
0.1$ sr and hence criteria more selective than a simple radio flux
density limit.  Using data from low resolution (i.e., a few
arcminutes) surveys we have applied spectral index and angular size
criteria to filter the 279 6C sources with $0.96 < S_{151} < 2.0$ in a
0.133\,sr patch of sky to a well defined sample of 34. VLA data show
that 27 of these objects (79\%) have true angular sizes less than the
target upper value of 15 arcsec. A combination of survey and VLA
spectral data show that 30 (88\%) objects have spectral indices at
(observed) 1\,GHz greater than a (nominal) target of 0.981. When
spectra were fitted to data available from the surveys, only one of
the 34 objects showed evidence for a curved spectrum. The inclusion of
flux densities from our VLA observations revealed that 10 of the 34
objects show significant concave spectral curvature.

\section*{Acknowledgments}

It is a pleasure to thank the staff of the VLA, especially Meri
Stanley and Rick Perley for their assistance; we are particularly
grateful to Barry Clark whose allocation of discretionary VLA time
allowed this project to be started.  The VLA is a facility of the
National Radio Astronomy Observatory operated by Associated
Universities, Inc., under co-operative agreement with the National
Science Foundation. The Texas catalogue was obtained using the
Einstein On-Line Service, Smithsonian Astrophysical Observatory.  This
research has made use of the NASA/IPAC Extragalactic Database, which
is operated by the Jet Propulsion Laboratory, Caltech, under contract
with the National Aeronautics and Space Administration.  We thank
David Spence help with the figures and Gavin Dalton and Clive
Davenhall for their expert computing assistance. We are particularly
grateful to Sally Hales who provided a preliminary version of the 6C
catalogue prior to publication. Warm thanks are due to Devinder Sivia
for providing some of the software used in the fitting of radio
spectral indices and for enlightening discussions.

\end{document}